\documentstyle[12pt]{article}
\makeatletter
\@addtoreset{equation}{section}
\renewcommand{\theequation}{\thesection.\arabic{equation}}
\makeatother
\makeatletter
\@addtoreset{equation}{subsection}

\def\RR{\displaystyle \mathop{R}}

\def\V{\displaystyle \mathop{V}}
\def\G1{\displaystyle \mathop{G}}
\def\e1{\displaystyle \mathop{e}}
\def\U1{\displaystyle \mathop{U}}

\def\n1{\displaystyle \mathop{\nu}}
\def\ps1{\displaystyle \mathop{\Psi}}

\def\p1{\displaystyle \mathop{p}}
\def\J1{\displaystyle \mathop{J}}

\def\O1{\displaystyle \mathop{O}}

\def\pr{\displaystyle \mathop{\partial}}
\def\lpr{\displaystyle \mathop{\overleftarrow{\partial}}}
\def\lD1{\displaystyle \mathop{\overleftarrow{D}}}
\def\lhp{\displaystyle \overleftarrow{\hat{p}}}
\def\hgam{\displaystyle \mathop{\hat{\gamma}}}

\def\hp1{\displaystyle \mathop{\hat{p}}}
\def\hps{\displaystyle \mathop{\hat{\Psi}}}

\def\bp{\displaystyle \mathop{\bar{\Psi}}}

\def\S{\displaystyle \sum}
\def\IIn{\displaystyle \int}
\def\FFr{\displaystyle \frac}
\def\Lm{\displaystyle \lim}

\def\sig1{\displaystyle \mathop{\sigma}}

\def\h1{\displaystyle \mathop{h}}
\def\H1{\displaystyle \mathop{H}}
\def\A1{\displaystyle \mathop{A}}

\def\D1{\displaystyle \mathop{D}}
\def\B1{\displaystyle \mathop{B}}
\def\L1{\displaystyle \mathop{L}}
\def\J1{\displaystyle \mathop{J}}
\def\A1{\displaystyle \mathop{A}}
\def\M1{\displaystyle \mathop{M}}
\def\g1{\displaystyle \mathop{g}}
\def\q1{\displaystyle \mathop{q}}
\def\x1{\displaystyle \mathop{x}}
\def\s1{\displaystyle \mathop{s}}
\begin{document}
\begin{center}
{\Large {\bf The Operator Manifold Formalism. II. \\
Physical Applications}}
\end{center}
{\begin{flushright}
G.T.Ter-Kazarian\\
{\small Byurakan Astrophysical Observatory, Armenia 378433}\\
{\small E-mail:gago@bao.sci.am}\\
{\small December 20, 1998}
\end{flushright}
\begin{abstract}
Within the operator manifold approach developed in the previous part
[1]
we consider some key problems of particle physics,
particularly, we derive the Gell-Mann-Nishijima relation and flavour group.
This scheme enables to conclude that the
leptons are particles with integer electric and leptonic 
charges and free of confinement, while quarks carry
fractional electric and baryonic charges and imply the 
confinement.
We consider the unified electroweak
interactions with small number of free parameters. 
We exploit the background of the local expanded symmetry
$SU(2)\otimes U(1)$ and P-violation.
The Weinberg mixing angle is shown to have fixed value 
at $30^{o}$.
Due to the Bose-condensation of relativistic fermion pairs
the Higgs bosons arise
on an analogy of the Cooper pairs in superconductivity.
Within the present microscopic approach we predict the Kobayashi-Maskawa 
quark flavour mixing; the appearance of the CP-violation phase;
derive the mass-spectrum of leptons and quarks, 
as well as other emerging particles, and also some useful relations between
their masses.
\end{abstract}

\section {Introduction}
\label {int}
Based on the operator manifold formalism (Part I) [1] 
in the present paper we develop the microscopic approach to the
problems of field theory.
Although the complete picture is largely beyond the scope of it, 
nevertheless some key problems are already solved. 
The proliferation of lepton and quark flavours prompts us within this 
framework to consider the fields as composites. 
Certainly, it may seem foolhardy to set up such picture
in the spacetime continuum. The difficulties here are 
well-known. The first problem is closely related to the fact that the 
expected mass differences of particles would be too large 
($\geq 1TeV$). 
Another problem concerns the transformations of particles. Our idea is
to remove these difficulties by employing the multiworld geometry.
The formalism of operator multimanifold yields the multiworld geometry 
decomposing into the spacetime continuum and internal worlds. 
Thus, all fields including the leptons and
quarks, along with the spacetime 
component have nontrivial composite internal multiworld 
structure. While the various
subquarks then are defined in the corresponding internal worlds.
The microscopic structure of leptons, quarks and other particles
are governed by the various possible conjunctions of subquarks
implying concrete symmetries.\\
This article is organized as follows: 
In an enlarged framework of the  operator multimanifold we define and 
clarify the conceptual basis of subquarks and their
characteristics stemming from the various symmetries of the internal worlds
(sec.2).
This scheme  enables an insight to the key concepts of particle physics 
(sec.3-9), and to conclude that the leptons are particles
with integer electric and leptonic charges and free of confinement, 
while the quarks carry fractional electric and baryonic 
charges and imply the confinement. We derive the Gell-Mann-
Nishijima relation and the flavour group.
The multiworld structure of primary field (sec.10,11) 
is described by the gauge invariant Lagrangian involving nonlinear 
fermion interactions of the components somewhat similar to
the theory by Heisenberg and
his co-workers
[10,11], but still it will be defined on the multiworld
geometry. From this Lagrangian the 
whole complexity of the leptons, quarks and their 
interactions arises. The number of free parameters in this 
Lagrangian is reduced to primary coupling constant of the nonlinear
interaction and gauge coupling, apart from  some additional primary mass 
constants which are not essential for the conclusion.
Based on it, we consider the unified electroweak interactions (sec.12-19).
It follows that contemporary phenomenological
standard model of electroweak interactions [4-8] is an approximation
to the suggested microscopic approach.
We exploit the background of the local symmetry
$SU^{loc}(2)\otimes U^{loc}(1)$ (sec.12), the
weak hypercharge and P (mirror symmetry)-violation (sec.13). 
The Weinberg mixing angle determining the symmetry reduction coefficient 
is shown to have a value fixed at $30^{o}$.
We develop the microscopic approach to the isospinor Higgs boson with 
self-interaction and Yukawa couplings (sec.14,15). It involves Higgs boson 
as the collective excitations of bound quasi-particle pair.   
Tracing a resemblance with the Cooper pairs [17-19],
within the framework of local gauge invariance of the theory 
incorporated with the phenomenon of P-violation in weak interactions 
we suggest a mechanism providing the Bose-condensation of 
relativistic fermion pairs.
Unobserved effects produced by ready made Higgs bosons are 
suppressed.
Finally we attempt to predict the 
mixing angles in the six-quark KM model (sec.17), 
the appearance of the CP-violation phase (sec.18),
derive the 
mass-spectrum of leptons and quarks and other emerging particles, 
as well as some useful relations between their masses (sec.19).

\section {Subquarks and Subcolour Confinement}
\label {sub}
Since our discussion within this section in many respects is similar to 
that of sec.3,4 in [1], here we will be brief.
According to our strategy we admit that the $\eta$-type
(fundamental) regular structure forms a stable system 
with infinite number of distorted ${}^{i}u$-type ordinary structures of 
different species $(i=1,\ldots,N)$. In this stable system the 
flat multimanifold $G_{N}$ is realized (eq.(4.3.6) in [1])
$$
G_{N}=\G1_{\eta}\oplus\G1_{u_{1}}\oplus\cdots
\oplus \G1_{u_{N}}.
$$
We assume that the distortion rotations
(${\G1_{u_{i}}}\stackrel{\theta}{\rightarrow} \widetilde{\G1_{u_{i}}}$)
through the angles
${}^{i}\theta_{+k}$ and ${}^{i}\theta_{-k}\quad k=1,2,3$ occur 
separately in the three dimensional internal spaces 
${\RR_{u_{i}}}_{+}^{3}$ and ${\RR_{u_{i}}}_{-}^{3}$ composing six dimensional
distorted submanifold $\widetilde{\G1_{u_{i}}}=
{\RR_{u_{i}}}_{+}^{3}\oplus{\RR_{u_{i}}}_{-}^{3}$. 
Furthermore, for the rotation angles we take the ansatz
\begin{equation}
\label{eq: R5.1}
{}^{i}\theta_{\pm k}=\delta_{0\,m^{c}_{i}}{}^{i}\theta_{\pm k}(\eta)+
(1-\delta_{0\,m^{c}_{i}})\,{}^{i}\theta_{\pm k}^{c}.
\end{equation}
Here $\delta$ is the Kronecker symbol, the angles 
${}^{i}\theta_{\pm k}(\eta)$ and ${}^{i}\theta_{\pm k}^{c}$ are
respectively local and global (specified by (c)), and
\begin{equation}
\label{eq: R5.2}
m^{c}_{i}\,{}^{i}\hps_{u}(0)\equiv\hat{p}_{u_{i}}\,{}^{i}\hps_{u}(0),
\end{equation}
where $\hat{p}_{u_{i}}=i\gamma^{(\lambda\alpha)}
{\pr_{u_{i}}}_{(\lambda\alpha)}$ eq.(4.2.5) [1], ${}^{i}\hps_{u}(0)$
is the plane wave function of state of regular ordinary structure.
Then, the distortion rotations are local in 
the internal world if $m^{c}_{i}=0$, and global otherwise.
As it is exemplified in sec.4 in [1], the laws apply 
in use the wave packets constructed by superposition of the link functions
of distorted ordinary structures furnished by generalized operators of
creation and annihilation as the expansion coefficients
\begin{equation}
\label{eq: R5.3}
{\hps_{u}}(\theta_{+})=
\S_{\pm s}\IIn\frac{d^{3}p_{u_{i}}}{{(2\pi)}^{3/2}}
\left( {}^{i}{\hgam_{u}}_{(+\alpha)}^{k}
{}^{i}{\ps1_{u}}^{(+\alpha)}({}^{i}\theta_{+k})+
{}^{i}{\hgam_{u}}_{(-\alpha)}^{k}
{}^{i}{\ps1_{u}}^{(-\alpha)}({}^{i}\theta_{+k})\right), 
\end{equation}
etc. The fields ${}^{i}{\ps1_{u}}(\theta_{+k})$
and ${}^{i}{\ps1_{u}}(\theta_{-k})$
are defined on the distorted internal spaces
${\RR_{u_{i}}}_{+}^{3}$ and ${\RR_{u_{i}}}_{-}^{3}$.
The generalized expansion coefficients in eq.(2.3) imply
\begin{equation}   
\label{eq: R5.4}
<\chi_{-}\mid \{ {}^{i}{\hgam_{u}}^{(+\alpha)}_{k}(p_{u_{i}},s_{i}),\,
{}^{j}{\hgam_{u}}_{(+\beta)}^{k'}(p'_{u_{j}},s'_{j})\}\mid\chi_{-}>=
-\delta_{ij}\delta_{kk'}\delta_{ss'}\delta_{\alpha\beta}\delta^{3}
({\vec{p}}_{u_{i}}- {\vec{p'}}_{u_{i}}).
\end{equation}
The condition eq.(4.3.5) in [1] of multiworld geometry realization
reduces to
\begin{equation}
\label{eq: R5.5}
\S_{i=1}^{N}\omega_{i}
\left[ \Lm_{{}^{i}\theta_{+}\rightarrow {}^{i}\theta_{-}}
{\G1_{u_{i}}}^{\theta}_{F}({}^{i}\theta_{+}-{}^{i}\theta_{-})\right]=
\Lm_{\eta_{f}\rightarrow\eta'_{f}}
\G1_{\eta}(\eta_{f}-\eta'_{f}),
\end{equation}
provided
$\omega_{i}=\FFr{u_{i}^{2}}{u^{2}}.$
Taking into account the expression of causal Green's function
for given $(i)$
\begin{equation}
\label{eq: R5.7}
\begin{array}{l}
{\G1_{u_{i}}}^{\theta}_{F}({}^{i}\theta_{+}-{}^{i}\theta_{-})=
-i\IIn\FFr{d^{3}p_{u_{i}}}{{(2\pi)}^{3/2}}\,
{{}^{i}\ps1_{u}}_{+p}({}^{i}\theta_{+k})\,
{}^{i}{\bar{\ps1_{u}}}_{+p}({}^{i}\theta_{-k})
\theta({}^{i}u^{0}_{+}-{}^{i}u^{0}_{-})+ \\ 
+i\IIn\FFr{d^{3}p_{u_{i}}}{{(2\pi)}^{3/2}}\,
{}^{i}{\bar{\ps1_{u}}}_{-p}({}^{i}\theta_{-k})\,
{}^{i}{\ps1_{u}}_{-p}({}^{i}\theta_{+k})
\theta({}^{i}u^{0}_{-}-{}^{i}u^{0}_{+}),
\end{array}
\end{equation}
in the case
$$
\begin{array}{l}
\Lm_{
\begin{array}{l}
u_{i_{1}}\rightarrow u_{i_{2}}\\
\eta_{f}\rightarrow\eta'_{f}
\end{array}
}
\left[ 
{\G1_{u_{i_{1}}}}_{F}(u_{i_{1}}-u_{i_{2}})
/
{\G1_{\eta}}_{F}(\eta_{f}-\eta'_{f})
\right]
= 
\Lm_{
\begin{array}{l}
u'_{i_{1}}\rightarrow u'_{i_{2}}\\
\eta_{f}\rightarrow\eta'_{f}
\end{array}
}
\left[ 
{\G1_{u_{i'_{1}}}}_{F}(u'_{i_{1}}-u'_{i_{2}})
/
{\G1_{\eta}}_{F}(\eta_{f}-\eta'_{f})
\right]
=\\
\cdots=inv,
\end{array}
$$
one gets
\begin{equation}
\label{eq: R5.16}
\S_{k}\,
{}^{i}{\ps1_{u}}({}^{i}\theta_{+k})\,
{}^{i}{\bar{\ps1_{u}}}({}^{i}\theta_{-k})=
\S_{k}\,
{}^{i}{\ps1_{u}}'({}^{i}\theta'_{+k})\,
{}^{i}{\bar{\ps1_{u}}}'({}^{i}\theta'_{-k})=\cdots = inv.
\end{equation}
So, in the context of multiworld geometry it is legitimate to substitute 
a term of quark $(q_{k})$ defined in sec.3 in [1] by
subquark $({}^{i}q_{k})$.
Everything said will then remain valid,
provided we make a simple change of quarks into subquarks, the 
colours into 
subcolours.
Hence, we may think of the function 
${}^{i}{\ps1_{u}}({}^{i}\theta_{+k})$ as being $u$-component of
bispinor field of subquark $({}^{i}q_{k})$ of species $(i)$
with subcolour $k$, and respectively 
${}^{i}{\bar{\ps1_{u}}}({}^{i}\theta_{-k})$ -conjugated bispinor field 
of antisubcolour $(k)$.
The subquarks and antisubquarks may be local $({}^{i}q_{k})$ or
global $({}^{i}q_{k}^{c})$.
Then, the subquark $({}^{i}q_{k})$ is the fermion with the 
half integer 
spin and subcolour degree of freedom. 
According to eq.(2.7), they obey a condition
\begin{equation}
\label{eq: R5.19}
\S_{k}\,
{}^{i}{q}_{kp}\,{}^{i}{\bar{q}}_{kp} =
\S_{k}\,
{}^{i}{q'}_{kp}\,{}^{i}{\bar{q'}}_{kp}=
\cdots=inv.
\end{equation}
To trace a resemblance with sec.3 in [1], the internal 
symmetry group
${}^{i}G=U(1), SU(2),$\\
$ SU(3)$ enables to introduce the gauge theory
in internal world with the subcolour charges as exactly conserved 
quantities. Thereto the subcolour transformation are implemented on
subquark fields right through local and global rotation matrices of
group ${}^{i}G$ in fundamental representation. 
Due to the Noether procedure the 
conservation of global charges ensued from the global 
gauge invariance of physical system, meanwhile reinforced requirement
of local gauge invariance may be satisfied as well by introducing the 
gauge fields with the values in Lie algebra ${}^{i}\hat{g}$ of group
${}^{i}G$.

\section {The Multiworld Structures: Leptons and Quarks}
\label {Struct}
For our immediate purpose, however, we shall consider the
collection of matter fields $\Psi(\zeta)$ with nontrivial internal structure
$
\Psi(\zeta)=\ps1_{\eta}(\eta)\,\,{}^{1}\ps1_{u}(\theta_{1})
\cdots\,\,{}^{N}\ps1_{u}(\theta_{N}).
$
We suppose that
the component ${}^{i}\ps1_{u}(\theta_{i})$ is made of product of 
some subquarks and antisubquarks
$
{}^{i}\ps1_{u}({}^{i}\theta)={}^{i}\ps1_{u}(\{ {}^{i}q \}
,\{ {}^{i}\bar{q} \})$. 
The fields of subquarks ${}^{i}q$ and antisubquarks
${}^{i}\bar{q}$ form the multiplets transforming by fundamental 
${}^{i}D(j)$ and contragradient ${}^{i}\bar{D}(j)$ irreducible
representations of group ${}^{i}G$. 
Hereinafter we admit that multiworld index $(i)$ will be running
only through $i=Q,W,B,$ $s,c,b,t$ specifying the internal worlds formally
taken to denote in following nomenclature: Q-world of electric 
charge; W-world of weak interactions;
B-baryonic world of strong interactions; the s,c,b,t  are
the worlds of strangeness, charm, bottom and top.
Also we admit that $m^{c}_{i}=0$ for $i=Q,W,B$ and
$m^{c}_{i}\neq 0$ for $i=s,c,b,t$, where $m^{c}_{i}$ is given by eq.(2.2).
According to the ansatz eq.(2.1), the distortion rotations 
in the worlds Q,W and B are local ${}^{i}\theta_{\pm k}(\eta)$, while
they are global  in the worlds s,c,b,t.  
Below we introduce the fields of leptons $(l)$ and 
quarks $(q_{f})$ with different flavours f=u,d,s,c,b,t. 
To develop some feeling for this problem and to avoid irrelevant 
complications, now we may temporarily skeletonize it by taking
the leptons have following multiworld structure:
\begin{equation}
\label{eq: R7.1}
l \equiv \Psi_{l}(\zeta)=\ps1_{\eta}(\eta)\ps1_{Q}(u_{Q})\ps1_{W}(u_{W}),
\end{equation}
while the quarks are in the form
\begin{equation}
\label{eq: R7.2}
q_{f} \equiv \Psi_{f}(\zeta)=\ps1_{\eta}(\eta)
\ps1_{Q}(u_{Q})\ps1_{W}(u_{W})\ps1_{B}(u_{B})q_{f}^{c},
\end{equation}
where the superscript $(c)$ specified the worlds in which rotations are 
global
\begin{equation}
\label{eq: R7.3}
\begin{array}{l}
q_{u}^{c}=q_{d}^{c}=1, \quad q_{s}^{c}={\ps1_{s}}^{c}(u_{s}),
\quad q_{c}^{c}={\ps1_{c}}^{c}(u_{c}),\quad
q_{b}^{c}={\ps1_{b}}^{c}(u_{b}),
\quad q_{t}^{c}={\ps1_{t}}^{c}(u_{t}).
\end{array}
\end{equation}
We can take this scheme as a starting point for our considerations.\\
We assume that none of these components is accidentally zero
and implies
\begin{equation}
\label{eq: R7.4}
{}^{i}{\bar{\ps1_{u}}}^{A}(\cdots,\theta_{i_{1}},\cdots\theta_{i_{n}},\cdots)\,\,
{}^{j}{\ps1_{u}}^{B}(\cdots,\theta_{i_{1}},\cdots\theta_{i_{n}},\cdots)=
\delta_{ij}\S_{l=i_{1},\ldots,i_{n}}f^{AB}_{il}{}\,\,^{i}\left(
\bar{q}_{l}q_{l}\right),
\end{equation}
namely, the contribution of 
each individual subquark ${}^{i}q_{l}$, into the component of given world 
($i$) is determined by the {\em partial formfactor}, (invariant 
charge) $f^{AB}_{il}$.
A fascinating opportunity has turned out to infer
these  formfactors if we assume that this contribution is
completely governed by the rules of  nonlinear processes. Then we may
utilize the equations of functional autoduality [12] adopted to describe a 
wide class of nonlinear processes with a different dynamical properties, 
including also the problems of renormalization in quantum field theory [13,14]
and radiative transfer [15].  
The significant feature of these equations is their universal character,
namely they are independent from the details of concrete physical problems. 
Drawing the analogy with radiative transfer in the 
problem of diffuse reflection and transmission of reduced 
incident flux $f$, which penetrates to the depth $\tau$, we consider, in 
general, a plane-parallel medium of finite thickness $\tau+\tau_{o}$ and ask 
for the intensity diffusing reflected and the intensity diffusing 
transmitted $f^{AB}_{il}\equiv \bar{f}$ below the surface $\tau_{0}$.
Tracing the implications of the principles of invariance [16] in radiative 
transfer, the nonlinear functional equations for the scattering and
transmission functions are derived in [15], which turned out to be identical 
to those of corresponding equations of renormalization group [17]. 
The basic functional equation governing the functions 
$\bar{f}$ may be written 
\begin{equation}
\label{eq: R7.5}
\bar{f}\left( \tau,\tau_{0},f\right)=\bar{f}\left(\tau-t,\tau_{0}-t,
\bar{f}\left(t,\tau_{0},f\right)
\right), 
\end{equation}
This equation is identical to functional equation of 
renormalization group for the massive invariant charge , where 
$\tau_{0} \rightarrow \ln \FFr{m^{2}}{\lambda^{2}}$ and 
$\tau =\ln \, x$ 
\begin{equation}
\label{eq: R7.6}
\bar{f}\left( x,y,f\right)=\bar{f}\left( x/\xi,y/\xi,
\bar{f}\left( \xi,y,f\right)
\right), 
\end{equation}
provided
$
\bar{f}\left( 1,y,f\right)=f.
$
Reviewing the notation  
$\lambda$ is taken to denote the normalization momentum,
$x=\FFr{p^{2}}{\lambda^{2}}$ and  $y=\FFr{m^{2}}{\lambda^{2}}$ 
are the dimensionless momentum and mass parameters.
The other case of massless invariant charge corresponds to the particular
problem of diffuse reflection and transmission by semi-infinite medium.  
The basic differential equation can be obtained by differentiating functional
equation with respect to $\xi$ and passing to the limit $\xi =1$. In the 
aftermath one gets the well-known equation of Ovsyannikov-K$\ddot{a}$llen-
Symanzik, the solution of which can be written only if the function 
$\beta=\left.\FFr{\partial\,\bar{f}\left( \xi,y,f\right)}
{\partial\, \xi}\right|_{\xi = 1}$ is known.\\
The explicit forms of the functions $\ps1_{i}(u_{i})$ determining 
microscopic structure of concrete fields will be important subject
for discussion in the next sections. Here, we just merely point out that 
due to concrete symmetries of internal worlds, the multiworld 
structure of leptons eq.(3.1) and quarks eq.(3.2) will come into being 
if only following two conditions would be satisfied, namely,
besides the multiworld geometry realization requirement eq.(2.5)
a condition of multiworld connections must be held too,
which will be discussed in section 5.

\section {The Rotational Modes}
\label {Mod}
We assign to  each distortion rotation mode in the three dimensional spaces
${\RR_{u_{i}}}_{+}^{3}$ and ${\RR_{u_{i}}}_{-}^{3}$ 
a scale $1/3$, namely
each of the subquarks associated with the rotations 
around the axes of given world carries the corresponding charge in the 
scale $1/3$ ; antisubquark carries respectively the $(-1/3)$ charge.
In the case of the worlds C=s,c,b,t, where distortion rotations are
global and diagonal with respect to axes 1,2,3, the physical system of
corresponding subquarks is invariant under the global transformations
$f^{(3)}_{C}(\theta^{c})$ of the global unitary group $SU^{c}_{3}$:
$$
f^{(3)}_{C}=\left(\matrix{
f^{c}_{11}  & 0  & 0 \cr 
0             & f^{c}_{22}  & 0 \cr
0             & 0  & f^{c}_{33}\cr
} \right)=
\exp \left\{ -\FFr{i}{3}
\left( \matrix{
\theta^{c}_{1}  & 0  & 0 \cr
0             & \theta^{c}_{2}  & 0 \cr
0             & 0  & \theta^{c}_{3}\cr
} \right)
\right\},
$$
where 
$
f^{(3)}_{C}\left( f^{(3)}_{C}\right)^{+}=1, \quad 
\|f^{(3)}_{C}\| = 1
$.
That is 
$
\theta^{c}_{1}+\theta^{c}_{2}+\theta^{c}_{3}=0.
$
The simplest possibility is
$\theta^{c}\equiv \theta^{c}_{1}=\theta^{c}_{2}$, then one gets
\begin{equation}
\label{eq: R7.6}
f^{(3)}_{C}=\exp \left\{ -\FFr{i}{3}
\left( \matrix{
1  & 0  & 0 \cr
0             & 1  & 0 \cr
0             & 0  & -2\cr
} \right)\theta^{c}
\right\}=
\exp\left( 
-i\FFr{\lambda_{8}}{\sqrt{3}}\theta^{c}
\right)=
e^{-iY^{c}\theta^{c}},
\end{equation}
provided with the operator of hypercharge $Y^{c}$ of diagonal group 
$SU^{c}_{i}$. 
If all worlds are involved, then
$
Y^{c} = s + c + b + t.
$
The case of Q-world, which makes a substantial contribution into the 
condition of multiworld connections will be discussed in detail in
next section. We only notice that conservation of each rotation mode in 
Q- and B- worlds, where the distortion rotations are local, means that
corresponding subquarks carry respectively the conserved charges
Q and B in the scale $1/3$, and antisubquarks - $(-1/3)$ charges.
It will be provided by including the matrix $\lambda_{8}$ as
the generator  with the others in the symmetries of corresponding worlds
(Q, B), and expressed in the invariance of the system of corresponding 
subquarks under the transformations of these symmetries.
The incompatibility relations eq.(3.5.5) in [1] for global 
distortion rotations in the worlds C=s,c,b,t reduced to
$$
f^{c}_{11}f^{c}_{22}=\bar{f}^{c}_{33},\quad
f^{c}_{22}f^{c}_{33}=\bar{f}^{c}_{11},\quad
f^{c}_{33}f^{c}_{11}=\bar{f}^{c}_{22},
$$
where 
$
\|f^{(3)}_{C}\| = f^{c}_{11} f^{c}_{22} f^{c}_{33}=1, \quad
f^{c}_{ii}\bar{f}^{c}_{ii}=1 \quad \mbox{for} \quad i=1,2,3.
$
This means that two subcolour singlets are available:
$
\left( q\bar{q}\right)^{c}_{i}=inv, 
\quad
\left(q_{1}q_{2}q_{3} \right)^{c}=inv,
$
carrying the charges
$
C_{\left( q\bar{q}\right)^{c}_{i}}=0 \quad 
C_{\left(q_{1}q_{2}q_{3} \right)^{c}}=1.
$
We make use of notation
$
\left( q\bar{q}\right)^{c}_{i}\equiv {}^{c}q_{i}{}^{c}\bar{q}_{i}, 
\quad \left(q_{1}q_{2}q_{3} \right)^{c}\equiv
{}^{c}q_{1}{}^{c}q_{2}{}^{c}q_{3} .
$
Including the baryonic charge into {strong hypercharge} 
\begin{equation}
\label{eq: R8.11}
Y = B + s + c + b + t,
\end{equation}
we conclude that the hypercharge Y is a
sum of all conserved rotation modes in the internal worlds B,s,c,b,t involved
in the multiworld geometry realization condition eq.(2.5) (see eq.(5.2)).

\section {Realization of Q-World and Gell-Mann-Nishijima Relation}
\label {Rel}
The symmetry of Q-world of electric charge is assumed to be a 
local unitary symmetry $diag\left( SU^{loc}(3)\right)$ is diagonal with 
respect to axes 1,2,3. The unitary unimodular matrix 
$f^{(3)}_{Q}$ of local distortion rotations takes the form
$$
f^{(3)}_{Q}=\left( \matrix{
f^{Q}_{11}  & 0  & 0 \cr
0             & f^{Q}_{22}  & 0 \cr
0             & 0  & f^{Q}_{33}\cr
} \right)=
Q_{1}e^{-i\theta_{1}}+Q_{2}e^{-i\theta_{2}}+
Q_{3}e^{-i\theta_{3}}=
e^{-i\vec{Q}\vec{\theta}}=e^{-i\lambda_{Q}\theta_{Q}},
$$
where $f^{(3)}_{Q}\left( f^{(3)}_{Q}\right)^{+}=1, \quad 
\|f^{(3)}_{Q}\| = 1$,
provided
$$
Q_{1}=\left( \matrix{
1  & 0  & 0 \cr
0  & 0  & 0 \cr
0  & 0  & 0\cr
} \right), \quad
Q_{2}=\left( \matrix{
0  & 0  & 0 \cr
0  & 1  & 0 \cr
0  & 0  & 0\cr
} \right), \quad
Q_{3}=\left( \matrix{
0  & 0  & 0 \cr
0  & 0  & 0 \cr
0  & 0  & 1\cr
} \right),
$$
and
$
\theta_{1}+\theta_{2}+\theta_{3}=0.
$
Taking into account the scale of rotation mode,
in the other than eq.(4.1) simple case 
$
\theta_{2}=\theta_{3}=-\FFr{1}{3}\theta_{Q}, 
$
it follows that
$
\theta_{1}=\FFr{2}{3}\theta_{Q} .
$
Among the generators of the group $SU(3)$ only the matrices
$\lambda_{3}$ and $\lambda_{8}$ are diagonal. Then the matrix $\lambda_{Q}$
may be written 
$\lambda_{Q}=\FFr{1}{2}\lambda_{3}+\FFr{1}{2\sqrt{3}}\lambda_{8}.
$
Making use of the corresponding operators of the group $SU(3)$
we arrive at Gell-Mann-Nishijima relation
\begin{equation}
\label{eq: R9.7}
Q= T_{3} + \FFr{1}{2}Y,
\end{equation}
where $Q=\lambda_{Q}$ is the generator of electric charge,
$T_{3}=\FFr{1}{2}\lambda_{3}$ is the third component of isospin 
$\vec{T}$, and $Y=\FFr{1}{\sqrt{3}}\lambda_{8}$ is the hypercharge.
The eigenvalues of 
these operators will be defined later on by considering the
symmetries and microscopic structures of fundamental fields. We 
think of operators $T_{3}$ and  $Y$ as the multiworld connection
charges and of relation eq.(5.1) as the condition of
the realization of multiworld connections. Thus, during realization of 
multiworld structure the symmetries of corresponding internal
worlds must be unified into more higher symmetry including also the 
$\lambda_{3}$ and $\lambda_{8}$. Meanwhile, the realization conditions of
multiworld structure are embodied in eq.(2.5) and eq.(5.1), provided by the 
conservation law of each rotational mode 
eq.(4.1) in the corresponding internal worlds involved into condition
eq.(2.5).
For example, in the case of quarks eq.(3.2) and eq.(3.3), the eq.(2.5)
reads
\begin{equation}
\label{eq: R9.8}
\S_{i=B,s,c,b,t}\omega_{i}{\G1_{i}}^{\theta}_{F}(0)={\G1_{\eta}}_{F}(0),
\end{equation}
and according to eq.(4.2), the Gell-Mann-Nishijima relation is 
written down
\begin{equation}
\label{eq: R9.9}
Q= T_{3} + \FFr{1}{2}(B+s+c+b+t).
\end{equation}
The other case of leptons eq.(3.1) closely related to
the realization of W-world of weak interactions will be discussed in detail
separately. The realization 
condition reduces to following:
\begin{equation}
\label{eq: R9.10}
{\G1_{Q}}^{\theta}_{F}(0)={\G1_{\eta}}_{F}(0), \quad u\equiv u_{Q},
\quad (i\equiv Q)
\end{equation}
and
\begin{equation}
\label{eq: R9.11}
Q= T_{3}^{w} + \FFr{1}{2}Y^{w},
\end{equation}
where $T_{3}^{w}$ and $Y^{w}$ are respectively the operators of third 
component of weak isospin $\vec{T}^{w}$ and weak hypercharge
(sec.6,12).
The incompatibility relations eq.(3.5.5) in [1]
for the local distortion transformations in Q-world lead to
$$
f^{Q}_{11}f^{Q}_{22}=\bar{f}^{Q}_{33},\quad
f^{Q}_{22}f^{Q}_{33}=\bar{f}^{Q}_{11},\quad
f^{Q}_{33}f^{Q}_{11}=\bar{f}^{Q}_{22},
$$
where 
$
f^{Q}_{ii}\bar{f}^{Q}_{ii}=1, \quad \mbox{for} \quad i=1,2,3,\quad
\|f^{(3)}_{Q}\| = f^{Q}_{11} f^{Q}_{22} f^{Q}_{33}=1.
$
This suggests two subcolour singlets
$
\left( q\bar{q}\right)^{Q}_{i}=inv,
\quad
\left(q_{1}q_{2}q_{3} \right)^{Q}=inv.
$
The corresponding electric charges are
$
Q_{\left( q\bar{q}\right)^{Q}_{i}}=0, \quad 
Q_{\left(q_{1}q_{2}q_{3} \right)^{Q}}=1.
$
Using orthogonal unit vectors $L_{i}$
the following subcolour singlets arise:
$
\left( q\bar{q}\right)^{Q}_{i}= L_{i}
\left( q\bar{q}\right)^{Q}$
and
$
\left(q_{1}q_{2}q_{3} \right)^{Q}_{i}= L_{i}
\left(q_{1}q_{2}q_{3} \right)^{Q},
$
that
\begin{equation}
\label{eq: R9.20}
\left(q_{1}q_{2}q_{3} \right)'^{Q}_{1}=L_{1}
\left( f^{Q}_{11} f^{Q}_{22} f^{Q}_{33} \right)
\left( q_{1}q_{2}q_{3} \right)^{Q}=
\left({\bf f}^{Q}_{11}\bar{{\bf f}}^{Q}_{11}q_{1} \right)q_{2}q_{3}=
\left( q_{1}q_{2}q_{3} \right)^{Q}_{1}=inv, 
\end{equation}
etc, where 
${\bf f}^{Q}_{ii}\bar{{\bf f}}^{Q}_{ii}=
L_{i}f^{Q}_{11} f^{Q}_{22} f^{Q}_{33}=
L_{i}f^{Q}_{ii}\bar{f}^{Q}_{ii}$ for given $i$.
According to it, in singlet combinations 
$\left( q_{1}q_{2}q_{3} \right)^{Q}_{i}$ and 
$\left( q\bar{q}\right)^{Q}_{i}$ only $i$-th subquark 
undergoes distortion transformations, which can be considered as the real 
dynamical field.

\section{The Symmetries of the W- and B-Worlds}
\subsection{The W-World}
\label {Wworld}
It will be seen in sec. 12 that the symmetry of 
W-world of weak interactions is $SU^{loc}(2)\otimes U^{loc}(1)$,
invoking local group of weak hypercharge $Y^{w}$ ($U^{loc}(1)$).
However, for the present it is worthwhile to restrict oneself 
by admitting that the symmetry of W-world is simply expressed by the group
of weak isospin  $SU^{loc}(2)$. Namely, we start by considering a case of
two dimensional distortion transformations through the angles 
$\theta_{\pm}$ around two arbitrary axes in the W-world. 
In accordance with the results of sec.3 in [1], the fields of subquarks and 
antisubquarks will come in doublets, which form the basis for fundamental 
representation of weak isospin group $SU^{loc}(2)$. The doublet states
are complex linear combinations of up and down states of weak isotopic 
spin. Three possible doublets of six subquark states are
$
\left( \begin{array}{c} q_{1}\\ q_{2}
\end{array} \right)^{w},\quad
\left( \begin{array}{c} q_{2}\\ q_{3}
\end{array} \right)^{w},\quad
\left( \begin{array}{c} q_{3}\\ q_{1}
\end{array} \right)^{w}.
$
\subsection{The B-World}
The B-world is responsible for strong interactions. The internal 
symmetry group is $SU^{loc}_{c}(3)$ enabling to introduce
gauge theory in subcolour space with subcolour charges as exactly 
conserved quantities (sec.3 in [1]).
The local distortion transformations implemented
on the subquarks $(q_{i})^{B},\quad i=1,2,3$ through a
$SU^{loc}_{c}(3)$ rotation matrix $U$ in the fundamental 
representation. Taking into account a conservation of rotation 
mode (sec.4), each subquark carries $(1/3)$ baryonic  
charge, while an antisubquark - $(-1/3)$ baryonic charge.

\section {The Microscopic Structure of Leptons: \\
Lepton Generations}
\label {Lept}
After the quantitative discussion of the properties of symmetries of internal
worlds, below we attempt to show how the known fermion fields of leptons and 
quarks fit into this scheme. 
In this section we start with the leptons.
Taking into account the eq.(3.1), eq(5.4), we 
consider six possible lepton fields forming three doublets of 
lepton generations
$
\left( \begin{array}{c} \nu_{e}\\ e
\end{array} \right),\quad
\left( \begin{array}{c} \nu_{\mu}\\ \mu
\end{array} \right),\quad
\left( \begin{array}{c} \nu_{\tau}\\ \tau
\end{array} \right),
$
where
\begin{equation}
\label{eq: R11.2}
\begin{array}{l}
\left\{ \begin{array}{l}
\nu_{e}\equiv {\ps1_{\eta}}_{\nu_{e}}(\eta)\, 
(q_{1}\bar{q_{1}})^{Q}(q_{1})^{w}=
L_{e}{\ps1_{\eta}}_{\nu_{e}}(\eta)\, (q\bar{q})^{Q}(q_{1})^{w},\\
e \equiv {\ps1_{\eta}}_{e}(\eta)\,
(\overline{q_{1}q_{2}q_{3}})^{Q}_{1}(q_{2})^{w}=
L_{e}{\ps1_{\eta}}_{e}(\eta)\, (\overline{q_{1}q_{2}q_{3}})^{Q}(q_{2})^{w},
\end{array} \right. 
\\
\left\{ \begin{array}{l}
\nu_{\mu}\equiv {\ps1_{\eta}}_{\nu_{\mu}}(\eta)\, 
(q_{2}\bar{q_{2}})^{Q}(q_{2})^{w}=
L_{\mu}{\ps1_{\eta}}_{\nu_{\mu}}(\eta)\, (q\bar{q})^{Q}(q_{2})^{w},\\
\mu \equiv {\ps1_{\eta}}_{\mu}(\eta)\,
(\overline{q_{1}q_{2}q_{3}})^{Q}_{2}(q_{3})^{w}=
L_{\mu}{\ps1_{\eta}}_{\mu}(\eta)\, (\overline{q_{1}q_{2}q_{3}})^{Q}(q_{3})^{w},
\end{array} \right. 
\\ 
\left\{ \begin{array}{l}
\nu_{\tau}\equiv  {\ps1_{\eta}}_{\nu_{\tau}}(\eta)\,
(q_{3}\bar{q_{3}})^{Q}(q_{3})^{w}=
L_{\tau}{\ps1_{\eta}}_{\nu_{\tau}}(\eta)\, (q\bar{q})^{Q}(q_{3})^{w},\\
\tau \equiv  {\ps1_{\eta}}_{\tau}(\eta)\,
(\overline{q_{1}q_{2}q_{3}})^{Q}_{3}(q_{1})^{w}=
L_{\tau}{\ps1_{\eta}}_{\tau}(\eta)\, (\overline{q_{1}q_{2}q_{3}})^{Q}(q_{1})^{w}
\end{array} \right. .
\end{array} 
\end{equation}
Here $e,\mu, \tau$ are electron, muon and tau meson, 
$\nu_{e},\nu_{\mu},\nu_{\tau}$ are corresponding neutrinos,
$L_{e}\equiv L_{1},L_{\mu}\equiv L_{2},L_{\tau}\equiv L_{3},$
are leptonic charges. The leptons carry leptonic charges 
as follows: the $e$ and $\nu_{e} \rightarrow L_{e}=1$;
$\mu$ and $\nu_{\mu} \rightarrow  L_{\mu}=1$;
$\tau$ and $\nu_{\tau} \rightarrow L_{\tau}=1$.
The leptonic charges are conserved in all interactions. 
The leptons carry also the weak isospins:
$T^{w}_{3}=\FFr{1}{2}$ for  $\nu_{e},\nu_{\mu},\nu_{\tau}$; and
$T^{w}_{3}=-\FFr{1}{2}$ for $e,\mu, \tau$. Corresponding electric charges
are as follows:
$
Q_{\nu_{e}}=Q_{\nu_{\mu}}=Q_{\nu_{\tau}}=0, \quad
Q_{e}=Q_{\mu}=Q_{\tau}=-1.
$
The Q-components $\ps1_{Q}(u_{Q})$ of lepton fields eq.(3.1) are made
of singlet combinations of subquarks in Q-world. They imply subcolour
confinement eq.(5.4). Then the multiworld geometry realization condition 
is already satisfied and leptons may emerge in free combinations without any
constraint. Thus, in suggested theory there are three possible
generations of six leptons with integer electric and leptonic charges
and being free of confinement.

\section {The Microscopic Structure of Quarks: \\
Quark Generations}
\label {Quark}
We assume that the microscopic structure of 18 possible multiworld quark 
fields is as follows:
\begin{equation}
\label{eq: R12.1}
\begin{array}{l}
\left\{ \begin{array}{l}
u_{i}\equiv {\ps1_{\eta}}_{u}(\eta)\,
(q_{1}q_{2})^{Q}(q_{1})^{w}(q_{i}^{B}),\\
d_{i} \equiv  {\ps1_{\eta}}_{d}(\eta)\,
(\bar{q}_{3})^{Q}(q_{2})^{w}(q_{i}^{B}),
\end{array} \right. 
\quad
\left\{ \begin{array}{l}
c_{i}\equiv {\ps1_{\eta}}_{c}(\eta)\,
(q_{2}q_{3})^{Q}(q_{2})^{w}(q_{i}^{B})(q_{c}^{c}),\\
s_{i} \equiv {\ps1_{\eta}}_{s}(\eta)\,
(\bar{q}_{1})^{Q}(q_{3})^{w}(q_{i}^{B})(\bar{q}_{s}^{c}),
\end{array} \right. 
\\
\left\{ \begin{array}{l}
t_{i}\equiv {\ps1_{\eta}}_{t}(\eta)\,
(q_{3}q_{1})^{Q}(q_{3})^{w}(q_{i}^{B})(q_{t}^{c}),\\
b_{i} \equiv {\ps1_{\eta}}_{b}(\eta)\,
(\bar{q}_{2})^{Q}(q_{1})^{w}(q_{i}^{B})(\bar{q}_{b}^{c}),
\end{array} \right. ,
\end{array} 
\end{equation}
where the subcolour index $(i)$ runs through $i=1,2,3$, the 
$(q_{f}^{c})$ are given in eq.(3.3).
Henceforth the subcolour index will be left implicit, but always a 
summation must be extended over all subcolours in B-world.
These fields form three possible doublets of weak isospin in the W-world
$
\left( \begin{array}{c} u \\ d
\end{array} \right),\quad
\left( \begin{array}{c} c \\ s
\end{array} \right),\quad
\left( \begin{array}{c} t \\ b
\end{array} \right).
$
The quark flavour mixing and similar issues are left for treatment 
in sec.17.
The corresponding electric charges of quarks
read
$
Q_{u}=Q_{c}=Q_{t}=\FFr{2}{3},\quad
Q_{d}=Q_{s}=Q_{b}=-\FFr{1}{3},
$
in agreement with the rules governing the multiworld connections 
eq.(5.5), where the electric charge difference of up and down quarks
implies
$
\Delta Q=\Delta T^{w}_{3}=1.
$
The explicit form of structure of 
$(q_{f}^{c})$ will be discussed in next section. Here we only notice 
that all components of $(q_{f}^{c})$ are made of singlet combinations of 
global subquarks in corresponding internal worlds. They obey a condition
of subcolour confinement. According to eq.(5.2), the subcolour confinement
condition for B-world still remains to be satisfied. Due to it the
total quark fields obey confinement.
Then quarks would not be free particles. Unwanted states (since not seen)
like quarks or diquarks etc. are eliminated by construction at the very
beginning. Thus, three quark generations of six possible quark 
Fermi fields exist. They carry fractional electric and baryonic charges
and imply a confinement.
Their other charges is left to be discussed below.
Although within considered schemes the subquarks are 
defined in the internal worlds, however due to eq.(2.1) the resulting 
$\eta$-components , which we are going to deal with to describe the leptons 
and quarks defined in the spacetime continuum, are completely affected by 
them. Actually, as it is seen in subsec.3.3 and 3.4 of [1]
the rotation through the angle $\theta_{+k}$ yields
a total subquark field 
$$
{q}_{k}(\theta)=\Psi(\theta_{+k})={\ps1_{\eta}}^{0}
\ps1_{u}(\theta_{+k})
$$
where ${\ps1_{\eta}}^{0}$ is a plane wave defined on $\G1_{\eta}$.
According to subsec.3.3 in [1] and eq.(2.1), one gets
$$
{q}_{k}(\theta(\eta))={\ps1_{\eta}}^{0}{\q1_{u}}_{k}(\theta(\eta))=
{\q1_{\eta}}_{k}(\theta(\eta)){\ps1_{u}}^{0},
\quad {\q1_{\eta}}_{k}(\theta(\eta))\equiv f_{(+)}(\theta_{+k}(\eta))
{\ps1_{\eta}}^{0}, 
$$
where ${\ps1_{u}}^{0}$ is a plane wave defined on $\G1_{u}$. The
${\q1_{\eta}}_{k}(\theta(\eta))$ is considered as the subquark field
defined on flat manifold $\G1_{\eta}$ with the same quantum numbers of
${\q1_{u}}_{k}(\theta(\eta))$.
Due to it, instead of the eq.(7.1) and eq.(8.1) we can consider on 
equal footing only the resulting $\eta$-components of leptons and quarks 
having the same structures, which
enable to return to the Minkowski spacetime continuum 
$\G1_{\eta}\rightarrow M^{4}$ (subsec.2.1 in [1]).

\section {The Flavour Group $SU_{f}(6)$}
\label {flavour}
We may think of the field
component $(q_{f}^{c})$ (f=u,d,s,c,b,t) eq.(3.3) associated with the 
global distortion rotations in the worlds s,c,b,t having following 
microscopic structure with corresponding global charges:
\begin{equation}
\label{eq: R13.1}
\begin{array}{l} 
q^{c}_{u}=q^{c}_{d}=1, \quad 
\bar{q}^{c}_{s}=\left( \overline{q_{1}^{c}q_{2}^{c}q_{3}^{c}} \right)^{s}, 
\quad s =-1;\quad
q^{c}_{c}=\left( q_{1}^{c}q_{2}^{c}q_{3}^{c} \right)^{c}, 
\quad c = 1;\\
\bar{q}_{b}^{c}=\left( \overline{q_{1}^{c}q_{2}^{c}q_{3}^{c}} \right)^{b}, 
\quad b = -1;\quad
q_{t}^{c}=\left( q_{1}^{c}q_{2}^{c}q_{3}^{c} \right)^{t}, 
\quad t = 1.
\end{array} 
\end{equation}
During the realization of multiworld structure the global symmetries
of internal worlds are unified into more higher symmetry including
the generators $\lambda_{3}$ and $\lambda_{8}$ (sec.5). 
This global group is the flavour group $SU_{f}(6)$ 
unifying all symmetries $SU^{c}_{i}$ of the worlds Q,B,s,c,b,t:
$
SU_{f}(6)\supset
SU_{f}(2)\otimes SU^{c}_{B}\otimes SU^{c}_{s}\otimes 
SU^{c}_{c}\otimes SU^{c}_{b}\otimes SU^{c}_{t}
$
Then the total symmetry reads
$
G_{tot}\equiv G^{loc}\otimes G^{glob}=G^{loc}\otimes SU_{f}(6),
$
provided
$
G^{loc}\equiv SU^{loc}(3) \otimes G^{loc}_{w},
$
where $G^{loc}_{w}$ is the local symmetry of the electroweak 
interactions (sec.12).  The other important aspects
of standard model are left for investigation in the next sections. 
However, below we proceed with further exposition of our approach
to consider a gauge invariant Lagrangian of primary field 
with multiworld structure and nonlinear fermion interactions of the 
components.

\section{The Primary Field}
\label{Fund}
All fields including the leptons eq.(7.1) and
quarks eq.(8.1), along with the spacetime components have also multiworld 
components made of the various subquarks defined in the corresponding 
internal worlds, namely, the internal components are consisted of 
distorted ordinary structures
\begin{equation}
\label{eq: R14.1}
\Psi(\theta)=\ps1_{\eta}(\eta)\ps1_{Q}(\theta_{Q})\ps1_{W}(\theta_{W})
\ps1_{B}(\theta_{B})\ps1_{C}(\theta^{c}).
\end{equation}
The components $\ps1_{Q}(\theta_{Q}), \ps1_{W}(\theta_{W}),
\ps1_{B}(\theta_{B})$ are primary massless bare Fermi fields, but the 
component $\ps1_{C}(\theta^{c})$ has a mass $m^{c}_{i}\neq 0$ eq.(2.2).
We assume that this field arises from 
{\em primary field} in lowest state $(s_{0})$ with the same
field components consisted of regular ordinary structures. 
It is motivated by the argument that
the  regular ordinary structures directly could not take part 
in link exchange processes with the $\eta$-type regular structure [1].
Therefore the primary field  defined on $G_{N}$ 
\begin{equation}
\label{eq: R14.2}
\Psi(0)=\ps1_{\eta}(\eta)\ps1_{Q}(0)\ps1_{W}(0)
\ps1_{B}(0)\ps1_{C}(0)
\end{equation}
serves as the ready made frame into which the distorted
ordinary structures of the same species should be involved.
We now apply the Lagrangian of this field possessed local gauge 
invariance 
\begin{equation}
\label{eq: R14.3}
\widetilde{L}_{0}(D)=
\FFr{i}{2} \{ \bar{\Psi}_{e}(\zeta)\,
{}^{i}\gamma
{\D1_{i}}\Psi_{e}(\zeta)-
{\D1_{i}}\bar{\Psi}_{e}(\zeta)\,
{}^{i}\gamma
\Psi_{e}(\zeta) \}.
\end{equation}
The latter is in the notation eq.(4.2.3) in [1], where
$
{\D1_{i}}=
{\pr_{i}}-ig{\bf \B1_{i}}
(\zeta), $
${\bf \B1_{i}}$ are gauge fields.
Since the components $\ps1_{B}$ and $\ps1_{C}$ will be of no consequence for
a discussion, then we temporarily leave them implicit, 
namely $i=Q,W$. 
The equation of primary field of multiworld structure with nonlinear fermion 
interactions of the components may be derived from an invariant action 
in terms of local gauge invariant Lagrangian,
which looks like Heisenberg theory [2,3]
\begin{equation}
\label{eq: R14.5}
\widetilde{L}(D)=\widetilde{L}_{0}(D)+\widetilde{L}_{I}+
\widetilde{L}_{B},
\end{equation}
provided by the Lagrangians of nonlinear fermion interactions of the 
components 
$
\widetilde{L}_{I}=\sqrt{2}\widetilde{O}_{1}\otimes L_{I}, 
$
and gauge field
$
\widetilde{L}_{B}=\sqrt{2}\widetilde{O}_{1}\otimes L_{B}.
$
The binding interactions are in the form
\begin{equation}
\label{eq: R14.7}
\begin{array}{l}
L_{I} = {\L1_{Q}}_{I} + {\L1_{W}}_{I},\quad 
{\L1_{Q}}_{I}=\FFr{\lambda}{4}({\J1_{Q}}_{L}{\J1_{Q}}_{R}^{+}
+{\J1_{Q}}_{R}{\J1_{Q}}_{L}^{+}), \quad 
{\L1_{W}}_{I}=\FFr{\lambda}{2}S_{W}S_{W}^{+}, \\
L_{B}=-\FFr{1}{2}Tr(\bf B\bar{\bf B})=-\FFr{1}{2}Tr
\left( {\bf \B1_{i}}
{\bf \B1_{i}} \right),
\end{array}
\end{equation}
where
$$
\begin{array}{l}
{\J1_{Q}}_{L,R}=\V_{Q} \mp \A1_{Q},\quad
{\V_{Q}}_{(\lambda\alpha)}=\bp_{Q}\gamma_{(\lambda\alpha)}\ps1_{Q},
\quad 
{\V_{Q}}_{(\lambda\alpha)}^{+}=
{\V_{Q}}^{(\lambda\alpha)}=
\bp_{Q}\gamma^{(\lambda\alpha)}\ps1_{Q} \\
{\A1_{Q}}_{(\lambda\alpha)}=\bp_{Q}\gamma_{(\lambda\alpha)}
\gamma_{5}\ps1_{Q},
\quad 
{\A1_{Q}}_{(\lambda\alpha)}^{+}=
{\A1_{Q}}^{(\lambda\alpha)}=
\bp_{Q}\gamma^{5}\gamma^{(\lambda\alpha)}\ps1_{Q},\quad
S_{W}=\bp_{W}\ps1_{W}, 
\end{array}
$$
$\gamma_{\mu}$ and
$\gamma_{5}=i\gamma_{0}\gamma_{1}\gamma_{2}\gamma_{3}$
are Dirac matrices.
According to Fiertz theorem 
the interaction Lagrangian 
$
{\L1_{Q}}_{I}=\FFr{\lambda}{2}(VV^{+} - AA^{+})
$
may be written
$
{\L1_{Q}}_{I}=-\lambda (S_{Q}S_{Q}^{+} - P_{Q}P_{Q}^{+}),
$ provided 
$
S_{Q}=\bp_{Q}\ps1_{Q}, \quad
P_{Q}=\bp_{Q}\gamma_{5}\ps1_{Q}.
$
Hence
\begin{equation}
\label{eq: R14.9}
\widetilde{L}(D)=\sqrt{2}\widetilde{O}_{1}\otimes L(D),
\quad
L(D)=\L1_{\eta}(\D1_{\eta})-
\L1_{Q}(\D1_{Q})-
\L1_{W}(\D1_{W}),
\end{equation}
where
$$
\begin{array}{l}
\L1_{\eta}(\D1_{\eta})=
{\L1_{\eta}}'\,^{(0)}_{0}(\D1_{\eta})-
\FFr{1}{2}Tr ({\bf \B1_{\eta}}\bar{\bf \B1_{\eta}}),
\quad
\L1_{Q}(\D1_{Q})=
{\L1_{Q}}'\,^{(0)}_{0}(\D1_{Q})-
{\L1_{Q}}_{I}-
\FFr{1}{2}Tr ({\bf \B1_{Q}}\bar{\bf \B1_{Q}}),
\\
\L1_{W}(\D1_{W})=
{\L1_{W}}'\,^{(0)}_{0}(\D1_{W})-
{\L1_{W}}_{I}-
\FFr{1}{2}Tr ({\bf \B1_{W}}\bar{\bf \B1_{W}}).
\end{array}
$$
Here
$$
\begin{array}{l}
{\L1_{\eta}}_{0}'\,^{(0)}=
\FFr{i}{2} \{ 
\bar{\Psi}\stackrel{\frown}{\gamma\D1_{\eta}}\Psi-
\bar{\Psi}\stackrel{\frown}{\gamma\lD1_{\eta}}\Psi
\}= {\ps1_{u}}^{+}
{\L1_{\eta}}_{0}^{(0)}
\ps1_{u},\\
{\L1_{u}}_{0}'\,^{(0)}=
\FFr{i}{2} \{ 
\bar{\Psi}\stackrel{\frown}{\gamma\D1_{u}}\Psi-
\bar{\Psi}\stackrel{\frown}{\gamma\lD1_{u}}\Psi
\}= {\ps1_{\eta}}^{+}
{\L1_{u}}_{0}^{(0)}
\ps1_{\eta},
\end{array}
$$
and 
$$
{\L1_{\eta}}_{0}^{(0)}=
\FFr{i}{2} \{ 
\bp_{\eta}\stackrel{\frown}{\gamma\D1_{\eta}}\ps1_{\eta}-
\bp_{\eta}\stackrel{\frown}{\gamma\lD1_{\eta}}\ps1_{\eta}
\},\quad
{\L1_{u}}_{0}^{(0)}=
\FFr{i}{2} \{ 
\bp_{u}\stackrel{\frown}{\gamma\D1_{u}}\ps1_{u}-
\bp_{u}\stackrel{\frown}{\gamma\lD1_{u}}\ps1_{u}
\}.
$$
The Lagrangian eq.(10.6) has the global 
$\gamma_{5}$ and local gauge symmetries. We consider 
only $\gamma_{5}$ symmetry in Q-world, namely  
${\bf \B1_{Q}}\equiv 0$.\\
According to the operator multimanifold formalism (subsec.4 in [1]),
it is the most important to fix the mass shell of 
the stable multiworld structure (eq.(4.2.1) in [1]). Thus,
at first we must take the variation of the Lagrangian eq.(10.3) with respect 
to primary field eq.(10.2) (eq.(4.2.2) in [1]), then 
switch on nonlinear fermion interactions of the components.
In other words we take the variation of the Lagrangian eq.(10.6) with
respect to the components on the fixed mass shell.
The equations of free field (${\bf B}=0$) of multiworld structure follow 
at once
\begin{equation}
\label{eq: R14.20}
\hat{p}\Psi_{e}(\zeta)=i\stackrel{\frown}{\gamma \partial}\Psi_{e}(\zeta)=
i\,{}^{i}\gamma
{\pr_{i}}\Psi_{e}(\zeta)=0,\quad
\bar{\Psi}_{e}\lhp=-i{\pr_{i}}\bar{\Psi}_{e}
\,{}^{i}\gamma=0,
\end{equation}
which lead to separate equations for the massless components
$\ps1_{\eta}$, $\ps1_{Q}$ and $\ps1_{W}$:
\begin{equation}
\label{eq: R14.21}
\gamma^{(\lambda\alpha)}
{\p1_{\eta}}_{(\lambda\alpha)}\ps1_{\eta}=
i\gamma
{\pr_{\eta}}\ps1_{\eta}=0, \quad
\gamma
{\p1_{Q}}\ps1_{Q}=
i\gamma
{\pr_{Q}}\ps1_{Q}=0, \quad
\gamma
{\p1_{W}}\ps1_{W}=
i\gamma{\pr_{W}}\ps1_{W}=0. 
\end{equation}
The important feature is that the field equations (10.7) remain invariant
under the substitution
${\ps1_{Q}}^{(0)}\rightarrow {\ps1_{Q}}^{(m)}$,
where
${\ps1_{Q}}^{(0)}$ and ${\ps1_{Q}}^{(m)}$
are respectively the massless and massive $Q$-component fields, to which
merely the substitution
${\ps1_{\eta}}^{(0)}\rightarrow {\ps1_{\eta}}^{(m)}$ is
corresponded.
Thus, in free state the massless field components
$\ps1_{\eta}$, $\ps1_{Q}$ and $\ps1_{W}$ are independent.
Due to eq.(10.8), the Lagrangian 
\begin{equation}
\label{eq: R14.22}
L_{0}'\,^{(0)}= 
{\ps1_{u}}^{+}{\L1_{\eta}}_{0}^{(0)}{\ps1_{u}}-
{\ps1_{\eta}}^{+}{\L1_{u}}_{0}^{(0)}\ps1_{\eta}=
{\ps1_{u}}^{+}{\L1_{\eta}}_{0}^{(0)}{\ps1_{u}}-
{\ps1_{\eta}}^{+}({\L1_{Q}}_{0}^{(0)}+{\L1_{W}}_{0}^{(0)})\ps1_{\eta}
\end{equation}
reduces to the following:
\begin{equation}
\label{eq: R14.21}
L_{0}'\,^{(0)}= 
{\L1_{\eta}}_{0}^{(0)}-{\L1_{u}}_{0}^{(0)}=
{\L1_{\eta}}_{0}^{(0)}-{\L1_{Q}}_{0}^{(0)}-{\L1_{W}}_{0}^{(0)}.
\end{equation}
Thus, we implement our scheme as follows:
starting with the reduced Lagrangian 
$L_{0}'\,^{(0)}$ of free field
we shall switch on nonlinear fermion interactions of the 
components. After a generation of nonzero mass of $\ps1_{Q}$ component
in Q-world (next sec.) look for the corresponding corrections via the 
eq.(10.9) to the reduced Lagrangian eq.(10.10) of free field.
These corrections mean the interaction between the components governed by the
eq.(10.7) and eq.(10.9), and do not at all imply a mass acquiring process
for the $\eta$-component (see eq.(11.6)).

\section{A Generation of Mass of Fermions in Q-World}
\label{Qrear}
We apply the well known Nambu-Jona-Lasinio model [18] to 
generate a fermion mass in the Q-world and
start from the chirality invariant total Lagrangian of field $\ps1_{Q}:
$ 
$
\L1_{Q}={\L1_{Q}}_{0}^{(0)}-{\L1_{Q}}_{I},
$
where primary field $\ps1_{Q}$ is  massless bare spinor implying 
$\gamma_{5}$ invariance, i.e. it is an eigenstate of chirality. However,
due to interaction a rearrangement of vacuum state caused a generation of 
nonzero mass of fermion such like to appearance of energy gap in 
superconductor [9-11]. The energy gap in BCS-Bogoliubov theory of
superconductivity is created by the electron-electron interaction of Cooper
pairs. Pursuing this analogy in [18,19] it was assumed that the
mass of Dirac quasi-particle excitation is due to some interaction between
massless bare fermions, which may be considered as a self-consistent
(Hartree-Fock) representation of it. This approach based on the main 
idea that due to a dynamical instability the field theory in general 
may admit also nontrivial solutions with less symmetry than the
initial symmetry of Lagrangian, Namely, it is considered such possibility
that the field equations may possess higher symmetry, while their solutions 
may reflect some asymmetries arisen due to fact that nonperturbative 
solutions to nonlinear equations do not in general possess the symmetry
of the equations themselves. In [18,19] the solution of massive fermion 
is obtained which lack the initial $\gamma_{5}$ symmetry of the Lagrangian.
On the analogy of Gor'kov's 
theory [20,21] it is shown that if one takes into account only the
qualitative dynamical effects connected with rearrangement of vacuum state,
in addition to the trivial solution of equation of massless fermion
a real Dirac quasi-particle will satisfy the equation with non-zero 
self-energy operator $\Sigma(p,m,\widetilde{\lambda},\Lambda)$ 
depending on mass $(m)$,
coupling constant $(\widetilde{\lambda})$ and cut-off 
$(\Lambda)$. In the mean time
$\widetilde{\lambda}=\lambda\Gamma(m,\widetilde{\lambda},\Lambda)$, where $\lambda$
is a bare coupling, $\Gamma$ is the vertex function. This theory leads to the 
expression of self-energy operator $\Sigma_{Q}$ for the field $\ps1_{Q}$. 
In lowest order it is quadratically divergent, but with a cutoff can be made 
finite.
Making use of passage $\G1_{Q}\rightarrow {\M1_{Q}}^{4}$ (subsec.2.1 in [1]), 
one shall proceed directly with the calculation.
In momentum space one gets [18,19]
\begin{equation}
\label{eq: R15.2}
\begin{array}{l}
\Sigma_{Q}=m_{Q}=-\FFr{8\lambda i}{{(2\pi)}^{4}}\IIn\FFr{m_{Q}}
{p^{2}_{Q}+m^{2}_{Q}-i\varepsilon}F(p_{Q},\Lambda)d^{4}p_{Q},
\end{array}
\end{equation}
where $F(p_{Q},\Lambda)$ is a cutoff factor, $m_{Q}=\mid \Delta_{Q}\mid$,
$\Delta_{Q}=4 \lambda <{\ps1_{Q}}_{R},{\ps1_{Q}}_{L}^{+}>$,
${\ps1_{Q}}_{L,R}=\FFr{1\mp \gamma_{5}}{2}\ps1_{Q}$, $<\cdots>$ specifies
the physical vacuum averaging. 
Besides of trivial solution  $m_{Q}=0$, this equation has also nontrivial 
solution determining $m_{Q}$ in terms of $\lambda$ and $\Lambda$.
Straightforward calculations with invariant cutoff yield the relation 
$
\FFr{2\pi^{2}}{\lambda\Lambda^{2}}=1-\FFr{m^{2}_{Q}}{\Lambda^{2}}
\ln \left( \FFr{\Lambda^{2}}{m^{2}_{Q}}+ 1 \right).
$
The latter is valid only if 
$\FFr{\lambda\Lambda^{2}}{2\pi^{2}}\simeq 1$. After a vacuum 
rearrangement the total Lagrangian of initial massless bare field
${\ps1_{Q}}^{0}$ gives rise to corresponding Lagrangian ${\L1_{Q}}^{(m)}$ 
of massive field ${\ps1_{Q}}^{(m)}:\quad$
$
\L1_{Q}={\L1_{Q}}_{0}^{(0)}-{\L1_{Q}}_{I}={\L1_{Q}}^{(m)}
$
describing Dirac particle 
$
(\gamma p_{Q}-\Sigma_{Q}){\ps1_{Q}}^{(m)}=0.
$
In lowest order 
$
\Sigma_{Q}=m_{Q}\ll \lambda^{-1/2}.
$
Within the refined theory of superconductivity, the collective 
excitations of quasi-particle pairs arise in addition to the individual
quasi-particle excitations when a quasi-particle accelerated in the
medium [11, 22-26]. This leads to the conclusion [18,19] that, in
general, the Dirac quasi-particle is only an approximate description
of an entire system with the collective excitations as the stable
or unstable bound quasi-particle pairs. In a simple approximation 
there arise CP-odd excitations of zero mass as well as CP-even
massive bound states of nucleon number zero and two.
Along the same line we must substitute 
in eq.(10.7) the massless field
$\Psi^{(0)}\equiv \ps1_{\eta}{\ps1_{Q}}^{(0)}\ps1_{W}$ by massive
field $\Psi^{(m)}\equiv \ps1_{\eta}{\ps1_{Q}}^{(m)}\ps1_{W}$. Then,
we obtain
\begin{equation}
\label{eq: R15.7}
\gamma p_{Q}{\ps1_{Q}}^{(m)}=
\Sigma_{Q}{\ps1_{Q}}^{(m)}, \quad
\gamma p_{W}{\Psi}^{(m)}=0,\quad
\gamma p_{\eta}{\Psi}^{(m)}=
(\gamma p_{Q}+\gamma p_{W}){\Psi}^{(m)}=
\Sigma_{Q}{\Psi}^{(m)}. 
\end{equation}
This applies following corrections to eq.(10.9):
\begin{equation}
\label{eq: R15.8}
\begin{array}{l}
{\L1_{\eta}}_{0}'\,^{(m)}=
{\ps1_{u}}^{+}{\L1_{\eta}}_{0}^{(m)}\ps1_{u}=
{\ps1_{u}}^{+}({\L1_{\eta}}_{0}^{(0)}-
\Sigma_{Q}\bp_{\eta}\ps1_{\eta})\ps1_{u} \rightarrow
{\L1_{\eta}}_{0}^{(0)}-
\Sigma_{Q}\bar{\Psi}\Psi, \\ 
{\L1_{Q}}_{0}'\,^{(m)}=
({\ps1_{\eta}}{\ps1_{W}})^{+}
{\L1_{Q}}_{0}^{(m)}
({\ps1_{\eta}}{\ps1_{W}})=
({\ps1_{\eta}}{\ps1_{W}})^{+}({\L1_{Q}}_{0}^{(0)}-
\Sigma_{Q}\bp_{Q}\ps1_{Q})({\ps1_{\eta}}{\ps1_{W}}) \rightarrow
{\L1_{Q}}_{0}^{(0)}-
\Sigma_{Q}\bar{\Psi}\Psi,
\end{array}
\end{equation}
where suffix $(m)$ in $\Psi^{(m)}$ is left implicit. 
Such redefinition 
${\ps1_{Q}}^{(0)}\rightarrow {\ps1_{Q}}^{(m)}$
leaves the structure of that part of Lagrangian eq.(10.10) involving
only the fields $\ps1_{\eta}$ and
$\ps1_{W}$ unchanged 
\begin{equation}
\label{eq: R15.9}
\begin{array}{l}
L_{0}={\L1_{\eta}}_{0}^{(0)}-{\L1_{W}}_{0}^{(0)}=
\left( {\L1_{\eta}}_{0}^{(0)}-
\Sigma_{Q}\bar{\Psi}\Psi \right)-
\left( 
{\L1_{W}}_{0}^{(0)}-
\Sigma_{Q}\bar{\Psi}\Psi
\right)=
{\L1_{\eta}}_{0}^{(m)}-{\L1_{W}}_{0}^{(m)},
\end{array}
\end{equation}
where the component $\ps1_{Q}$ is left implicit.
The gauge invariant Lagrangian eq.(10.6)
takes the form
\begin{equation}
\label{eq: R15.10}
L(D)=\L1_{\eta}(\D1_{\eta})-
\L1_{W}(\D1_{W}),
\end{equation}
where upon combining and rearranging relevant terms we separate the
Lagrangians
\begin{equation}
\label{eq: R15.11}
\L1_{\eta}(\D1_{\eta})=
\FFr{i}{2} \{ 
\bp_{\eta}\stackrel{\frown}{\gamma\D1_{\eta}}\ps1_{\eta}-
\bp_{\eta}\stackrel{\frown}{\gamma\lD1_{\eta}}\ps1_{\eta}
\}-f_{Q}\bar{\Psi}\Psi -\FFr{1}{2}Tr({\bf \B1_{\eta}}\bar{\bf \B1_{\eta}})
\end{equation}
\begin{equation}
\label{eq: R15.12}
\L1_{W}(\D1_{W})=
\FFr{i}{2} \{ 
\bp_{W}\stackrel{\frown}{\gamma\D1_{W}}\ps1_{W}-
\bp_{W}\stackrel{\frown}{\gamma\lD1_{W}}\ps1_{W}
\}-\Sigma_{Q}\bar{\Psi}\Psi-
\FFr{\lambda}{2}S_{W}{S_{W}}^{+}-
\FFr{1}{2}Tr({\bf \B1_{W}}\bar{\bf \B1_{W}}),
\end{equation}
provided
$
f_{Q}\equiv \Sigma_{Q}(p_{Q},m_{Q},\lambda,\Lambda),\quad 
\Psi=\ps1_{\eta}\ps1_{W}. 
$
\\
The eq.(11.5) is the Lagrangian that we shall be concerned 
within the following.
The self-energy operator $\Sigma_{Q}$ takes into account the 
mass-spectrum of all expected collective excitations, which 
arise as the poles of the function $\Sigma_{Q}$.
At
$\Sigma_{Q} \simeq m_{Q}\ll \lambda,$
in eq.(11.7) we may use the approximation
$
\Sigma_{Q}\bar{\Psi}\Psi=\Sigma_{Q}{\ps1_{\eta}}^{+}\left( 
\bp_{W}\ps1_{W}\right)\ps1_{\eta}\simeq \Sigma_{Q}\bp_{W}\ps1_{W}.
$
The Lagrangians eq.(11.6) and eq.(11.7) will be further evaluated.

\section {The Electroweak Interactions: the P-Violation}
\label{Symm}
The standard model of the unified $SU(2)\otimes U(1)$ gauge theory 
of weak and electromagnetic interactions [4-8] and the colour $SU(3)$ gauge 
theory of strong interactions [27-36] have became generally acceptable 
paradigm as the view on particles and interactions in the various models of 
the same class. Their phenomenological success, which suggests
that the effective gauge group at ordinary energies is
$SU(3)\otimes SU(2)\otimes U(1)$ seemed to be proven for certain
by many experiments. 
Below we extend our approach to the pertinent concepts and ideas 
of the unified electroweak interactions.
We start with the idea that the local rotations
in W-world are occurred at very beginning only around two arbitrary axes 
(sec.6), 
namely
$
Dim W^{loc}_{(2)} = N_{(q^{w}_{1},q^{w}_{2})}=2,
$
where N is a subquark number, {\em Dim} is a dimension of local rotations.
The subquarks come up in doublets forming the basis of fundamental 
representation of weak isospin group $SU(2)$
$
q^{w}_{(2)}=
\left(\matrix{
q_{1}\cr
q_{2} \cr}\right)^{w}.
$
The transformations $U$ of local group $SU^{loc}(2)$ are implemented upon the
left- and right-handed fields
$
q_{L,R}({\vec{T}}^{w})=\FFr{1 \mp \gamma_{5}}{2}
q^{w}_{(2)}({\vec{T}}^{w}).
$
If P-symmetry holds, one has
$
q'_{L,R}=U \,q_{L,R}, \quad U\in SU^{loc}(2).
$
But, under such circumstances  
the weak interacting particles could not be realized, because the
condition of multiworld connections eq.(5.5) 
still not yet satisfied.
A simple way of effecting a reconciliation is to assume that
during a realization of weak interacting fermions the local
rotations in W-world always occur around all three axes. 
That is, instead of initial symmetry, we admit the spanning
$
W^{loc}_{(2)}\rightarrow W^{loc}_{(3)}
$
where
$
Dim W^{loc}_{(3)} = 3 \neq N_{(q^{w}_{1},q^{w}_{2})}=2.
$
Taking into account that at the very beginning all subquark fields 
in W-world are massless,
we cannot rule out the possibility that they are transformed independently.
On the other hand, when this situation prevails the spanning 
$
W^{loc}_{(2)}\rightarrow W^{loc}_{(3)}
$
must be occurred compulsory
in order to provide a necessary background for the condition eq.(5.5)
to be satisfied.
The most likely attitude here is that doing away
this shortage the subquark fields $q_{L_{1}},q_{L_{2}},q_{R_{1}},$
and $q_{R_{2}}$ tend to give rise to triplet.
The three dimensional effective space 
$W^{loc}_{(3)}$ will then arise
\begin{equation}
\label{eq: R16.7}
W^{loc}_{(2)} \ni  q^{w}_{(2)}\,
({\vec{T}}^{w}=\FFr{1}{2})
\rightarrow q^{w}_{(3)}=
\left(\matrix{
q_{R}({\vec{T}}^{w}=0)\cr
\cr
q_{L}({\vec{T}}^{w}=\FFr{1}{2})\cr}
\right)= 
\left(\matrix{
q^{w}_{3}\cr
q^{w}_{1}\cr
q^{w}_{2}\cr}
\right) \equiv
\left(\matrix{
q_{R_{2}}\cr
q_{L_{1}}\cr
q_{L_{2}}\cr}
\right)\in W^{loc}_{(3)}.
\end{equation}
The latter holds if violating initial P-symmetry 
the components $q_{R_{1}},q_{R_{2}}$ still remain in isosinglet 
states, namely the components $q_{L}$ are forming isodoublet while
$q_{R}$ is a isosinglet:
$
q_{L}\,({\vec{T}}^{w}=\FFr{1}{2}),\quad
q_{R}\,({\vec{T}}^{w}=0).
$
So, the mirror symmetry is broken.
Corresponding local transformations are implemented upon triplet
$
{q^{w}}'_{(3)}=f_{W}^{(3)}q^{w}_{(3)},
$
where the unitary matrix of three dimensional local rotations reads
$$
f^{(3)}_{W}=\left( \matrix{
f_{33}  & 0  & 0 \cr 
0       & f_{11}  & f_{12} \cr
0       & f_{21}  & f_{22} \cr
} \right)^{w}.
$$
Making use of incompatibility relations eq.(3.5.5) in [1] one gets
\begin{equation}
\label{eq: R16.11}
\|f^{(3)}_{W}\|=f_{33}(f_{11}f_{22}-f_{12}f_{21})=f_{33}
\varepsilon_{123}\varepsilon_{123}
\|f^{(3)}_{W}\| f_{33}^{*},
\end{equation}
or
$
f_{33}f_{33}^{*}=1.
$
That is
$
f_{33}=e^{-i\beta}.
$
While
$$
\|f^{(2)}_{W}\|=f_{11}f_{22}-f_{12}f_{21}=
\|f^{(3)}_{W}\| f_{33}^{*}=
\|f^{(3)}_{W}\| e^{i\beta},
$$
and due to condition $\|f^{(3)}_{W}\|=1$ it reads
$
\|f^{(2)}_{W}\|=e^{i\beta}\neq 1.
$
Thus, the initial symmetry $SU^{loc}(2)$ is broken. Restoring it
the fields $q_{L}$ must be undergone to additional transformations
$$
f^{(2)}_{W}\rightarrow{f^{(2)}_{W}}'=\left( \matrix{
f_{11}e^{-i\frac{\beta}{2}}  & f_{12}e^{-i\frac{\beta}{2}} \cr
f_{21}e^{-i\frac{\beta}{2}}  & f_{22}e^{-i\frac{\beta}{2}} \cr
} \right)^{w}
$$
in order to satisfy the unimodularity condition of matrix of the group
$SU^{loc}(2)$, namely
$
\|{f^{(2)}_{W}}'\|=\| f^{(2)}_{W}\| e^{-i\beta}=1, \quad
{f^{(2)}_{W}}'\in SU^{loc}(2).
$
Then, the expanded group of local rotations in W-world arises
\begin{equation}
\label{eq: R16.18}
f^{(3)}_{exp}=\left( \matrix{
e^{-i\beta}  & 0  & 0 \cr
0  & f_{11}e^{-i\frac{\beta}{2}}  & f_{12}e^{-i\frac{\beta}{2}} \cr
0  & f_{21}e^{-i\frac{\beta}{2}}  & f_{22}e^{-i\frac{\beta}{2}} \cr
} \right)\in SU^{loc}(2)\otimes U^{loc}(1),
\end{equation}
where
$
U=e^{-i\vec{T}^{w}\vec{\theta^{w}}}\in SU^{loc}(2), \quad
U_{1}=e^{-iY^{w}\theta_{1}}\in U^{loc}(1).
$
Here $U^{loc}(1)$ is the group of weak hypercharge $Y^{w}$ taking
the following values for left- and right-handed subquark fields:
$
q_{R}:Y^{w}=0,-2,\quad q_{L}:Y^{w}=-1.
$
Whence 
$
q'^{w}_{(3)}=f_{exp}^{(3)}q^{w}_{(3)},
$
namely
$$
q'_{L}= {\displaystyle e^{-i\vec{T}^{w}\vec{\theta}^{w}-iY^{w}_{L}\theta_{1}}}
q_{L}, \quad
q'_{R}={\displaystyle e^{-iY^{w}_{R}\theta_{1}}}q_{R}.
$$
\section{The Reduction Coefficient and the Weinberg Mixing Angle}
\label{Reduc}
The realization of weak interacting 
particles always has incorporated with the spanning eq.(12.1).
This implies P-violation in W-world expressed in the
reduction of initial symmetry group of local transformations of
right-handed components $q_{R}$, namely 
\begin{equation}
\label{eq: R17.1}
\left[ SU(2)\right]_{R}\rightarrow \left[ U(1)\right]_{R},
\end{equation}
where subscript $(R)$ specified the transformations implemented upon
right-handed components. The invariance of physical system of the fields
$q_{R}$ under initial group $\left[ SU(2)\right]_{R}$ may be
realized as well by introducing non-Abelian massless vector gauge fields
${\bf A}=\vec{T}^{w}\vec{A}$ with the values in Lie algebra 
of the group $\left[ SU(2)\right]_{R}$. 
Under a reduction eq.(13.1) the coupling 
constant $(g)$ changed into $(g')$ specifying the interaction strength
between $q_{R}$ and the Abelian gauge field $B$ associated 
with the local group $\left[ U(1)\right]_{R}$. Thereto
$
g = g'\tan \theta_{w},
$
where $\theta_{w}$ is the Weinberg mixing angle, in terms of which the
reduction coefficient reads
$
r_{p}=\FFr{g-g'}{g+g'}=\FFr{1-\tan \theta_{w}}{1+\tan \theta_{w}}.
$
To define the $r_{p}$ we consider the interaction vertices corresponding
to the groups $\left[ SU(2)\right]_{R}:$
$
g {\bf A}\bar{q}_{R}\gamma\FFr{\bf \tau}{2}q_{R}
\quad$
and $\quad \left[ U(1)\right]_{R}:\quad $
$
g' B\bar{q}_{R}\gamma\FFr{Y^{w}}{2}q_{R}.
$
To notice that the matrix $\FFr{\lambda_{8}}{2}$ is in the same 
normalization scale as each of the matrices $\FFr{\lambda_{i}}{2}\quad 
(i=1,2,3):$
$
Tr\left( \FFr{\lambda_{8}}{2}\right)^{2}=
Tr\left( \FFr{\lambda_{i}}{2}\right)^{2}=\FFr{1}{2}
$
the vertex scale reads
$
(\mbox{Scale})_{SU(2)}=g \FFr{\lambda_{3}}{2},
$
which is equivalent to $g \FFr{\lambda_{8}}{2}$.
It is obvious that per generator scale could not be changed under the
reduction eq.(13.1), i.e.
$
\FFr{(\mbox{Scale})_{SU(2)}}{N_{SU(2)}}=
\FFr{(\mbox{Scale})_{U(1)}}{N_{U(1)}},
$
where $N_{SU(2)}$ and $N_{U(1)}$ are the numbers of generators 
respectively in the groups $SU(2)$ and $U(1)$. 
Hence
$
(Scale)_{U(1)}=\FFr{1}{3}(Scale)_{SU(2)}.
$
Stated somewhat differently, the normalized vertex
for the group $\left[ U(1)\right]_{R}$  reads
$
\FFr{1}{3}g B\bar{q}_{R}\gamma\FFr{\lambda_{8}}{2}q_{R}.
$
In comparing the coefficients can then be equated 
$
\FFr{g'}{g} = \tan \theta_{w}=\FFr{1}{\sqrt{3}},
$
and
$
r_{p}\simeq 0.27.
$
We may draw a statement that during the realization of multiworld structure
the spanning eq.(12.1) compulsory occurred, which is the source of
P-violation in W-world incorporated with the reduction eq.(13.1). The
latter is characterized by the Weinberg mixing angle 
with the value fixed at $30^{0}$.

\section {The Emergence of Composite Isospinor-Scalar 
Meson}
\label{Mes}
The field $q^{w}_{(2)}$ is the W-component of total field
$
q_{(2)}={\q1_{\eta}}_{(2)}{\q1_{W}}_{(2)}(\equiv
{\ps1_{\eta}}_{(2)}{\ps1_{W}}_{(2)}),
$
where the field component $\q1_{Q}(\equiv \ps1_{Q})$ is left implicit.
Instead of it, below we introduce the additional suffix $(Q=0,\pm)$ specifying 
electric charge of the field.
At the beginning there is an absolute 
symmetry between the components
$
q_{1}={\q1_{\eta}}_{1}{\q1_{W}}_{1}$ and
$
q_{2}={\q1_{\eta}}_{2}{\q1_{W}}_{2}$
Hence, left-
and right-handed components of fields may be written
$$
q_{1L}={\q1_{\eta}}_{1L}^{(0)}{\q1_{W}}_{1L}^{(-)}, \quad
q_{2L}={\q1_{\eta}}_{2L}^{(-)}{\q1_{W}}_{2L}^{(0)},\quad
q_{1R}={\q1_{\eta}}_{1R}^{(0)}{\q1_{W}}_{1R}^{(-)}, \quad
q_{2R}={\q1_{\eta}}_{2R}^{(-)}{\q1_{W}}_{2R}^{(0)}.
$$
On the example of one lepton generation $e$ and $\nu$ we shall
exploit the properties of these fields. An implication
of other fermion generations will be straightforward.
Then
$$
\begin{array}{l}
{\q1_{\eta}}_{L}=
\left(\matrix{
{\q1_{\eta}}_{1L}^{(0)}\cr
\cr
{\q1_{\eta}}_{2L}^{(-)}\cr}
\right)\equiv L =
\left(\matrix{
\nu_{L}\cr
e^{-}_{L}\cr}
\right), \quad
{\q1_{\eta}}_{R}=\left({\q1_{\eta}}_{1R}^{(0)},\,{\q1_{\eta}}_{2R}^{(-)}
\right)\equiv
R =\left(\nu_{R},e^{-}_{R}\right),\\\\
{\q1_{W}}_{L}=
\left(\matrix{
{\q1_{W}}_{2L}^{(0)}\cr
\cr
{\q1_{W}}_{1L}^{(-)}\cr}
\right),\quad
{\q1_{W}}_{R}=\left({\q1_{W}}_{1R}^{(-)},\,{\q1_{W}}_{2R}^{(0)}\right).
\end{array}
$$
We evaluate the term $f_{Q}\bar{\Psi}\Psi$ in the Lagrangian eq.(11.6)
as follows:
$$
\bar{\Psi}\Psi=\Psi^{+}_{L}\Psi_{R}+\Psi^{+}_{R}\Psi_{L}
\equiv q^{+}_{L}q_{R}+q^{+}_{R}q_{L},
$$
provided
$$
\begin{array}{l}
q^{+}_{L}q_{R}={\q1_{\eta}}_{L}^{+}{\q1_{W}}_{L}^{+}
{\q1_{W}}_{R}{\q1_{\eta}}_{R}=
{\bar{\q1_{\eta}}}_{L}\left( 
\gamma^{0}{\q1_{W}}_{L}^{+}{\q1_{W}}_{R}
\right){\q1_{\eta}}_{R}=\bar{L}\varphi R,\\
q^{+}_{R}q_{L}={\q1_{\eta}}_{R}^{+}{\q1_{W}}_{R}^{+}
{\q1_{W}}_{L}{\q1_{\eta}}_{L}=
{\bar{\q1_{\eta}}}_{R}\left( 
\gamma^{0}{\q1_{W}}_{R}^{+}{\q1_{W}}_{L}
\right){\q1_{\eta}}_{L}=\bar{R}\varphi^{+} L.
\end{array}
$$
Hence, for appropriate values of the parameters, this term causes
$$
\bar{\Psi}\Psi=\bar{q}_{(2)}q_{(2)}=\bar{L}\varphi R+\bar{R}\varphi^{+} L,
$$
where the isospinor-scalar meson field $\varphi$ reads
$$
\varphi \equiv  
\gamma^{0}{\q1_{W}}_{L}^{+}{\q1_{W}}_{R},\quad
\varphi^{+}\equiv 
\gamma^{0}{\q1_{W}}_{R}^{+}{\q1_{W}}_{L}.
$$
A calculation gives
$$
\varphi=
\left(\matrix{
\varphi_{1}\cr
\varphi_{2}\cr}
\right), \quad
\varphi_{1}\equiv \left( {\q1_{W}}_{1L}^{(-)}\right)^{+}
{\q1_{W}}_{R},\quad
\varphi_{2}\equiv \left( {\q1_{W}}_{2L}^{(0)}\right)^{+}
{\q1_{W}}_{R},\quad
\varphi^{+}=\left( \varphi^{+}_{1},\varphi^{+}_{2} \right). 
$$
Thereupon 
$$
\varphi_{1e}\equiv \left( {\q1_{W}}_{1L}^{(-)}\right)^{+}
{\q1_{W}}_{2R}^{(0)}\equiv \varphi^{(+)},\quad
\varphi_{2e}\equiv \left( {\q1_{W}}_{2L}^{(0)}\right)^{+}
{\q1_{W}}_{2R}^{(0)}\equiv \varphi^{(0)},
$$
and
$$
\varphi_{1\nu}\equiv \left( {\q1_{W}}_{1L}^{(-)}\right)^{+}
{\q1_{W}}_{1R}^{(-)}\equiv c_{\nu}\left(\varphi^{(0)}\right)^{+},\quad
\varphi_{2\nu}\equiv \left( {\q1_{W}}_{2L}^{(0)}\right)^{+}
{\q1_{W}}_{1R}^{(-)}\equiv -c_{\nu}\varphi^{(-)},
$$
where the charge conjugated field $\varphi_{c}$ is defined 
$
\left( \varphi_{c}\right)_{i}=\varphi^{*\,k}\varepsilon_{ik},
$
$c_{\nu}$ is a constant. Then, the composite 
isospinor-scalar meson reads
$$
\varphi_{e}\equiv \varphi=
\left(\matrix{
\varphi^{(+)}\cr
\cr
\varphi^{(0)}\cr}
\right), \quad
\varphi_{\nu}= c_{\nu} \varphi_{c}=
c_{\nu}\left(\matrix{
\left(\varphi^{(0)}\right)^{+}\cr
\cr
-\varphi^{(-)}\cr}
\right).
$$
This field is a scalar as far as it is 
invariant under Lorentz transformations
$$
q_{R}\rightarrow 
\exp \left[ \FFr{i}{2}
\vec{\sigma}(\vec{\theta}-i\vec{\varphi}) \right]q_{R},\quad
q_{L}\rightarrow 
\exp \left[ \FFr{i}{2}
\vec{\sigma}(\vec{\theta}+i\vec{\varphi}) \right]q_{L},
$$
where $\vec{\theta}$ is ordinary rotations, $\vec{\varphi}$ is the boost.\\
In accordance with eq.(5.5), the isospinor-scalar
meson carries following weak hypercharge
$
\varphi:Y^{w}=1.
$
Thus, the term $-f_{Q}\bar{\Psi}\Psi$ arisen in the total Lagrangian
of fundamental fermion field eq.(11.5) accommodates the Yukawa 
couplings between the fermions and corresponding isospinor-scalar mesons 
in fairy conventional form
\begin{equation}
\label{eq: R18.17}
-f_{Q}\bar{\Psi}\Psi=
-f_{e}\left( \bar{L}\varphi e_{R}+\bar{e_{R}}\varphi^{+} L\right)-
f_{\nu}\left( \bar{L}\varphi_{c} \nu_{R}+\bar{\nu_{R}}\varphi^{+}_{c} 
L\right),
\end{equation}
where
$
f_{e}\equiv f_{Q}=\Sigma_{Q}, \quad f_{\nu}\equiv c_{\nu}f_{Q}.
$
The rules regarding to this change apply the gauge invariant Lagrangian of
isospinor-scalar $\varphi$-meson to be added to
total Lagrangian eq.(11.5).
The $\varphi$-meson undergoes following gauge 
transformations:
$
\varphi'(\eta,u_{W})=f^{(3)}_{exp}(\theta(\eta))\varphi(u_{W}),
$
where $\eta\in \G1_{\eta}$ and $u_{W}\in \G1_{W}$.
It is due to fact that the $\varphi$-meson carries isospin
and hypercharge. Making use of the local transformations implemented 
upon the components $q_{L}$ and $q_{R}$, one gets this 
transformation rule.

\renewcommand{\theequation}{\thesubsection.\arabic{equation}}
\section{The Higgs Boson}
\label{Higgs}
A common feature of gauge theories is that introducing of
ready made multiplets of zero mass scalar bosons as predicted by
Goldstone theorem [37,38]. Such bosons tend to restore the primary
symmetry [38,39]. In Higgs relativistic model [40,41] these 
unwanted massless Goldstone particles are eliminated by coupling in
gauge fields. To break the gauge symmetry down and leading to masses of the
fields , one needs in general, several kinds of spinless Higgs mesons, with 
conventional Yukawa couplings to fermion currents and transforming by an
irreducible representation of gauge group. The Higgs theory like [37]
involves these bosons as the ready made fundamental elementary fields, which
entails various difficulties. 
Within our approach the 
self-interacting isospinor-scalar Higgs bosons arise 
as the collective modes of excitations of bound quasi-particle iso-pairs. 

\subsection{The Bose Condensate of Iso-Pairs}
\label{Cond}
The ferromagnetism [42],
Bose superfluid [43] and BCS-Bogoliubov model of superconductivity
[9-11] are characterized by the condensation phenomenon leading
to the symmetry-breaking ground state. It is particularly helpful to 
remember that in BCS-Bogoliubov theory the
importance of this phenomenon resides in the possibility suggested by 
Cooper [44] that in the case of an arbitrary weak interaction the pair, 
composed of two mutually interacting electrons above the quiescent Fermi 
sea, remains in a bound state. The electrons filling the Fermi sea do not 
interact with the pair and in the same time they block the levels below
the Fermi surface. The superconductive phase arises due to effective
attraction between electrons occurred by exchange of virtual phonons [45].
In BCS microscopic theory of superconductivity instead of bound states, with 
inception by Cooper, one has a state with strongly correlated electron pairs 
or condensed state in which the pairs form the condensate. The energy
of a system in the superconducting state is smaller than the energy in the
normal state described by the Bloch individual-particle model. The energy gap
arises is due to existence of the binding energy of a pair as a collective 
effect, the width of which is equal to twice the binding energy. According
to Pauli exclusion principle, only the electrons situated in the spherical 
thin shell near the Fermi surface can form bound pairs in which they have
opposite spin and momentum. The binding energy is maximum at absolute zero 
and decreases along the temperature increasing because of the disintegration 
of pairs.
Pursuing the analogy with these ideas in outlined here approach a serious 
problem is to find out the eligible mechanism 
leading to the formation of pairs, somewhat like Cooper mechanism, but
generalized for relativistic fermions, of course in absence of any lattice.
We suggest this mechanism in the framework of gauge invariance 
incorporated with the P-violation phenomenon in W-world.
To trace a maximum resemblance to the 
superconductivity theory,
within this section it will be advantageous to describe our approach
in terms of four dimensional Minkowski space ${\M1_{W}}^{4}$ corresponding
to internal W-world:
$\G1_{W}\rightarrow {\M1_{W}}^{4}$(subsec.2.1 in [1]). 
Although we shall leave the suffix $(W)$  implicit, but it goes without 
saying that all results obtained within this section refer to W-world.
According to previous section, we consider the 
isospinor-scalar $\varphi$-meson arisen in W-world
$$
\varphi(x)=\gamma^{0}{\Psi_{L}}^{+}(x)\Psi_{R}(x),
$$
where $x \in M^{4}$ is a point of W-world. The following notational
conventions will be employed throughout
$$
{\q1_{W}}_{L}\equiv {\ps1_{W}}_{L}(\x1_{W})\rightarrow \Psi_{L}(x),
\quad {\M1_{W}}^{4}\rightarrow M^{4},\quad
{\q1_{W}}_{R}\equiv {\ps1_{W}}_{R}(\x1_{W})\rightarrow \Psi_{R}(x),
$$
where 
$
\Psi_{R}(x)=\gamma(1+\vec{\sigma}\vec{\beta})\Psi_{L}(x), \quad 
\vec{\beta}=\FFr{\vec{v}}{c},\quad
\Psi_{L}(x)=\gamma(1-\vec{\sigma}\vec{\beta})\Psi_{R}(x), \quad
\gamma =\FFr{E}{m},
$
provided by the spin $\vec{\sigma}$, energy $E$ and velocity
$\vec{v}$ of particle. In terms of Fourier integrals
\begin{equation}
\label{eq: R19.1.4}
\Psi_{L}(x)=\FFr{1}{(2\pi)^{4}}\IIn\Psi_{L}(p_{L})
{\displaystyle e^{ip_{L}x}} d^{4}p_{L},
\quad
\Psi_{R}(x)=\FFr{1}{(2\pi)^{4}}\IIn\Psi_{R}(p_{R})
{\displaystyle e^{ip_{R}x}} d^{4}p_{R},
\end{equation}
it is readily to get
\begin{equation}
\label{eq: R19.1.5}
\varphi(k)=\IIn\varphi(x)
{\displaystyle e^{-ikx}}d^{4}x=
\gamma^{0}\IIn\FFr{d^{4}p_{L}}{(2\pi)^{4}}
{\Psi_{L}}^{+}(p_{L})\Psi_{R}(p_{L}+k)=
\gamma^{0}\IIn\FFr{d^{4}p_{R}}{(2\pi)^{4}}
{\Psi_{L}}^{+}(p_{R}-k)\Psi_{R}(p_{R})
\end{equation}
provided by conservation law of fourmomentum
$
k = p_{R}-p_{L},
$
where $k=k(\omega, \vec{k})$, $\,p_{L,R}=p_{L,R}(E_{L,R}, \vec{p}_{L,R})$.
Our arguments on Bose-condensation are based on the local gauge invariance 
of the theory incorporated with the P-violation in weak interactions.
The rationale for this approach is readily forthcoming from the 
consideration of gauge transformations of the fields eq.(15.1.2) under
the P-violation in W-world
$$
\Psi'_{L}(x)=U_{L}(x)\Psi_{L}(x), \quad
\Psi'_{R}(x)=U_{R}(x)\Psi_{R}(x),
$$
where the Fourier expansions carried out over corresponding {gauge quanta}
with wave fourvectors $q_{L}$ and $q_{R}$
\begin{equation}
\label{eq: R19.1.8}
U_{L}(x)=\IIn\FFr{d^{4}q_{L}}{(2\pi)^{4}}
{\displaystyle e^{iq_{L}x}} U_{L}(q_{L}), \quad
U_{R}(x)=\IIn\FFr{d^{4}q_{R}}{(2\pi)^{4}}
{\displaystyle e^{iq_{R}x}} U_{L}(q_{R}),
\end{equation}
and
$
U_{L}(x)\neq U_{R}(x).
$
They induce the gauge transformations implemented upon $\varphi$-field
$
\varphi'(x) =U(x)\varphi(x).
$
The matrix of {\em induced gauge transformations} may be written down
in terms of {\em induced gauge quanta}
\begin{equation}
\label{eq: R19.1.11}
U(x)\equiv U^{+}_{L}(x)U_{R}(x)=
\IIn\FFr{d^{4}q}{(2\pi)^{4}}
{\displaystyle e^{iq x}} U(q), 
\end{equation}
where $q=-q_{L}+q_{R}, \quad q(q^{0},\vec{q})$. In momentum space
one gets
\begin{equation}
\label{eq: R19.1.12}
\begin{array}{l}
\varphi'(k')=\IIn\FFr{d^{4}q}{(2\pi)^{4}}
U(q) \varphi(k'-q)=
\IIn\FFr{d^{4}k}{(2\pi)^{4}}
U(k'-k) \varphi(k).
\end{array}
\end{equation}
Conservation of fourmomentum requires that
$
k'=k + q.
$
According to eq.(15.1.2) and eq.(15.1.5), we have
$$
-p'_{L}+p'_{R}=-p_{L}+p_{R} + q =-p''_{L}+p_{R}=
-p_{L}+p''_{R},
$$
where 
$
p''_{L}=p_{L}-q,\quad
p''_{R}=p_{R}+q.
$
Whence the wave vectors of fermions imply the conservation law
$
\vec{p}_{L}+\vec{p}_{R}=\vec{p}''_{L}+\vec{p}''_{R},
$
characterizing the scattering process of two fermions with {\em effective
interaction caused by the mediating induced gauge quanta}.  We suggest the 
mechanism for the effective attraction between the fermions 
in the following manner: Among all induced gauge 
transformations with miscellaneous gauge quanta we distinguish 
a special subset with the induced gauge quanta of the frequencies
belonging to finite region characterized by the maximum frequency
$\FFr{\widetilde{q}}{\hbar}\quad (\widetilde{q}=max\{q^{0}\})$
greater than the frequency of inducing oscillations fermion force
$
\FFr{\bar{E_{L}}-\bar{E''_{L}}}{\hbar}< \FFr{\widetilde{q}}{\hbar}.
$
To the extent that this is a general phenomenon, we can expect under 
this condition the effective attraction 
(negative interaction) arisen between the fermions caused by exchange of
virtual induced gauge quanta if only {\em the forced oscillations of
these quanta} occur in the same phase with the oscillations of
inducing force (the oscillations of fermion density).
In view of this we may think of isospinor $\Psi_{L}$ and isoscalar
$\Psi_{R}$ fields as the fermion fields composing the iso-pairs with the 
same conserving net momentum $\vec{p}={\vec{p}}_{L}+{\vec{p}}_{R}$
and opposite spin, for which the maximum number of negative matrix
elements of operators composed by corresponding creation and annihilation
operators 
$
a^{+}_{\vec{p}''_{R}}\, a_{\vec{p}_{R}}\,
a^{+}_{\vec{p}''_{L}}\, a_{\vec{p}_{L}}
$
(designated by the pair wave vector $\vec{p}$)
may be obtained for coherent ground state with
$
\vec{p}=\vec{p}_{L}+\vec{p}_{R}=0.
$
In the mean time the interaction potential reads
\begin{equation}
\label{eq: R19.1.20}
V=\S_{\vec{p}''_{R},\, \vec{p}''_{L},\, \vec{p}_{R},\, \vec{p}_{L}}
\left( a^{+}_{\vec{p}_{L}} \right)^{+}a^{+}_{\vec{p}''_{R}}
\left( a^{+}_{\vec{p}''_{L}} \right)a_{\vec{p}_{R}}=
\S_{\vec{p}''_{R},\, \vec{p}''_{L},\, \vec{p}_{R},\,\vec{p}_{L}}
a^{+}_{\vec{p}''_{R}}\, a^{+}_{\vec{p}''_{L}}\,
a_{\vec{p}_{R}}\, a_{\vec{p}_{L}},
\end{equation}
implying the attraction between the fermions situated in the spherical
thin shall near the Fermi surface
\begin{equation}
\label{eq: R19.1.20}
V_{\vec{p}\vec{p}''}=\left\{ \matrix{
-V\quad \mbox{at} \quad \mid E_{\vec{p}}-E_{F}\mid \leq \widetilde{q},
\quad \mid E_{\vec{p}''}-E_{F}\mid \leq \widetilde{q},\cr
0 \quad \mbox{otherwise}.\cr}
\right. 
\end{equation}
The fermions filled up the Fermi sea block
the levels below Fermi surface. Hence, the fermions are in 
superconductive state if the condition eq.(15.1.7) holds. Otherwise, they are in {\em normal state} described
by Bloch individual particle model. Thus, the Bose-condensate arises
in the W-world as the collective mode of excitations of bound
quasi-particle iso-pairs described by the same wave function in the
superconducting phase
$
\Psi=<\Psi_{L}\Psi_{R}>,
$
where $<\cdots>$ is taken to denote the vacuum averaging. The vacuum of the 
W-world filled up by such iso-pairs at absolute zero $T=0$.\\
We make a final observation that 
$\Psi_{R}\Psi_{R}^{+}=n_{R}$ is a scalar density number of
right-handed particles. Then it readily follows:
\begin{equation}
\label{eq: R19.1.22}
(\Psi_{L}\Psi_{R})^{+}(\Psi_{L}\Psi_{R})=\Psi_{R}^{+}\Psi_{L}^{+}
\Psi_{L}\Psi_{R}=
\FFr{1}{n_{R}}
\Psi_{R}^{+}\gamma^{0}(\gamma^{0}\Psi_{L}^{+}\Psi_{R})
(\Psi_{R}^{+}\Psi_{L}\gamma^{0})
\gamma^{0}\Psi_{R}=\varphi\varphi^{+},
\end{equation}
where
$
\mid \Psi\mid^{2}=
<\varphi\varphi^{+}>=\mid <\varphi>\mid^{2}\equiv 
\mid \varphi \mid^{2}.
$
It is convenient to abbreviate the $<\varphi>$ by the 
symbol $\varphi $. The eq.(15.1.8) shows that
the $\varphi$-meson actually arises as the collective mode of excitations of
bound quasi-particle iso-pairs.

\subsection{The Non-Relativistic Approximation}
\label{Appr}
In the approximation to non-relativistic limit
$(\beta \ll 1, \quad \Psi_{L}\simeq \Psi_{R}, \quad \gamma^{0}\rightarrow 1)$
by making use of Ginzburg-Landau's phenomenological theory [46] it is
straightforward to write down the free-energy functional for the order
parameter in equilibrium superconducting phase in presence of
magnetic field. The self-consistent coupled GL-equations are differential 
equations like Schr\"{o}dinger and Maxwell equations, which relate
the spatial variation of the order parameter $\Psi$ to the vector 
potential $\vec{A}$ and the current ${\vec{j}}_{s}$. In the papers
[20,21], by means of thermodynamic Green's functions in well defined
limit Gor'kov was able to show that GL-equations are a consequence of
the BCS-Bogoliubov microscopic theory of superconductivity. The theoretical
significance of these works resides in the microscopic interpretation
of all physical parameters of GL-theory. Subsequently these ideas were 
extended to lower temperatures by others [47-49] using a requirement
that the order parameter and vector potential vary slowly over distances of the 
order of the coherence length and that the electrodynamics be local
(London limit). Namely, the validity of derived GLG-equations is restricted
to the temperature $T$, such $T_{c}-T\ll T_{c}$ and to the local
electrodynamics region $q\xi_{0}\ll 1$, where $T_{c}$ is transition 
temperature, $\xi_{0}$ is coherent length characterizing the spatial extent 
of the electron pair correlations, $q$ are the wave numbers of magnetic
field $\vec{A}$. Thereto the most important order parameter $\Psi$, 
mass $m_{\Psi}$ and coupling constant $\lambda_{\Psi}$ figuring in 
GLG-equations read
\begin{equation}
\label{eq: R19.2.2}
\begin{array}{l}
\Psi(\vec{r})=
\FFr{(7\zeta(3)N)^{1/2}}
{4\pi k_{B}T_{c}}\Delta(\vec{r}),\quad
\Delta(T)=
\simeq 3.1k_{B}T_{c}\left( 1-\FFr{T}{T_{c}}\right)^{1/2},\quad
\xi_{0}=\simeq 0.18\FFr{\hbar v_{F}}{k_{B}T_{c}},\\
m^{2}_{\Psi}=
1.83\FFr{\hbar^{2}}{m}\FFr{1}{\xi^{2}_{0}}
\left( 1-\FFr{T}{T_{c}}\right),\quad
\lambda^{2}_{\Psi}=
1.4\FFr{1}{N(0)}\left( \FFr{\hbar^{2}}{2m\xi^{2}_{0}}\right)^{2}
\FFr{1}{(k_{B}T_{c})^{2}}.
\end{array}
\end{equation}
Reviewing the notation $\Delta(\vec{r})$ is the energy gap, 
$e^{*}=2e$ is the effective charge, $N(0)$ is the state density at Fermi
surface, $N$ is the number of particles per unit volume in normal mode,
$v_{F}$ is the Fermi velocity, $m\equiv \Sigma_{Q}=f_{Q}$ is the mass of
fermion field. The transition temperature
relates to gap at absolute zero $\Delta_{0}$ [9]. 
The estimate for the pair size at $v_{F}\sim 10^{8}cm/s,
\quad T_{c}\sim 1$ gives [50]
$
\xi_{0} \simeq 10^{-4}cm.
$
To set up the total Lagrangian we must add a 
$SU(2)$ multiplet of spinless meson fields that couples to the gauge fields 
in a gauge invariant way. According to eq.(15.2.1) these 
$\varphi$-meson fields acquire vacuum expectation values, which 
subsequently gives the gauge fields a mass. This Lagrangian of
self-interacting iso-pairs spinless scalar $\varphi$-meson at 
degenerate vacuum of W-world is given 
$
L_{\varphi}(D)=L_{\varphi}^{0}(D)-V(\mid \varphi\mid),
$
provided by 
$
V(\mid \varphi\mid)=\FFr{1}{2}\lambda^{2}_{\varphi}
\left(\mid \varphi\mid^{2}-\FFr{\eta_{\varphi}^{2}}{2}\right)^{2},
$ and
$
\eta_{\varphi}^{2}=\FFr{m^{2}_{\varphi}}{\lambda^{2}_{\varphi}},\quad
m^{2}_{\varphi}\equiv m^{2}_{\Psi},
\quad
\lambda^{2}_{\varphi}\equiv \lambda^{2}_{\Psi}.
$
In accordance with a standard method one may introduce the real-valued scalar 
field $\chi(x)$ describing the excitation in the neighbourhood of stable
vacuum $\FFr{\eta_{\varphi}}{\sqrt{2}}:\quad
\varphi'(x)=\FFr{1}{\sqrt{2}}\left(\eta_{\varphi}+\chi(x)\right),
$
which reflects the spontaneous breakdown of symmetry in W-world.

\subsection{The Relativistic Treatment}
\label{Rel}
We start with total Lagrangian eq.(11.7) of self-interacting
fermion field in W-world, which is arisen from the Lagrangian eq.(10.6)
of primary fundamental field after the rearrangement of the vacuum
of Q-world
\begin{equation}
\label{eq: R19.3.1}
\begin{array}{l}
\L1_{W}(x)=
\FFr{i}{2} \{ 
\bp_{W}(x)\gamma^{\mu}{\pr_{W}}_{\mu}\ps1_{W}(x)-
\bp_{W}(x)\gamma^{\mu}{\lpr_{W}}_{\mu}\ps1_{W}(x)\}-
m\bp_{W}(x)\ps1_{W}(x)-\\
-\FFr{\lambda}{2}\bp_{W}(x)\left(\bp_{W}(x)\ps1_{W}(x)\right)\ps1_{W}(x).
\end{array}
\end{equation}
Here, $m =\Sigma_{Q}$ is the self-energy operator of the fermion field 
component in Q-world, the suffix $(W)$ just was put forth in 
illustration of a point at issue. For the sake of simplicity, we also admit 
${\bf\B1_{W}}(x)=0$, but of course one is free to restore the gauge field 
${\bf \B1_{W}}(x)$ whenever it should be needed. 
In lowest order the relation
$m\equiv m_{Q}\ll \lambda^{-1/2}$ holds.
The Lagrangian eq.(15.3.1) leads to the field equations
\begin{equation}
\label{eq: R19.3.2}
\begin{array}{l}
(\gamma p-m)\Psi(x)-\lambda \left(\bar{\Psi}(x)\Psi(x) \right)
\Psi(x)=0,\\ 
\bar{\Psi}(x)(\gamma \overleftarrow{p}+m)+
\lambda \bar{\Psi}(x)\left(\bar{\Psi}(x)\Psi(x) \right)=0,
\end{array}
\end{equation}
where the indices have been suppressed as usual.
At non-relativistic limit the function $\Psi$ reads
$
\Psi \rightarrow e^{imc^{2}t}\Psi,
$
and Lagrangian eq.(15.3.1) leads to Hamiltonian used in [20]. 
Our discussion will be in close analogy with that of [20]. 
We make use of the Gor'kov's technique and evaluate the equations 
(15.3.2) in following manner: 
The spirit of the calculation will be to treat interaction between 
the particles as being absent everywhere except the thin spherical shell
$2\widetilde{q}$ near the Fermi surface. The Bose
condensate of bound particle iso-pairs occurred at zero momentum.
The scattering processes between the particles are absent. 
We consider the matrix elements 
$$
\begin{array}{l}
<T \left(\Psi_{\alpha}(x_{1})\Psi_{\beta}(x_{2})
\bar{\Psi}_{\gamma}(x_{3})\bar{\Psi}_{\delta}(x_{4})
\right)>=-<T \left(\Psi_{\alpha}(x_{1})\bar{\Psi}_{\gamma}(x_{3})\right)>
<T \left(\Psi_{\beta}(x_{2})\bar{\Psi}_{\delta}(x_{4}) \right)>\\
+<T \left( \Psi_{\alpha}(x_{1})\bar{\Psi}_{\delta}(x_{4})\right)>
<T \left(\Psi_{\beta}(x_{2})\bar{\Psi}_{\gamma}(x_{3})\right)>+ 
<T \left( \Psi_{\alpha}(x_{1})\Psi_{\beta}(x_{2})\right)>\times\\
<T \left( \bar{\Psi}_{\gamma}(x_{3})\bar{\Psi}_{\delta}(x_{4})\right)>,
\end{array}
$$
where
$$
\begin{array}{l}
<T \left( \Psi_{\alpha}(x_{1})\Psi_{\beta}(x_{2})\right)>
<T \left( \bar{\Psi}_{\gamma}(x_{3})\bar{\Psi}_{\delta}(x_{4})\right)>
=\\
<T \left( \Psi^{+}_{\gamma}(x_{3})\gamma^{0}\Psi^{+}_{\delta}(x_{4})\right)>
<T \left( \gamma^{0} \Psi_{\alpha}(x_{1})\Psi_{\beta}(x_{2})\right)>,
\end{array}
$$
also introduce the functions
\begin{equation}
\label{eq: R19.3.11}
\begin{array}{l}
<N \mid
T \left( \gamma^{0} \Psi(x)\Psi(x')\right)
\mid N+2>=e^{-2i\mu't}F(x-x'), \\ 
<N +2\mid
T \left( \Psi^{+}(x)\gamma^{0}\Psi^{+}(x')\right)
\mid N>=e^{2i\mu't}F^{+}(x-x'),
\end{array}
\end{equation}
and 
\begin{equation}
\label{eq: R19.3.12}
\begin{array}{l}
<N \mid
T \left( \Psi_{L}(x)\Psi_{R}(x')\right)
\mid N+2>=e^{-2i\mu't}F_{LR}(x-x'), \\ 
<N +2\mid
T \left( \Psi^{+}_{L}(x)\Psi^{+}_{R}(x')\right)
\mid N>=e^{2i\mu't}F^{+}_{LR}(x-x'),\\ 
<N \mid
T \left( \Psi_{R}(x)\Psi_{L}(x')\right)
\mid N+2>=e^{-2i\mu't}F_{RL}(x-x'), \\ 
<N +2\mid
T \left( \Psi^{+}_{R}(x)\Psi^{+}_{L}(x')\right)
\mid N>=e^{2i\mu't}F^{+}_{RL}(x-x').
\end{array}
\end{equation}
Thereupon
$$
F(x-x')=
\left(\matrix{
F_{LR}(x-x')\cr
F_{RL}(x-x')\cr}
\right),\quad
F^{+}(x-x')=
\left(
F^{+}_{LR}(x-x'),F^{+}_{RL}(x-x')
\right).
$$
Here, $\mu'=\mu + m$, $\mu$ is a chemical potential. 
We omit a prime over $\mu$, but should
understand under it $\mu + m$. 
Let us now make use of Fourier integrals
$$
G_{\alpha\beta}(x-x')=
\IIn\FFr{d\omega\,d\vec{p}}{(2\pi)^{4}}
G_{\alpha\beta}(p)
{\displaystyle e^{i\vec{p}(\vec{x}-\vec{x'})-i\omega (t-t')}}
$$
etc, which render the equation (15.3.2) easier to handle in momentum space
\begin{equation}
\label{eq: R19.3.18}
\begin{array}{l}
(\gamma p-m)G(p)-i\lambda \gamma^{0}F(0+)\bar{F}(p)=
1, \\ 
\bar{F}(p)(\gamma p+m-2\mu \gamma^{0})+
i\lambda \bar{F}(0+)G(p)=0,
\end{array}
\end{equation}
where
$
F_{\alpha\beta}(0+)=e^{2i\mu t}
<T \left( \gamma^{0} \Psi_{\alpha}(x)\Psi_{\beta}(x)\right)>=
\Lm_{x \rightarrow x'(t \rightarrow t')}F_{\alpha\beta}(x-x').
$
Next we substitute 
$$
(\gamma p-m)=(\omega' -\xi_{p})\gamma^{0},\quad
(\gamma p+m-2\mu \gamma^{0})=\gamma^{0}(\omega' +\xi_{p}^{+}),
$$
where
$$
\begin{array}{l}
\omega' =\omega -\mu'=\omega -m -\mu ,\quad
\xi_{p}=(\vec{\gamma}\vec{p}+m)\gamma^{0}-\mu'=
(\vec{\gamma}\vec{p}+m)\gamma^{0}-m-\mu, \\
\xi_{p}^{+}=\gamma^{0}({\vec{\gamma}}^{+}\vec{p}+m) -m-\mu,
\end{array}
$$
and omit a prime over $\omega'$ for the rest of this section.
We employ
$$
F(0+)=-JI, \quad 
I=\left( \matrix{
\hspace{0.3cm} 0 \quad 1\cr
-1 \quad 0 \cr
}\right),\quad
F^{+}(0+)F(0+)=-J^{2}I^{2}=J^{2}, 
$$
and
$\widehat{\omega} +\widehat{\xi}_{p}=
\gamma^{0}(\omega +\xi_{p})$.
The gap function $\Delta$ reads
$
\Delta^{2}=\lambda^{2} J^{2},
$
where
$
J=\IIn\FFr{d\omega\,d\vec{k}}{(2\pi)^{4}}F^{+}(p).
$
Making use of standard rules [43], one may pass over the poles. 
This method allows oneself to extend the study up to limit of
temperatures, such that $T_{c}-T\ll T_{c}$, by making use of thermodynamic 
Green's function.
So
\begin{equation}
\label{eq: R19.3.32}
\begin{array}{l}
F^{+}(p)=-i\lambda J (\omega-\xi_{p}+i\delta)^{-1}
(\omega+\xi_{p}-i\delta)^{-1} 
-\FFr{\pi\Delta}{\varepsilon_{p}}n(\varepsilon_{p})
\{ \delta(\omega-\varepsilon_{p})+
\delta(\omega+\varepsilon_{p})\}, 
\\
G(p)=\gamma^{0}\{u_{p}^{2}(\omega-\xi_{p}+i\delta)^{-1}+
v_{p}^{2}(\omega+\xi_{p}-i\delta)^{-1}+ 
2\pi i n(\varepsilon_{p})
[ u_{p}^{2}\delta(\omega-\varepsilon_{p})-\\
v_{p}^{2}\delta(\omega+\varepsilon_{p})]\},
\end{array}
\end{equation}
where 
$u_{p}^{2}=\FFr{1}{2}\left( 1+ \FFr{\xi_{p}}{\varepsilon_{p}}\right),
\quad
v_{p}^{2}=\FFr{1}{2}\left( 1- \FFr{\xi_{p}}{\varepsilon_{p}}\right),\quad
\varepsilon_{p}=(\xi_{p}^{2}+\Delta^{2}(T))^{1/2}.
$
and
$n(\varepsilon_{p})$ is the usual Fermi function
$
n(\varepsilon_{p})=\left( \exp\FFr{\varepsilon_{p}}{T} +1\right)^{-1}.
$
Then 
\begin{equation}
\label{eq: R19.3.35}
1=\FFr{\mid \lambda \mid}{2(2\pi)^{3}}
\IIn d\vec{k}\, \FFr{1-2n(\varepsilon_{k})}{\varepsilon_{k}(T)}
\quad \left( \mid \xi_{p}\mid
< \widetilde{q} \right),
\end{equation}
determining the energy gap $\Delta$ as a function of $T$. According to
eq.(15.3.7), the $\Delta(T)\rightarrow 0$ at $T\rightarrow T_{c}\sim
\Delta(0)$ [9].
\subsection{Self-Interacting Potential of Bose-Condensate}
\label{Pot}
To go any further in exploring the form and significance of obtained
results it is entirely feasible to include 
the generalization of the equations (15.3.5) in presence of spatially 
varying magnetic field with vector potential $\vec{A}(\vec{r})$, which is
straightforward $(t\rightarrow \tau=it)$
\begin{equation}
\label{eq: R19.4.1}
\begin{array}{l}
\left\{ 
-\gamma^{0}\FFr{\partial}{\partial \tau}-
i\vec{\gamma}\left(\FFr{\partial}{\partial \vec{r}}-ie\vec{A}(\vec{r}) 
\right)-m +\gamma^{0}\mu 
\right\}
G(x,x')+\gamma^{0}\Delta(\vec{r})\bar{F}(x,x')=
\delta (x-x'), \\ 
\bar{F}(x,x')
\left\{ 
\gamma^{0}\FFr{\partial}{\partial \tau}+
i\vec{\gamma}\left(\FFr{\partial}{\partial \vec{r}}+ie\vec{A}(\vec{r}) 
\right)-m +\gamma^{0}\mu 
\right\}-
\Delta^{*}(\vec{r})\gamma^{0}G(x,x')=0,
\end{array}
\end{equation}
where the thermodynamic Green's function [51,52] is used and
the energy gap function is in the form
$$
\Delta^{*}(\vec{r})=\lambda F^{+}(\tau,\vec{r};\,\tau,\vec{r})=
\left(
\Delta^{*}_{LR}(\vec{r}),\Delta^{*}_{RL}(\vec{r})
\right), \quad
F(x,x')=
\left(\matrix{
F_{LR}(x,x')\cr
F_{RL}(x,x')\cr}
\right),
$$
This function is logarithmically divergent, but with a cutoff of
energy of interacting fermions at the spatial distances in order
of $\FFr{\hbar v}{\widetilde{\omega}}$ can be made finite, 
where $\widetilde{\omega} \equiv \FFr{\widetilde{q}}{\hbar}$.
If one uses the Fourier components of functions $G(x,x')$ and
$F(x,x')$
\begin{equation}
\label{eq: R19.4.3}
G(\vec{r},\vec{r}';u)=T\S_{n}e^{-i\omega u}
G_{\omega}(\vec{r},\vec{r}'), \quad
G_{\omega}(\vec{r},\vec{r}')=\FFr{1}{2}\IIn^{1/T}_{-1/T}e^{i\omega u}
G(\vec{r},\vec{r}';u)d\,u,
\end{equation}
where $u=\tau - \tau'$, $\omega$ is the discrete index $\omega =
\pi T(2n + 1), \quad n=0,\pm 1, \ldots$, then the eq.(15.4.1) reduces
to
\begin{equation}
\label{eq: R19.4.4}
\begin{array}{l}
\left\{ 
i\omega\gamma^{0}-
i\vec{\gamma}\left(\vec{\partial}_{\vec{r}}-ie\vec{A}(\vec{r}) 
\right)-m +\gamma^{0}\mu 
\right\}
G_{\omega}(\vec{r},\vec{r}')+
\gamma^{0}\Delta(\vec{r})
\bar{F_{\omega}}(\vec{r},\vec{r}')=
\delta (\vec{r}-\vec{r}'), \\ 
\bar{F_{\omega}}(\vec{r},\vec{r}')
\left\{ 
-i\omega\gamma^{0}+
i\vec{\gamma}\left(\vec{\partial}_{\vec{r}}+ie\vec{A}(\vec{r}) 
\right)-m +\gamma^{0}\mu 
\right\}-
\Delta^{*}(\vec{r})\gamma^{0}G_{\omega}(\vec{r},\vec{r}')=0,
\end{array}
\end{equation}
where the gap function is defined by
$
\Delta^{*}(\vec{r})=\lambda T\S_{n}
F^{+}_{\omega}(\vec{r},\vec{r}').
$
The Bloch individual particle Green's function
$\widetilde{G}_{\omega}(\vec{r},\vec{r}')$ for the fermion in
normal mode is written
\begin{equation}
\label{eq: R19.4.6}
\left\{ 
i\omega\gamma^{0}-
i\vec{\gamma}\left(\vec{\partial}_{\vec{r}}-ie\vec{A}(\vec{r}) 
\right)-m +\gamma^{0}\mu 
\right\}
\widetilde{G}_{\omega}(\vec{r},\vec{r}')=
\delta (\vec{r}-\vec{r}'), 
\end{equation}
or the adjoint equation
\begin{equation}
\label{eq: R19.4.7}
\left\{ 
i\omega\gamma^{0}+
i\vec{\gamma}\left(\vec{\partial}_{\vec{r}'}+ie\vec{A}(\vec{r}') 
\right)-m +\gamma^{0}\mu 
\right\}
\widetilde{G}_{\omega}(\vec{r},\vec{r}')=
\delta (\vec{r}-\vec{r}'). 
\end{equation}
By means of eq.(15.4.5) the eq.(15.4.3) gives rise to
\begin{equation}
\label{eq: R19.4.8}
G_{\omega}(\vec{r},\vec{r}')=\widetilde{G}_{\omega}(\vec{r},\vec{r}')-
\IIn\widetilde{G}_{\omega}(\vec{r},\vec{s})
\gamma^{0}\Delta(\vec{s})
\bar{F_{\omega}}(\vec{s},\vec{r}')d^{3}s,
\end{equation}
and
\begin{equation}
\label{eq: R19.4.9}
\bar{F_{\omega}}(\vec{r},\vec{r}')=
\IIn\widetilde{G}_{\omega}(\vec{s},\vec{r}')
\Delta^{*}(\vec{s})\gamma^{0}
\widetilde{G}_{-\omega}(\vec{s},\vec{r})d^{3}s.
\end{equation}
The gap function $\Delta(\vec{r})$ as well as 
$\bar{F_{\omega}}(\vec{r},\vec{r}')$ are small ones at close neighbourhood
of transition temperature $T_{c}$ and varied slowly over a coherence
distance. This approximation, which went into the derivation of 
equations,  meets our interest in eq..(15.4.6), eq.(15.4.7).
Using standard procedure one may readily express them in power series of $\Delta$ and $\Delta^{*}$ by keeping
only the terms in $\bar{F_{\omega}}(\vec{r},\vec{r}')$ up to the cubic
and in $G_{\omega}(\vec{r},\vec{r}')$ - quadratic order in 
$\Delta$. After averaging over the polarization of particles the
following equation coupling $\Delta(\vec{r})$ and 
$\vec{A}(\vec{r})$ ensued:
\begin{equation}
\label{eq: R19.4.10}
\begin{array}{l}
\overline{\Delta^{*}(\vec{r})}=\lambda T\S_{n}
\IIn 
\overline{\widetilde{G}_{\omega}(\vec{r},\vec{r}')
\widetilde{G}_{-\omega}(\vec{r},\vec{r}')}
\Delta^{*}(\vec{r}')d^{3}r'
-\\ 
\lambda T\S_{n}
\IIn\IIn\IIn
\overline{\widetilde{G}_{\omega}(\vec{s},\vec{r})
\widetilde{G}_{-\omega}(\vec{s},\vec{l})
\widetilde{G}_{\omega}(\vec{m},\vec{l})
\widetilde{G}_{-\omega}(\vec{m},\vec{r})}
\Delta(\vec{s})\Delta^{*}(\vec{l})\Delta^{*}(\vec{m})
d^{3}s \, d^{3}l \, d^{3}m.
\end{array}
\end{equation}
It is worthwhile to determine the function
$\widetilde{G}_{\omega}(\vec{r},\vec{r}')$. 
In the absence of applied magnetic field, i.e. $\vec{A}(\vec{r})=0$
the eq.(15.4.4) reduces to
\begin{equation}
\label{eq: R19.4.11}
\left\{ 
(i\omega+\mu )\gamma^{0}-
\vec{\gamma}\vec{p} -m  
\right\}
\widetilde{G}^{0}_{\omega}(\vec{r},\vec{r}')=
\delta (\vec{r}-\vec{r}').
\end{equation}
It is well to study at this point certain properties of the solution
which we shall continually encounter
\begin{equation}
\label{eq: R19.4.12}
\widetilde{G}^{0}_{\omega}(\vec{r},\vec{r}')=
\FFr{1}{2m}\left\{ 
(i\omega+\mu )\gamma^{0}-
\vec{\gamma}\vec{p} +m  
\right\}
\widetilde{G}^{0}_{G\omega}(\vec{r}-\vec{r}'),
\end{equation}
where the function $\widetilde{G}^{0}_{G\omega}(\vec{r}-\vec{r}')$
satisfies the equation
\begin{equation}
\label{eq: R19.4.13}
\FFr{1}{2m}\left( q^{2}+\Delta
\right)
\widetilde{G}^{0}_{G\omega}(\vec{r}-\vec{r}')=
\delta (\vec{r}-\vec{r}'),
\end{equation}
provided
$
q^{2}=(i\omega+\mu )^{2}-m^{2}=2im\Omega +p^{2}_{0}
$
and
$
\Omega =\omega\FFr{\mu}{m}, \quad
p^{2}_{0}=\mu^{2}-m^{2}.
$
At $\mu \gg \mid \omega\mid$ one has $\FFr{p^{2}_{0}}{2m}
\gg \mid \Omega\mid$, and
\begin{equation}
\label{eq: R19.4.16}
\FFr{1}{2m}
\left\{
2im\Omega +p^{2}_{0}+\Delta
\right\}
\widetilde{G}^{0}_{G\omega}(\vec{r}-\vec{r}')=
\delta (\vec{r}-\vec{r}'),
\end{equation}
the solution of which reads
$$
\widetilde{G}^{0}_{G\omega}(\vec{r}-\vec{r}')=
-\FFr{m}{2\pi R}\exp(iqR),
$$
where
$$
q=sgn\, \Omega \,p_{0}+i\FFr{\mid \Omega \mid}{v}, \\
R=\mid \vec{r}-\vec{r}' \mid, \quad
sgn\, \Omega=\FFr{\Omega}{\mid \Omega \mid}.
$$
In approximation to non-relativistic limit $\vec{p}\rightarrow 0$
this Green's function reduces to\\
$\widetilde{G}^{0}_{G\omega}(\vec{r}-\vec{r}')$ used in [20].
So, our discussion is consistent with 
well-known [20].
Making use of Fourier integrals we readily get
$$
\widetilde{G}^{0}_{\omega}(\vec{p})=
\FFr{\mu \gamma^{0}+\vec{\gamma}\vec{p} +m }
{q^{2}-\vec{p}^{2}+i0}=
\widehat{\gamma}\,\widetilde{G}^{0}_{G\omega}(\vec{p}),
$$
where 
$
\widehat{\gamma}=\FFr{1}{2m}
(\mu \gamma^{0}+\vec{\gamma}\vec{p} +m ).
$
One has
$$
\widetilde{G}^{0}_{G\omega}(\vec{p})=
\widetilde{G}^{0}_{G\Omega}(\vec{p})=
\FFr{1}{i\Omega -\xi}=\widetilde{G}^{0}_{G\Omega}(-\vec{p}),
$$
where as usual $\xi=\FFr{\vec{p}}{2m}-\FFr{\vec{p}_{0}}{2m}$.
The Green's function $\widetilde{G}_{\omega}(\vec{r},\vec{r}')$ in presence
of magnetic field differs from 
$\widetilde{G}_{\omega}^{0}(\vec{r}-\vec{r}')$ 
only by phase multiplier [20]
$$
\widetilde{G}_{\omega}(\vec{r},\vec{r}')=
\exp\left\{
\FFr{ie}{c}\left(\vec{A}(\vec{r}), \vec{r}-\vec{r}' \right)
\right\}
\widetilde{G}_{\omega}^{0}(\vec{r}-\vec{r}').
$$
The technique now is to expand a second term in right-hand side of 
the eq.(15.4.8)
up to the terms quadratic in $(\vec{r}-\vec{r}')$.
After calculations it transforms 
\begin{equation}
\label{eq: R19.4.40}
\begin{array}{l}
\left\{
\left(i\hbar\vec{\nabla}+\FFr{e^{*}}{c}\vec{A} \right)^{2}+
\FFr{2m}{\nu}\left[ 
\FFr{2\pi^{2}}{\lambda m p_{0}}\left(\FFr{\mu}{m} \right)^{2}
\left( \FFr{\mu}{m}-1\right)+
\left(\FFr{\mu}{m} \right)^{2}
\left( \FFr{T}{T_{c\mu}}-1\right)+
\right. \right.\\ 
\left. \left.
\FFr{2}{N}\mid \Psi(\vec{r})\mid^{2}
\right]
\right\}\Psi(\vec{r})=0,
\end{array}
\end{equation}
where $\nu=\FFr{7\zeta(3)mv_{F}^{2}}{24(\pi k_{B}T_{c})^{2}}$ and 
$T_{c\mu}= \FFr{m}{\mu}T_{c}$.
Succinctly 
\begin{equation}
\label{eq: R19.4.41}
\left\{
\vec{p}_{A}^{2}-\FFr{1}{2}m_{\Psi}^{2}+\FFr{1}{4}\lambda_{\Psi}^{2}
\mid \Psi(\vec{r})\mid^{2}
\right\}\Psi(\vec{r})=0,
\end{equation}
provided
\begin{equation}
\label{eq: R19.4.41}
\begin{array}{l}
m_{\Psi}^{2}(\lambda,T,T_{c\mu})=
\FFr{24}{7\zeta(3)}\left(\FFr{\hbar}{\xi_{0}} \right)^{2}
\left(\FFr{\mu}{m} \right)^{2}
\left[
1-\FFr{T}{T_{c\mu}}-\left( \FFr{\mu}{m}-1\right)
\ln\FFr{2\widetilde{\omega}}{\Delta_{0}}
\right],\\ 
\lambda_{\Psi}^{2}(\lambda,T_{c})=
\FFr{96}{7\zeta(3)}\left(\FFr{\hbar}{\xi_{0}} \right)^{2}
\FFr{1}{N}, \quad 
\Psi(\vec{r})=\Delta(\vec{r})\FFr{\left(7\zeta(3)N\right)^{1/2}}
{4\pi k_{B}T_{c}}.
\end{array}
\end{equation}
According to the eq.(15.4.15), the 
magnitude of the relativistic effects, however, is found to be greater to
account for the large contribution to
the values $m_{\Psi}^{2}$ and $\lambda_{\Psi}^{2}$. Thereby the transition
temperature decreases inversely by the relativistic factor
$\FFr{\mu}{m}$.
A spontaneous breakdown of symmetry of ground state occurs at
$
\eta_{\Psi}^{2}(\lambda,T < T_{c\mu}) > 0,
$
where
$
\eta_{\Psi}^{2}(\lambda,T,T_{c\mu})=\FFr{m_{\Psi}^{2}}{\lambda_{\Psi}^{2}}.
$
\\
As far as 
$
\Psi=\left(
\matrix{
\Psi_{LR}\cr
\Psi_{RL}\cr
}
\right), \quad
\Delta=\left(
\matrix{
\Delta_{LR}\cr
\Delta_{RL}\cr
}
\right),
$
and 
$
\Psi_{LR}=\Psi_{RL},\quad
\Delta_{LR}=\Delta_{RL},
$
then the eq.(15.4.14) splits into the couple of equations for
$\Psi_{LR}$ and $\Psi_{RL}$. Subsequently, a Lagrangian of the
$\varphi$
will be arisen  with
the corresponding values of mass 
$m_{\Psi}^{2}\equiv m_{\varphi}^{2}$
and coupling constant 
$\lambda_{\Psi}^{2}\equiv\lambda_{\varphi}^{2}$.

\subsection{The Four-Component Bose-Condensate in Magnetic Field}
\label{Four}
Now we are going to derive the equation of four-component 
bispinor field of Bose-condensate, where due to self-interaction the 
spin part of it is vanished. We start with 
{\em the nonsymmetric state $\Delta_{LR}\neq \Delta_{RL}$}, 
where $\Psi_{LR}$ and $\Psi_{RL}$ are two eigenstates of chirality 
operator $\gamma_{5}$. In standard representation
\begin{equation}
\label{eq: R19.5.1}
\begin{array}{l}
\Psi=\FFr{1}{\sqrt{2}}
\left(
\matrix{
1 & 1\cr
1 & -1\cr
}
\right)
\left(
\matrix{
\Psi_{LR}\cr
\Psi_{RL}\cr
}
\right)
=\FFr{1}{\sqrt{2}}
\left(
\matrix{
\Psi_{LR}+\Psi_{RL}\cr
\Psi_{LR}-\Psi_{RL}\cr
}
\right)
\equiv
\left(
\matrix{
\Psi_{1}\cr
\Psi_{2}\cr
}
\right),\\ \\
\Delta
=
\left(
\matrix{
\Delta_{1}\cr
\Delta_{2}\cr
}
\right)=
\FFr{1}{\sqrt{2}}
\left(
\matrix{
\Delta_{LR}+\Delta_{RL}\cr
\Delta_{LR}-\Delta_{RL}\cr
}
\right), \\ \\
\gamma_{5}\Psi_{LR}=\Psi_{LR}, \quad
\gamma_{5}\Psi_{RL}=-\Psi_{RL}, \quad
\Delta_{LR}\neq \Delta_{RL}.
\end{array}
\end{equation}
The eq.(15.4.14) enables to postulate the equation of four-component 
Bose-condensate in magnetic field and equilibrium state 
\begin{equation}
\label{eq: R19.5.2}
i\hbar \FFr{\partial \Psi}{\partial t}=
\left\{
c\vec{\alpha}\left(\vec{p}+ \FFr{e^{*}}{c}\vec{A} \right)+\beta mc^{2}+
M(F)+L(F)\mid \Psi\mid^{2}
\right\}\Psi=0,
\end{equation}
or succinctly
\begin{equation}
\label{eq: R19.5.3}
\left( \gamma p_{A} -m \right)\Psi=0.
\end{equation}
This is in standard notation
$$
\begin{array}{l}
\gamma^{0}=\beta=\left(
\matrix{
1 &0\cr
0 & -1\cr
}
\right), \quad \vec{\alpha}=\gamma^{0}\vec{\gamma}=
\left(
\matrix{
0 & \vec{\sigma}\cr
\vec{\sigma} & 0\cr
}
\right),\quad F=F_{\mu\nu}\sigma^{\mu\nu}=inv, 
\\ 
\sigma^{\mu\nu}=\FFr{1}{2}\left[\gamma^{\mu},\gamma^{\nu} \right],\quad
\FFr{i}{2}e^{*}F=-e^{*}\vec{\Sigma}\vec{H},\quad
\vec{H}=rot \vec{A},\quad 
F_{\mu\nu}=(0,\vec{H}), \quad \vec{\Sigma}=
\left(
\matrix{
\vec{\sigma} &0\cr
0 &\vec{\sigma}\cr
}
\right),\\
p_{A}=(p_{A0},\vec{p}_{A}), \quad 
\vec{p}_{A}=i\hbar\vec{\nabla}+
\FFr{e^{*}}{c}\vec{A},\quad p_{A0}=-\left(M(F)+L(F)\mid \Psi\mid^{2} 
\right).
\end{array}
$$
The $M(F)$ and $L(F)$ are some functions depending upon the invariant F,
which will be determined under the requirement that the second-order 
equations ensued from the eq.(15.5.3) must match onto eq.(15.4.14).
Defining the functions $M(F)$ and $L(F)$: 
$$
\begin{array}{l}
M(F)=\left(M^{2}_{0}+ \FFr{i}{2}e^{*}F \right)^{1/2}, \quad
M_{0}=\left( m^{2}+\FFr{1}{2}m^{2}_{\varphi} \right)^{1/2},\quad
L(F)=-\FFr{\lambda^{2}_{\varphi}}{8M(F)}, \\ 
L_{0}=-\FFr{\lambda^{2}_{\varphi}}{8M_{0}},
M(F)L(F)=M_{0}L_{0}=-\FFr{1}{8}\lambda^{2}_{\varphi},
\end{array}
$$
and taking into account an approximation fitting our interest that the gap
function is small at close neighbourhood of transition temperature,
one gets
\begin{equation}
\label{eq: R19.5.8}
\left\{ \vec{p}_{A}^{2} +m^{2}-\left(M_{0}+L_{0}\mid \Psi\mid^{2} 
\right)^{2} \right\}\Psi(\vec{r})\equiv
\left\{ \vec{p}_{A}^{2}-\FFr{1}{2}m^{2}_{\varphi}+
\FFr{1}{4}\lambda^{2}_{\varphi}\mid \Psi\mid^{2} 
 \right\}\Psi(\vec{r})=0.
\end{equation}
This has yet another 
important consequence. At $\Delta_{LR}\neq 0$ and
imposed constraint
$
\left( m+M(F) + L(F)\mid \Psi\mid^{2}\right)_{F\rightarrow 0}
\rightarrow 0
$
we have
\begin{equation}
\label{eq: R19.5.12}
\Delta_{2}=
\FFr{1}{\sqrt{2}}(\Delta_{LR}-\Delta_{RL})=0, \quad
\Psi_{2}=0.
\end{equation}
So, {\em the $\mid \Psi_{0}\mid $ is the gap function symmetry-restoring 
value}
$$
\Delta_{2}\left( \mid \Psi_{0}\mid^{2}=\FFr{m+M_{0}}{-L_{0}}\right)
=0,
\quad
\Delta_{LR}\left( \mid \Psi_{0}\mid^{2}\right)=
\Delta_{RL}\left( \mid \Psi_{0}\mid^{2}\right),
$$
where, according to eq.(15.5.4), one has
\begin{equation}
\label{eq: R19.5.15}
V\equiv \left[m^{2}-\left( M_{0}+L_{0}\mid \Psi\mid^{2}\right)^{2}\right]
\Psi^{2} =\left[-\FFr{1}{2}m^{2}_{\varphi}+
\FFr{1}{4}\lambda^{2}_{\varphi}\mid \Psi\mid^{2}\right]
\Psi^{2},
\end{equation}
and 
\begin{equation}
\label{eq: R19.5.16}
V\left( \mid \Psi_{0}\mid^{2}=\FFr{m+M_{0}}{-L_{0}}=
\FFr{1}{2}\eta^{2}_{\varphi}(\lambda,T,T_{c\mu})
\right)=0.
\end{equation}
It leads us to the conclusion that the field
of symmetry-breaking Higgs boson must be counted off from the
$\Delta_{LR}=\Delta_{RL}$ symmetry-restoring value of Bose-condensate
$
\mid \Psi_{0}\mid=\FFr{1}{\sqrt{2}}\eta_{\varphi}(\lambda,T,T_{c\mu})
$
as the point of origin describing the excitation in the neighbourhood
of stable vacuum eq.(15.5.7).\\
We may write down the Lagrangian corresponding to the eq.(15.5.3)
$$
L_{\Psi}=\FFr{1}{2}\left\{ 
\bar{\Psi} \gamma p_{A} \Psi -\bar{\Psi} \gamma \overleftarrow{p}_{A}\Psi
\right\}-m\bar{\Psi}\Psi.
$$
The gauge invariant Lagrangian eq.(11.7) takes the form
\begin{equation}
\label{eq: R19.5.19}
\L1_{W}(\D1_{W})=\FFr{i}{2}
\left\{ 
\bp_{W} \gamma \D1_{W} \ps1_{W} -
\bp_{W} \gamma \overleftarrow{\D1_{W}}\ps1_{W}
\right\}- 
\bp_{W}\left\{ 
m +\gamma^{0}\left[M(F) + L(F)\mid \Psi\mid^{2}\right] 
\right\}\ps1_{W}.
\end{equation}
At the symmetry-restoring point, this Lagrangian 
can be replaced by
$$
\L1_{W}(\D1_{W})\rightarrow {\L1_{W}}_{1}(\D1_{W})=
\FFr{1}{2}\left( \D1_{W}{\ps1_{W}}_{1} \right)^{2} -
\V_{W}\left(\mid {\ps1_{W}}_{1} \mid^{2} \right),
$$
provided 
$$
\V_{W}\left(\mid {\ps1_{W}}_{1} \mid^{2} \right)=
-\FFr{1}{2}m^{2}_{\varphi}{\ps1_{W}}_{1}^{2}+
\FFr{1}{4}\lambda^{2}_{\varphi}{\ps1_{W}}_{1}^{4}.
$$
Taking into account the eq.(15.1.8), in which
$
\mid {\ps1_{W}}_{1}\mid^{2}=\mid \varphi\mid^{2}=
\FFr{1}{2}\mid \eta_{\varphi} +\chi\mid^{2},
$
one gets
\begin{equation}
\label{eq: R19.5.23}
{\L1_{W}}_{\varphi}(\D1_{W})=
\FFr{1}{2}\left( \D1_{W}\varphi \right)^{2} -
\V_{W}\left(\mid \varphi \mid^{2} \right),
\quad
\V_{W}\left(\mid \varphi \mid^{2} \right)=
-\FFr{1}{2}m^{2}_{\varphi}\varphi^{2} +
\FFr{1}{4}\lambda^{2}_{\varphi}\varphi^{4}.
\end{equation}
The average value of well-defined current source expressed in terms of 
spinless field $\Psi$ eq.(15.5.3) is given by
$
\left.\vec{j}(\vec{r}) \right|_{\vec{\Sigma}=0}.
$
According to eq.(15.4.6) and eq.(15.4.7), one has
\begin{equation}
\label{eq: R19.6.6}
\begin{array}{l}
\left.\delta G(x,x')\right|_{x=x'}=
-T\S_{\omega}
\IIn\widetilde{G}_{G\Omega}(\vec{r},\vec{s})
\widetilde{G}_{G\Omega}(\vec{l},\vec{r})
\widetilde{G}_{G-\Omega}(\vec{l},\vec{s})
\Delta^{*}(\vec{l})\Delta(\vec{s})d^{3}s\,d^{3}l.
\end{array}
\end{equation}
Thus 
\begin{equation}
\label{eq: R19.6.11}
\left.\vec{j}(\vec{r}) \right|_{\vec{\Sigma}=0}=
\simeq \FFr{m}{4(m-M_{0}-L_{0}\mid \Psi\mid^{2})}
\left( \gamma^{0}+\FFr{\mu}{m}\right)
\left\{ 
\left( \gamma^{0}+\FFr{\mu}{m} \right)^{2}+3\beta^{2}_{F}
\right\}
\vec{j}_{G}(\vec{r}),
\end{equation}
provided
\begin{equation}
\label{eq: R19.6.10}
\vec{j}_{G}(\vec{r})=\left( \FFr{m}{\mu}\right)^{3}
\left\{ 
\FFr{-ie^{*}\hbar}{2m}
\left(\Psi\FFr{\partial \Psi^{*}}{\partial \vec{r}}- 
\Psi^{*}\FFr{\partial \Psi}{\partial \vec{r}}\right)-
\FFr{{e^{*}}^{2}}{mc}\vec{A} | \Psi |^{2}
\right\}.
\end{equation}
Below we write the eq.(15.5.3)
in the form on close analogy of the elementary excitations in 
superconductivity model described by coherent mixture of electrons and
holes near the Fermi surface [10, 11, 53]. In chirial representation
$
\Psi=
\left(
\matrix{
\Psi_{LR}\cr
\Psi_{RL}\cr
}
\right)
$
one has
$$
\begin{array}{l}
p_{A0}\Psi_{LR}=\vec{\sigma}\vec{p}_{A}\Psi_{LR}+m\Psi_{RL}, \quad 
p_{A0}\Psi_{RL}=-\vec{\sigma}\vec{p}_{A}\Psi_{RL}+m\Psi_{LR}, \\ 
p_{A0}=\pm \left(\vec{p}_{A}^{2}+m^{2} \right)^{1/2}.
\end{array}
$$
The two states of quasi-particle are separated in energy by 
$2\mid p_{A0} \mid$. In the ground state all quasi-particles should be
in lower (negative) energy states. It would take a finite 
energy $2\mid p_{A0} \mid\geq 2m$ to excite a particle to the upper
state (the case of Dirac particle). Thus, one may assume that the energy
gap parameter $m$ is also due to some interaction between massless
bare fermions [18,19].
\subsection{Extension to Lower Temperatures}
\label{Ext}
It is worth briefly recording the question of whether or not it is 
possible to extend the ideas of former approach to lower temperatures as 
it was investigated in the case of Gor'kov's theory by others [47-49].
Here, as usual we admit that the order parameter and vector potential vary 
slowly over distances of the order of the coherence length. 
We restrict ourselves to the London limit and the derivation of equations
will be proceeded by iterating to a low order giving only the leading 
terms. Taking into account the eq.(15.4.1), eq.(15.4.3), eq.(15.4.6) 
and eq.(15.4.7), it is straightforward to derive the separate integral
equations for $G$ and $F^{+}$ in terms of $\Delta$, $\Delta^{*}$
and $\widetilde{G}$.
We introduce
\begin{equation}
\label{eq: R19.7.8}
K_{\Omega}(\vec{r},\vec{s})=\delta (\vec{r}-\vec{s})
\left\{
i\Omega + \FFr{1}{2m} \left(\FFr{\partial}{\partial \vec{s}} -
 i\FFr{e}{c}\vec{A}(\vec{s}) \right)^{2}+\mu_{0}
\right\}+
\Delta(\vec{r})\widehat{\gamma}_{A}^{2}(\vec{s})
\widetilde{G}_{-\Omega}(\vec{s},\vec{r})
\Delta^{*}(\vec{s}),
\end{equation}
provided by $\mu_{0}=\FFr{p_{0}^{2}}{2m}$,
and
$$
F^{+}_{\omega}(\vec{s},\vec{r}')=\widehat{\gamma}_{A}(\vec{s})
F^{+}_{\Omega}(\vec{s},\vec{r}'),\quad
\widehat{\gamma}_{A}(\vec{s})=\FFr{1}{2m}\{
\gamma^{0}(i\omega+\mu)-\vec{\gamma}{\vec{p}}_{A}(\vec{s})+m\}.
$$
We write down the coupled equations in the form
\begin{equation}
\label{eq: R19.7.10}
\IIn d^{3}s\, K_{\Omega}(\vec{r},\vec{s})G_{\Omega}(\vec{s},\vec{r}')=
\delta (\vec{r}-\vec{r}'),
\end{equation}
and
\begin{equation}
\label{eq: R19.7.11}
\IIn d^{3}s\, F_{\Omega}^{+}(\vec{s},\vec{r}')K_{-\Omega}(\vec{s},\vec{r})=
\Delta^{*}(\vec{r})\widehat{\gamma}_{A}(\vec{r}')
\widetilde{G}_{\Omega}(\vec{r},\vec{r}'),
\end{equation}
The mathematical structure of obtained equations are closely similar to
that studied by [47,54,55] in somewhat different context. So, adopting 
their technique we introduce sum and difference coordinates, and
Fourier transform with respect to the difference coordinates as follows:
\begin{equation}
\label{eq: R19.7.12}
K_{\Omega}(\vec{p},\vec{R})\equiv \IIn d^{3}(r-s)e^{-i\vec{p}
(\vec{r}-\vec{s})}K_{\Omega}(\vec{r},\vec{s})
\end{equation}
with $\vec{R}\equiv \FFr{1}{2}(\vec{r}+\vec{s})$. We also involve similar 
expansions for all other functions. Then the eq.(15.6.2) and
eq.(15.6.3) reduce to following:
\begin{equation}
\label{eq: R19.7.13}
\begin{array}{l}
\Theta\left[K_{\Omega}(\vec{p},\vec{R}) 
G_{\Omega}(\vec{p}',\vec{R}')
\right]=1,\\ 
\Theta\left[
F^{+}_{\Omega}(\vec{p}',\vec{R}')
K_{-\Omega}(-\vec{p},\vec{R})
\right]=
\Theta\left[
\Delta^{*}(\vec{R})\widehat{\gamma}_{A}(\vec{p}',\vec{R}')
\widetilde{G}_{\Omega}(\vec{p}',\vec{R}')
\right]
\end{array}
\end{equation}
provided by the standard differential operator of finite order
defined under the requirement that it produces the Fourier transform of
the matrix product of two functions when it operates on the transforms
of the individual functions [54,55]
\begin{equation}
\label{eq: R19.7.14}
\Theta\equiv \Lm_{\matrix{
\vec{R}'\rightarrow \vec{R}\cr
\vec{p}'\rightarrow \vec{p}\cr
}}\exp\left[ 
\FFr{i}{2}\left( 
\FFr{\partial}{\partial\vec{R}}\FFr{\partial}{\partial\vec{p}'}-
\FFr{\partial}{\partial\vec{p}}\FFr{\partial}{\partial\vec{R}'}
\right)
\right].
\end{equation}
One gets
\begin{equation}
\label{eq: R19.7.15}
\begin{array}{l}
K_{\Omega}(\vec{p},\vec{R})=i\Omega - \epsilon (\vec{p},\vec{R})+
\Lm_{\vec{R}',\vec{R}''\rightarrow \vec{R}}
\exp\left[ 
\FFr{i}{2}\FFr{\partial}{\partial\vec{p}}
\left( \FFr{\partial}{\partial\vec{R}'}-
\FFr{\partial}{\partial\vec{R}''}\right)
\right]\times \\ 
\times
\Delta^{*}(\vec{R}')\widehat{\gamma}_{A}(\vec{p},\vec{R}')
\widehat{\gamma}_{A}(-\vec{p},\vec{R})\Delta^{*}(\vec{R}''),
\end{array}
\end{equation}
where it is denoted
$
\epsilon (\vec{p},\vec{R})\equiv \FFr{1}{2m}\left( 
\vec{p}-\FFr{e}{c}\vec{A}(\vec{R})
\right)^{2}-\mu_{0}.
$
To obtain resulting expressions we shall proceed with further 
calculations, 
but shall forbear to write them out as they are so standard. 
There is only one thing to be noticed about the integration. That is,
due to the angular
integration in momentum space, as mentioned above, the terms linear in the
vector $\vec{p}$ will be vanished , as well as the integration over the 
energies removes the linear terms in $\epsilon (\vec{p})$. So, we may
expand the quantities in eq.(15.6.5) according to the degree of
inhomogenity somewhat like it we have done in equation (15.4.8) of
gap function $\Delta^{*}(\vec{r})$, which in mixed 
representation transforms to the following:
\begin{equation}
\label{eq: R19.7.17}
\Delta^{*}(\vec{R})=
T\S_{\omega}\IIn\FFr{d^{3}p}{(2\pi)^{3}}
F^{+}_{\omega}(\vec{p},\vec{R})
=T\S_{\omega}\IIn\FFr{d^{3}p}{(2\pi)^{3}}
\widehat{\gamma}_{A}(\vec{p},\vec{R})
F^{+}_{\Omega}(\vec{p},\vec{R}).
\end{equation}
The approximation was used to obtain the function $F^{+}_{\Omega}$
must be of one order higher 
$F^{+}_{\Omega}\simeq F^{(0)+}_{\Omega}+F^{(1)+}_{\Omega}+
F^{(2)+}_{\Omega}$ than that for function 
$\widetilde{G}_{\Omega}\simeq \widetilde{G}_{\Omega}^{(0)}+
\widetilde{G}_{\Omega}^{(1)}$. Employing an iteration method of solution
one replaces $K\rightarrow \widetilde{K},\quad 
G\rightarrow \widetilde{G}$ in eq.(15.6.5) and puts $\Theta^{(0)}=1,\quad
\widetilde{K}^{(1)}=0, \quad \widetilde{G}^{(1)}=0$. Hence 
$\widetilde{G}=\widetilde{G}^{(0)}$. \\
The resulting equation
for gap function is similar to those occurring in [47], although 
not identical. The sole difference is that in the resulting equation we use
the expressions of $\Omega$ and $\xi$. With this replacement the  
equation reads 
\begin{equation}
\label{eq: R19.7.19}
\begin{array}{l}
\Delta^{*}=
T\S_{\omega}\IIn\FFr{d^{3}p}{(2\pi)^{3}}
\left\{ 
\FFr{\Delta^{*}}{\Omega^{2} +\xi^{2}}+
\left[ \left( \FFr{\partial}{\partial \vec{R}}+\FFr{2ie}{c}\vec{A} 
\right)^{2}\Delta^{*}\right.\right.
+\Delta\left( 
\left( \FFr{\partial}{\partial \vec{R}}+\FFr{2ie}{c}\vec{A} 
\right)\Delta^{*}\right)^{2}\FFr{\partial}{\partial|\Delta |^{2}}+\\ 
+\left.\left.\FFr{\Delta^{*}}{3}\FFr{\partial^{2}|\Delta |^{2}}
{\partial \vec{R}^{2}}
\FFr{\partial}{\partial|\Delta |^{2}}+
\FFr{\Delta^{*}}{6}\left(
\FFr{\partial |\Delta |^{2}}{\partial \vec{R}}\right)^{2}
\FFr{\partial^{2}}{\partial\left(|\Delta |^{2}\right)^{2}}
\right]\FFr{p^{2}/6m^{2}}{\left( \Omega^{2}+ \xi^{2}\right)^{2}}
\right\},
\end{array}
\end{equation}
where 
$
\xi(\vec{p},\vec{R})\equiv \left[\epsilon^{2}(\vec{p})+
|\Delta(\vec{R}) |^{2} \right]^{1/2}.
$
The average value of the operator of current
density at $T\sim T_{c}$ follows at once
\begin{equation}
\label{eq: R19.7.21}
\begin{array}{l}
\vec{j}_{G}(\vec{R})=\left(\overline{\gamma^{0}{\widehat{\gamma}}^{3}} 
\right)_{\Sigma=0}\FFr{2e}{m}
\left\{ -\FFr{i}{2}\left( \Delta^{*}(\vec{R})
\FFr{\partial \Delta (\vec{R}) }{\partial \vec{R}}-
\FFr{\partial \Delta^{*} (\vec{R}) }{\partial \vec{R}}\Delta(\vec{R})
\right)-\right.\\ 
\left.-\FFr{2e}{c} |\Delta (\vec{R})|^{2} \vec{A}(\vec{R})\right\}
T\S_{\omega}\IIn\FFr{d^{3}p}{(2\pi)^{3}}
\FFr{p^{2}/3m}{\left( \Omega^{2}+ \xi^{2}(\vec{p},\vec{R})\right)^{2}},
\end{array}
\end{equation}
where 
$\left(\overline{\gamma^{0}{\widehat{\gamma}}^{3}} 
\right)_{\Sigma=0}=\FFr{1}{8}\left( \gamma^{0}+\FFr{\mu}{m}\right)
\left\{ \left( \gamma^{0}+\FFr{\mu}{m}\right)^{2}+3\beta_{F}^{2}
\right\}$. At $\Delta\ll \pi k_{B}T$ and $\vec{A}$ is independent of
position the eq.(15.6.9) and eq.(18.6.10) lead back to the equations
(15.4.13) and (15.5.11).
Actually, from such results it is then easy by ordinary manipulations
to investigate the pertinent
physical problem in several particular cases, but a separate calculation
for each case would be needed.

\renewcommand{\theequation}{\thesection.\arabic{equation}}

\section{The Lagrangian of Electroweak Interactions}
\label{lagr}

The results obtained within the previous sections enable us to trace 
unambiguously rather general scheme of unified electroweak interactions.
Below we remind some features
allowing  us to write down the final Lagrangian of electroweak 
interactions.\\
1. During the realization of multiworld connections of weak interacting 
fermions the P-violation compulsory occurred in
W-world incorporated with the symmetry reduction eq.(13.1)
characterized by the Weinberg mixing angle with
the fixed value at $30^{0}$. This gives rise to 
the local symmetry $SU(2)\otimes U(1)$, under which the
left-handed fermions transformed as six independent doublets, 
while the right-handed fermions transformed as twelve 
independent singlets.\\
2. Due to vacuum rearrangement in Q-world the Yukawa couplings arise 
between the fermion fields and corresponding isospinor-scalar $\varphi$-
meson in conventional form.\\
3. In the framework of suggested mechanism providing the effective 
attraction between the relativistic fermions caused by the exchange of 
the mediating induced gauge quanta in W-world, the 
self-interacting isospinor-scalar Higgs bosons arise 
as Bose-condensate. Then we must add to the total
Lagrangian a $SU(2)$ multiplet of spinless $\varphi$-meson fields
coupled to the gauge fields in a gauge invariant way. We involve the 
Lagrangian of $\varphi$-meson with the degenerate 
vacuum of W-world, where the symmetry-breaking Higgs 
boson is counted off from the gap symmetry restoring
value as the point of origin.
In view of this a Lagrangian ensues from the eq.(11.5)-
eq.(11.7), which is now invariant under
local symmetry $SU(2)\otimes U(1)$, where a set of gauge fields are coupled 
to various multiplets of fields among which is also a multiplet of 
Higgs boson. Subsequently, we separate a part of Lagrangian 
containing only the fields defined on four dimensional Minkowski 
flat spacetime continuum $M^{4}$.
The resulting Lagrangian reads 
\begin{equation}
\label{eq: R20.1}
\begin{array}{l}
L=-\FFr{1}{2}TrG_{\mu\nu}G^{\mu\nu}-\FFr{1}{4}F_{\mu\nu}F^{\mu\nu}+
i\bar{L}\hat{D}L +i\bar{e}_{R}\hat{D}e_{R}+ 
i\bar{\nu}_{R}\hat{D}\nu_{R}+|D_{\mu}\varphi |^{2}-\\
\FFr{1}{2}\lambda^{2}_{\varphi}\left( |\varphi |^{2}-
\FFr{1}{2}\eta^{2}_{\varphi}\right)^{2}- 
f_{e}\left(\bar{L} \varphi e_{R}+\bar{e}_{R} \varphi^{+}L \right)
-f_{\nu}\left(\bar{L} \varphi_{c} \nu_{R}+\bar{\nu}_{R} 
\varphi^{+}_{c}L \right)+\\ 
\mbox{similar terms for other fermion generations}.
\end{array}
\end{equation}
The $\varphi$ is taken to denote Higgs boson according to redefinition
$
\varphi (x)\equiv \varphi(x,\x1_{W})=f_{exp}(x)\varphi(\x1_{W}), 
$
where the coordinates $\x1_{W}\in {\M1_{W}}^{4}$ are left implicit.
The gauge fields ${\bf A}_{\mu}(x)$ and $B_{\mu}(x)$ associate 
respectively with the groups $SU(2)$ and $U(1)$,
where the gauge covariant curls are $F_{\mu\nu}, G_{\mu\nu}.$
The corresponding gauge covariant derivatives are in standard form.
One took into account corresponding values of the operators
${\bf T}$ and $Y$ for left- and right-handed fields, and for isospinor
$\varphi$-meson. The Yukawa coupling constants $f_{e}$ and
$f_{\nu}$ are inserted in eq.(14.1), but in the case of quarks,
according to eq.(2.2) and eq.(10.9), it respectively must be changed 
into $f_{e}\rightarrow f_{q} \equiv f_{Q}+m^{c}_{i}$.
In standard scenario a gauge invariance of the Lagrangian is broken when
the $\varphi$-meson fields acquire a 
vacuum expectation value. Thereby the mass $m_{\varphi}$ and coupling 
constant $\lambda_{\varphi}$ are in the form eq.(15.4.15). 
The spontaneous breakdown of symmetry is vanished at 
$\eta_{\varphi}^{2}(\lambda, T > T_{c\mu}) < 0$.
When this doublet obtains a vacuum expectation value, three of the
gauge fields acquire masses.
These fields will mediate the weak interactions. Consequently, a remaining 
massless gauge field will be identified as the photon field coupled to the electric current.
The microscopic structure
of these fields reads
$$
\begin{array}{l}
W^{+}\equiv {\ps1_{\eta}}_{W}(\eta)\, 
(q_{1}q_{2}q_{3})^{Q}(q\bar{q})^{w}, \quad
W^{-}\equiv {\ps1_{\eta}}_{W}(\eta)\, 
(\overline{q_{1}q_{2}q_{3}})^{Q}(\bar{q}q)^{w}, \\
Z^{0}\equiv {\ps1_{\eta}}_{Z}(\eta)\, 
(q\bar{q})^{Q}(q\bar{q})^{w},\quad
A \equiv {\ps1_{\eta}}_{A}(\eta)\, 
(q\bar{q})^{Q}.
\end{array}
$$
The expressions of the masses $m_{W}$ and $m_{Z}$ are changed if the Higgs
sector is built up more compoundly. 
Due to Yukawa couplings the fermions acquire the masses after 
symmetry-breaking. The mass of electron reads
$
m_{e}=\FFr{\eta_{\varphi}}{\sqrt{2}}f_{e}
$
etc. One gets for the leptons
$
f_{e}:f_{\mu}:f_{\tau}=m_{e}:m_{\mu}:m_{\tau}.
$
This mechanism does not disturb the
renormalizability of the theory  [56,57].
In approximation to lowest order
$
f=\Sigma_{Q}\simeq m_{Q}\ll \lambda^{-1/2}\quad
\left( \lambda^{-1}=\FFr{mp_{0}}{2\pi^{2}}
\ln\FFr{2\widetilde{\omega}}{\Delta_{0}}\right),
$
the Lagrangian
eq.(16.1) leads to Lagrangian of phenomenological
standard model. 
At $f\sim 10^{-6}, \quad \lambda\ll 10^{12}$.
In this case the phenomenological standard model survives. 
The self-energy operator $f=\Sigma_{Q}$ takes into
account the mass-spectrum of expected various collective excitations of
bound quasi-particle pairs produced by higher-order interactions, which 
arise as the poles of the function $\Sigma_{Q}$ eq.(11.1). 

\section{Quark flavour Mixing and the Cabibbo Angle}
\label{Cabib}

An implication of quark generations into this scheme will be carried out
in the same manner. Namely, under the group $SU(2)\otimes U(1)$ the 
left-handed quarks transform as three doublets, while the right-handed
quarks transform as independent singlets except of following 
differences:\\
1. The values of weak-hypercharge of quarks are changed is due to their
fractional electric charges
$
q_{L}:Y^{w}=\FFr{1}{3},\quad u_{R}:Y^{w}=\FFr{4}{3},\quad
d_{R}:Y^{w}=-\FFr{2}{3}$
etc.\\
2. All Yukawa coupling constants have nonzero values.\\
3. An appearance of quark mixing and Cabibbo angle, which
is unknown in the scope of standard model.\\
4. An existence of CP-violating phase in unitary matrix of quark
mixing. We shall discuss it in the next 
section.\\
Below, we attempt to give an
explanation to quark mixing and Cabibbo angle.
Here for simplicity, we consider this problem on the example of four quarks 
$u,d,s,c$. The further implication of all quarks would complicate 
the problem only in algebraic sense.
Let consider four left-handed quarks forming a $SU(2)_{L}$ doublets mixed
with Cabibbo angle
$
\left(\matrix{
u'\cr
d\cr}\right)_{L}, \quad \mbox{and}\quad
\left(\matrix{
c'\cr
s\cr}\right)_{L},
$
where $u'=u\cos \theta + c\sin \theta, \quad
c'=-u\sin \theta + c\cos \theta$.
One must distinguish two kind of fermion states:
an eigenstate of gauge interactions, i.e. the fields of $u'$
and $c'$; an eigenstate of mass-matrices, i.e the fields of
$u$ and $c$. The qualitative properties
of Cabibbo flavour mixing could be understood in terms of Yukawa couplings.
Unlike the case of leptons, where the Yukawa couplings are
characterized  by two constants $f_{e}$ and $f_{\mu}$, the interaction
of Higgs boson with $u'$ and $c'$ is due to following three terms:
$$
\begin{array}{l}
\FFr{1}{\sqrt{2}}f_{u'}(\bar{u}'_{L}u'_{R}+\bar{u}'_{R}u'_{L})
(\eta +\chi)=
\FFr{1}{\sqrt{2}}f_{u'}(\bar{u}'u')(\eta +\chi), \\
\FFr{1}{\sqrt{2}}f_{c'}(\bar{c}'_{L}c'_{R}+\bar{c}'_{R}c'_{L})
(\eta +\chi)=
\FFr{1}{\sqrt{2}}f_{c'}(\bar{c}'c')(\eta +\chi), \\ 
\FFr{1}{\sqrt{2}}f_{u'c'}(\bar{c}'_{L}u'_{R}+\bar{c}'_{R}u'_{L}+
\bar{u}'_{R}c'_{L}+\bar{u}'_{L}c'_{R})
(\eta +\chi)=
\FFr{1}{\sqrt{2}}f_{u'c'}(\bar{c}'u'+\bar{u}'c')(\eta +\chi).
\end{array}
$$
Throughout a small part our discussion will be a standard [58], 
except one point: instead of mixing of fields 
$d'$ and $s'$ we consider a quite equivalent mixing of $u'$ and $c'$.
The last expression may be diagonalized by means of rotation
right through Cabibbo angle. In the sequel one gets
$
m_{u}\bar{u} u+m_{c}\bar{c} c,
$
where $m_{u}$ and $m_{c}$ are masses of quarks $u$ and $c$. Straightforward
comparison of two states gives
$$
\begin{array}{l}
m_{u}=\FFr{1}{\sqrt{2}}(f_{u'}\cos^{2}\theta+f_{c'}\sin^{2}\theta-
2f_{u'c'}\cos\theta\sin\theta)\eta, \\ 
m_{c}=\FFr{1}{\sqrt{2}}(f_{u'}\sin^{2}\theta+f_{c'}\cos^{2}\theta+
2f_{u'c'}\cos\theta\sin\theta)\eta, \\ 
\tan 2\theta=\FFr{2f_{u'c'}}{f_{c'}-f_{u'}}\neq 0.
\end{array}
$$
Similar formulas can be worked out for the other mixing.
Thus, the nonzero value of Cabibbo angle arises due to nonzero coupling
constant $f_{u'c'}$. Then the problem is to calculate all coupling
constants $f_{u'c'}$,$f_{c't'}$, and $f_{t'u'}$ generating three Cabibbo 
angles
$$
\tan 2\theta_{2}=\FFr{2f_{u'c'}}{f_{c'}-f_{u'}},\quad
\tan 2\theta_{3}=\FFr{2f_{c't'}}{f_{t'}-f_{c'}},\quad
\tan 2\theta_{1}=\FFr{2f_{t'u'}}{f_{u'}-f_{t'}}.
$$
Taking into account the explicit form of Q-components of quark fields 
eq.(8.1)
$$
{\ps1_{Q}}_{u'}=(q_{1}q_{2})^{Q}, \quad
{\ps1_{Q}}_{c'}=(q_{2}q_{3})^{Q}, \quad
{\ps1_{Q}}_{t'}=(q_{3}q_{1})^{Q}, 
$$
also eq.(11.2) and eq.(14.1), we may write down
\begin{equation}
\label{eq: R21.8}
f_{u'}\rightarrow \FFr{1}{2}\left\{
{\bp_{u}}_{u'}\hat{p}_{u}{\ps1_{u}}_{u'}-
\left( {\bp_{u}}_{u'}\overleftarrow{\hat{p}}_{u} \right) {\ps1_{u}}_{u'}
\right\}
=\left( \Sigma_{Q}^{1}+\Sigma_{Q}^{2}\right)
{\bp_{u}}_{u'}{\ps1_{u}}_{u'}\rightarrow 
\left( \Sigma_{Q}^{1}+\Sigma_{Q}^{2}\right), 
\end{equation}
where for given $(i)$ one has 
$
\hat{p}_{Q}\,q^{Q}_{i}=\Sigma_{Q}^{i}\,q^{Q}_{i}.
$
In analogy, the $f_{c'}$ and $f_{u'c'}$ imply
\begin{equation}
\label{eq: R21.10}
f_{c'}\rightarrow \FFr{1}{2}\left\{
{\bp_{u}}_{c'}\hat{p}_{u}{\ps1_{u}}_{c'}-
\left( {\bp_{u}}_{c'}\overleftarrow{\hat{p}}_{u} \right) {\ps1_{u}}_{c'}
\right\}\rightarrow 
\left( \Sigma_{Q}^{2}+\Sigma_{Q}^{3}+m^{c}_{c}\right), 
\end{equation}
and 
\begin{equation}
\label{eq: R21.11}
\begin{array}{l}
f_{u'c'}\rightarrow \FFr{1}{4}\left\{
{\bp_{Q}}_{u'}\hat{p}_{Q}{\ps1_{Q}}_{c'}+
{\bp_{Q}}_{c'}\hat{p}_{Q}{\ps1_{Q}}_{u'}-
\left( {\bp_{Q}}_{u'}\overleftarrow{\hat{p}}_{Q} \right) {\ps1_{Q}}_{c'}-
\left( {\bp_{Q}}_{c'}\overleftarrow{\hat{p}}_{Q} \right) {\ps1_{Q}}_{u'}
\right\}=\\ \\
=\FFr{1}{2}\left\{
\left( \Sigma_{Q}^{1}+\Sigma_{Q}^{2}\right){\bp_{Q}}_{u'}{\ps1_{Q}}_{c'}+
\left( \Sigma_{Q}^{2}+\Sigma_{Q}^{3}\right)
{\bp_{Q}}_{c'}{\ps1_{Q}}_{u'}
\right\}.
\end{array}
\end{equation}
In accordance with eq.(3.4), one has
$$
\begin{array}{l}
{\bp_{Q}}_{u'}{\ps1_{Q}}_{c'}=
(\overline{q_{1}q_{2}})^{Q}(q_{2}q_{3})^{Q}=
f^{u'c'}_{Q2}(\bar{q}_{2}q_{2})^{Q},
\\
{\bp_{Q}}_{c'}{\ps1_{Q}}_{u'}=
(\overline{q_{2}q_{3}})^{Q}(q_{1}q_{2})^{Q}=
f^{c'u'}_{Q2}(\bar{q}_{2}q_{2})^{Q},
\end{array}
$$
where 
$$f^{c'u'}_{Q2}(\bar{q}_{2}q_{2})^{Q}=\left(
f^{u'c'}_{Q2}(\bar{q}_{2}q_{2})^{Q}
\right)^{*}=f^{c'u'}_{Q2}(\bar{q}_{2}q_{2})^{Q}\equiv
\bar{f}_{2}.$$
The partial formfactors $\bar{f}\equiv f^{AB}_{il}$ imply the 
functional equation of renormalization 
group eq.(3.6). 
Hence
$
f_{u'c'}
=\FFr{\bar{f}_{2}}{2}
\left( \Sigma_{Q}^{1}+\Sigma_{Q}^{3}+
2\Sigma_{Q}^{2}\right)
$ and
\begin{equation}
\label{eq: R21.16}
\begin{array}{l}
\tan 2\theta_{2}=
\FFr{\bar{f}_{2}\left( \Sigma_{Q}^{1}+\Sigma_{Q}^{3}+2\Sigma_{Q}^{2}
\right)}
{\Sigma_{Q}^{3}-\Sigma_{Q}^{1}+m^{c}_{c}},\quad
\tan 2\theta_{3}=
\FFr{\bar{f}_{3}\left( \Sigma_{Q}^{2}+\Sigma_{Q}^{1}+2\Sigma_{Q}^{3}
\right)}
{\Sigma_{Q}^{1}-\Sigma_{Q}^{2}+m^{c}_{t}-m^{c}_{c}},\\
\tan 2\theta_{1}=
\FFr{\bar{f}_{1}\left( \Sigma_{Q}^{3}+\Sigma_{Q}^{2}+2\Sigma_{Q}^{1}
\right)}
{\Sigma_{Q}^{2}-\Sigma_{Q}^{3}-m^{c}_{t}},
\end{array}
\end{equation}
where a rest of $\bar{f}_{i}$ reads
$\bar{f}_{3}\equiv f^{c't'}_{Q3}=f^{t'c'}_{Q3}$ and
$\bar{f}_{1}\equiv f^{t'u'}_{Q1}=f^{u't'}_{Q1}$.
So, the Q-components of the quark fields $u',c'$ and $t'$
contain at least one identical subquark, due to which in
eq.(3.4) the partial formfactors 
$\bar{f}_{i}$ have nonzero values causing a
quark mixing with the Cabibbo angles eq.(17.4). Therefore, 
the unimodular orthogonal group of global rotations arises, and the quarks 
$u',c'$ and $t'$ come up in doublets 
$(u',c')$,$(c',t')$, and $(t',u')$. For the leptons
these formfactors equal zero 
$\bar{f}_{i}^{lept}\equiv 0$, because of
eq.(7.1), namely the lepton mixing is absent.
In conventional notation 
$
\left(\matrix{
u'\cr
d\cr}\right)_{L}, 
\left(\matrix{
c'\cr
s\cr}\right)_{L},
\left(\matrix{
t'\cr
b\cr}\right)_{L}\rightarrow
\left(\matrix{
u\cr
d'\cr}\right)_{L}, 
\left(\matrix{
c\cr
s'\cr}\right)_{L},
\left(\matrix{
t\cr
b'\cr}\right)_{L},
$
which gives rise to
$
f_{u'c'}\rightarrow f_{d's'},\quad
f_{c't'}\rightarrow f_{s'b'},\quad
f_{t'u'}\rightarrow f_{b'd'},\quad
f_{u'}\rightarrow f_{d'},\quad 
f_{c'}\rightarrow f_{s'},\quad 
f_{t'}\rightarrow f_{b'},\quad 
f_{d}\rightarrow f_{u},\quad
f_{s}\rightarrow f_{c},\quad
f_{b}\rightarrow f_{t}.
$
\section{The Appearance of the CP-Violating Phase}
\label{Phase}

The required magnitude of the CP-(or time reversal T- related to CP by CPT 
theorem [63]) violating complex parameter $\varepsilon$
[64] depends upon the specific choice of theoretical model for
explaining the $ K^{0}_{2}\rightarrow 2\pi $ decay.
From the experimental data it is somewhere
$
| \varepsilon |\simeq 2.3\times 10^{-3}.
$
In the framework of Kobayashi-Maskawa (KM) parametrization of unitary matrix
of quark mixing [65], this parameter may be expressed in terms of 
three Eulerian angles of global rotations in the three dimensional quark 
space and one phase parameter. Below we attempt to derive the KM-matrix with
an explanation given to an appearance of the CP-violating 
phase.
We recall that during the realization of multiworld structure 
the P-violation 
compulsory occurred in the W-world provided by the spanning
eq.(12.1). The three dimensional effective space 
$W^{loc}_{v}(3)$ arises as follows:
\begin{equation}
\label{eq: R22.3}
\begin{array}{l}
W^{loc}_{v}(3)\ni q^{(3)}_{v}=
\left( \matrix{
q^{w}_{R}(\vec{T}=0)\cr
\cr
q^{w}_{L}(\vec{T}=\FFr{1}{2})\cr
}\right)\equiv\\ \\
\equiv
\left( \matrix{
u_{R},d_{R}\cr
\cr
\left( \matrix{
u'\cr
d\cr
}\right)_{L}
\cr
}\right),
\left( \matrix{
c_{R},s_{R}\cr
\cr
\left( \matrix{
c'\cr
s\cr
}\right)_{L}
\cr
}\right),
\left( \matrix{
t_{R},b_{R}\cr
\cr
\left( \matrix{
t'\cr
b\cr
}\right)_{L}
\cr
}\right)
\equiv
\left( \matrix{
q^{w}_{3}\cr
\cr
\left( \matrix{
q^{w}_{1}\cr
q^{w}_{2}\cr
}\right)
\cr
}\right),
\left( \matrix{
q^{w}_{1}\cr
\cr
\left( \matrix{
q^{w}_{2}\cr
q^{w}_{3}\cr
}\right)
\cr
}\right),
\left( \matrix{
q^{w}_{2}\cr
\cr
\left( \matrix{
q^{w}_{3}\cr
q^{w}_{1}\cr
}\right)
\cr
}\right),
\end{array}
\end{equation}
where the subscript $(v)$ formally specifies a vertical direction of
multiplet, the subquarks $q^{w}_{\alpha} (\alpha=1,2,3)$ associate with 
the local rotations around corresponding axes of three dimensional
effective space $W^{loc}_{v}(3)$. The local gauge transformations
$f^{v}_{exp}$ are implemented upon the multiplet 
${q'}^{(3)}_{v}=f^{v}_{exp} q^{(3)}_{v}$, where 
$f^{v}_{exp}\in SU^{loc}(2)\otimes U^{loc}(1)$. 
If for the moment we leave it intact and make a closer examination of the 
composition of middle row in eq.(18.1), then we distinguish the other 
symmetry arisen along the horizontal line $(h)$. 
The situation is exactly similar to that discussed in
sec.12:
due to the specific structure of W-world implying the condition 
of realization of multiworld connections eq.(5.5) with 
$\vec{T}\neq 0, \quad Y^{w}\neq 0$, the subquarks $q^{w}_{\alpha}$
tend to be compulsory involved into triplet. They form
one ``doublet''  $\vec{T}\neq 0$ and one singlet $Y^{w}\neq 0$. Then the
quarks $u'_{L},c'_{L}$ and $t'_{L}$ form
a $SO^{gl}(2)$ ``doublet'' and a $U^{gl}(1)$ singlet 
\begin{equation}
\label{eq: R22.4}
\begin{array}{l}
\left( \left( u'_{L},c'_{L}\right)t'_{L}\right)\equiv
\left( \left( q^{w}_{1},q^{w}_{2}\right)q^{w}_{3}\right)\equiv
q^{(3)}_{h}\in W^{gl}_{h}(3),\\ 
\left( u'_{L},\left( c'_{L},t'_{L}\right)\right)\equiv
\left( q^{w}_{1},\left( q^{w}_{2},q^{w}_{3}\right)\right),\quad
\left( \left( t'_{L},u'_{L}\right)c'_{L}\right)\equiv
\left( \left( q^{w}_{3},q^{w}_{1}\right),q^{w}_{2}\right).
\end{array}
\end{equation}
Here $W^{gl}_{h}(3)$ is the three dimensional effective space in
which the global rotations occur. They are implemented upon the
triplets through the transformation matrix $f^{h}_{exp}$: 
$
{q'}^{(3)}_{h}=f^{h}_{exp} q^{(3)}_{h},
$
which reads (eq.(18.2))
$$
f^{h}_{exp}=\left( \matrix{
f_{33} &0 &0\cr
0 & c & s\cr
0 & -s & c\cr
}\right)
$$
in the notation $c=\cos \theta, \quad s=\sin \theta$. This 
implies the incompatibility relation eq.(3.5.5) [1], namely
\begin{equation}
\label{eq: R22.7}
\|f^{h}_{exp}\|=f_{33}(f_{11}f_{22}-f_{12}f_{21})=
f_{33}\varepsilon_{123}\varepsilon_{123}\|f^{h}_{exp}\|f^{*}_{33}.
\end{equation}
That is
$
f_{33}f^{*}_{33}=1,
$
or
$
f_{33}=e^{i\delta}
$
and $\|f^{h}_{exp}\|=1$.
The general rotation in $W^{gl}_{h}(3)$ is described by Eulerian three angles
$\theta_{1},\theta_{2},\theta_{3}$. If we put the arisen phase only in the 
physical sector then a final KM-matrix of quark flavour mixing would result
\begin{equation}
\label{eq: R22.10}
\begin{array}{l}
\left( \bar{u}_{L},\bar{c}_{L},\bar{t}_{L}\right)V_{K-M}
\left( \matrix{
d\cr
s\cr
b\cr
}\right)\equiv
\left( \bar{u}'_{L},\bar{c}'_{L},\bar{t}'_{L}\right)
\left( \matrix{
d\cr
s\cr
b\cr
}\right)\equiv
\left( \bar{u}_{L},\bar{c}_{L},\bar{t}_{L}\right)
\left( \matrix{
d'\cr
s'\cr
b'\cr
}\right)=\\ 
=
\left( \bar{u}_{L},\bar{c}_{L},\bar{t}_{L}\right)
\left( \matrix{
1 &0 &0\cr
0 &c_{2} &s_{2}\cr
0 &-s_{2} &c_{2}\cr
}\right)
\left( \matrix{
c_{1} &s_{1} &0\cr
-s_{1} &c_{1} &0\cr
0 &0 &e^{i\delta}\cr
}\right)
\left( \matrix{
1 &0 &0\cr
0 &c_{3} &s_{3}\cr
0 &-s_{3} &c_{3}\cr
}\right)
\left( \matrix{
d\cr
s\cr
b\cr
}\right),
\end{array}
\end{equation}
where
$$
\left( \bar{u}'_{L},\bar{c}'_{L},\bar{t}'_{L}\right)\equiv
\left( \bar{u}_{L},\bar{c}_{L},\bar{t}_{L}\right)V_{K-M}, \quad
\left( \matrix{
d'\cr
s'\cr
b'\cr
}\right)\equiv
V_{K-M}
\left( \matrix{
d\cr
s\cr
b\cr
}\right).
$$
The CP-violating parameter $\varepsilon$ 
approximately is written [58, 62]
$
\varepsilon\sim s_{1}s_{2}s_{3}\sin \delta\neq 0.
$
While the spanning
$W^{loc}_{v}(2)\rightarrow W^{loc}_{v}(3)$ eq.(18.1) underlies 
the P-violation and the expanded symmetry
$G^{loc}_{v}(3)=SU^{loc}(2)\otimes U^{loc}(1)$, the 
CP-violation stems from the similar
spanning $W^{gl}_{h}(2)\rightarrow W^{gl}_{h}(3)$ eq.(18.2) 
with the expanded global symmetry group.

\section{The Mass-Spectrum of Leptons and Quarks}
\label{Mass}
The mass-spectrum of leptons and quarks
stems from their internal multiworld structure eq.(7.1) and eq.(8.1)
incorporated with the quark mixing eq.(17.4). We start a discussion 
with the leptons. It might be worthwhile 
to adopt a simple viewpoint on Higgs sector that
$
\eta^{l}\equiv \eta_{e}=\eta_{\mu}=\eta_{\tau}=\eta_{\nu_{e}}=
\eta_{\nu_{\mu}}=\eta_{\nu_{\tau}}.
$
According to eq.(5.6), a real dynamical field in the
state $(q_{1}q_{2}q_{3})^{Q}_{i}$ is the field of subquark 
$q^{Q}_{i}$ providing a mass of this state, namely
$
m^{Q}_{(q_{1}q_{2}q_{3})^{Q}_{i}}\equiv \Sigma_{Q}^{i}
$. Whence the explicit expressions of the 
masses of leptons, in accordance with eq.(7.1)), read
$
m_{i}=\FFr{\eta^{l}}{\sqrt{2}}f_{i}\equiv 
\FFr{\eta^{l}}{\sqrt{2}}\Sigma_{Q}^{i},
$
where
$
\Sigma_{Q}^{1}:\Sigma_{Q}^{2}:\Sigma_{Q}^{3}=
m_{e}:m_{\mu}:m_{\tau},
$
and (eq.(14.1))
$
m_{\nu_{i}}=\FFr{\eta^{l}}{\sqrt{2}}f_{\nu_{i}}\equiv 
\FFr{\eta^{l}}{\sqrt{2}}c_{\nu}\Sigma_{Q}^{i}.
$
In the case of quarks, we admit in analogy that
$
\eta^{q}\equiv \eta_{u}=\eta_{d}=\eta_{s}=\eta_{c}=
\eta_{b}=\eta_{t}.
$
Taking into account the eq.(9.1) and eq.(2.2), one may write down
the expressions of the primary mass parameters of 
states $q^{c}_{i} (i=s,c,b,t)$
$
m^{c}_{s}=m^{c}_{(\overline{q_{1}q_{2}q_{3}})^{c}_{s}}, \quad
m^{c}_{c}=m^{c}_{(q_{1}q_{2}q_{3})^{c}_{c}},\quad 
m^{c}_{b}=m^{c}_{(\overline{q_{1}q_{2}q_{3}})^{c}_{b}},\quad 
m^{c}_{t}=m^{c}_{(q_{1}q_{2}q_{3})^{c}_{t}}.
$
According to eq.(8.1), we derive the masses of quarks\\
\begin{equation}
\label{eq: R23.8}
\begin{array}{l}
m_{d}=\FFr{\eta^{q}}{\sqrt{2}}f_{d}=
\FFr{\eta^{q}}{\sqrt{2}}\Sigma^{3}_{Q},\quad
m_{s}=\FFr{\eta^{q}}{\sqrt{2}}f_{s}=
\FFr{\eta^{q}}{\sqrt{2}}\left(\Sigma^{1}_{Q}+m^{c}_{s}\right),\quad
m_{b}=\FFr{\eta^{q}}{\sqrt{2}}f_{b}=
\FFr{\eta^{q}}{\sqrt{2}}\left(\Sigma^{2}_{Q}+m^{c}_{b}\right),\\
m_{u}=\FFr{\eta^{q}}{\sqrt{2}}\left\{ 
\left( \Sigma^{1}_{Q}+ \Sigma^{2}_{Q} \right)\cos^{2} \theta_{2}
+\left( \Sigma^{2}_{Q}+ \Sigma^{3}_{Q} +m^{c}_{c}\right)\sin^{2} 
\theta_{2} -
\FFr{\bar{f}_{2}}{2}
\left( \Sigma^{1}_{Q}+ \Sigma^{3}_{Q} + \right.\right.\\
\left.\left.
2\Sigma^{2}_{Q}\right)
\sin 2\theta_{2}
\right\}=
\FFr{\eta^{q}}{\sqrt{2}}\left\{ 
\left( \Sigma^{1}_{Q}+ \Sigma^{2}_{Q} \right)\cos^{2} \theta_{1}
+\left( \Sigma^{3}_{Q}+ \Sigma^{1}_{Q} +m^{c}_{t}\right)\sin^{2} 
\theta_{1} + \FFr{\bar{f}_{1}}{2}
\left( \Sigma^{3}_{Q}+ \Sigma^{2}_{Q} + \right.\right.\\
\left.\left.
2\Sigma^{1}_{Q}\right)
\sin 2\theta_{1}
\right\},\\
m_{c}=\FFr{\eta^{q}}{\sqrt{2}}\left\{ 
\left( \Sigma^{2}_{Q}+ \Sigma^{3}_{Q} +m^{c}_{c} \right)\cos^{2} 
\theta_{2}
+\left( \Sigma^{1}_{Q}+ \Sigma^{2}_{Q} \right)\sin^{2} 
\theta_{2} +
\FFr{\bar{f}_{2}}{2}
\left( \Sigma^{1}_{Q}+ \Sigma^{3}_{Q} + \right.\right.\\
\left.\left.
2\Sigma^{2}_{Q}\right)
\sin 2\theta_{2}
\right\}=
\FFr{\eta^{q}}{\sqrt{2}}\left\{ 
\left( \Sigma^{2}_{Q}+ \Sigma^{3}_{Q} +m^{c}_{c} \right)\cos^{2} 
\theta_{3}
+\left( \Sigma^{3}_{Q}+ \Sigma^{1}_{Q} +m^{c}_{t}\right)\sin^{2} 
\theta_{3} -
\FFr{\bar{f}_{3}}{2}
\left( \Sigma^{2}_{Q}+ \right.\right.\\
\left.\left.
\Sigma^{1}_{Q} +2\Sigma^{3}_{Q}\right)
\sin 2\theta_{3}
\right\},\\
m_{t}=\FFr{\eta^{q}}{\sqrt{2}}\left\{ 
\left( \Sigma^{3}_{Q}+ \Sigma^{1}_{Q} +m^{c}_{t} \right)\cos^{2} 
\theta_{3}
+\left( \Sigma^{2}_{Q}+ \Sigma^{3}_{Q}+m^{c}_{c} \right)\sin^{2} 
\theta_{3} +
\FFr{\bar{f}_{3}}{2}
\left( \Sigma^{2}_{Q}+ \Sigma^{1}_{Q} + \right.\right.\\
\left.\left.
2\Sigma^{3}_{Q}\right)
\sin 2\theta_{3}
\right\}=
\FFr{\eta^{q}}{\sqrt{2}}\left\{ 
\left( \Sigma^{3}_{Q}+ \Sigma^{1}_{Q} +m^{c}_{t} \right)\cos^{2} 
\theta_{1}
+\left( \Sigma^{1}_{Q}+ \Sigma^{2}_{Q} \right)\sin^{2} 
\theta_{1} -
\FFr{\bar{f}_{1}}{2}
\left( \Sigma^{3}_{Q}+ \Sigma^{2}_{Q} + \right.\right.\\
\left.\left.
2\Sigma^{1}_{Q}\right)
\sin 2\theta_{1}
\right\},
\end{array}
\end{equation}
where in view of eq.(17.4) the  Cabibbo angles imply
\begin{equation}
\label{eq: R21.16}
\begin{array}{l}
\tan 2\theta_{2}=
\FFr{\bar{f}_{2}\left(m_{e}+m_{\tau}+2m_{\mu}
\right)}
{m_{\tau}-m_{e}+\bar{m}_{c}},\quad
\tan 2\theta_{3}=
\FFr{\bar{f}_{3}\left( m_{\mu}+m_{e}+2m_{\tau}
\right)}
{m_{e}-m_{\mu}+\bar{m}_{t}-\bar{m}_{c}},\\ 
\tan 2\theta_{1}=
\FFr{\bar{f}_{1}\left( m_{\tau}+m_{\mu}+2m_{e}
\right)}
{m_{\mu}-m_{\tau}-\bar{m}_{t}}.
\end{array}
\end{equation}
This is legitimate for
$
\bar{m}_{i}=\FFr{\eta^{l}}{\sqrt{2}}m^{c}_{i}, \quad i=s,c,b,t.
$
By making use of eq.(19.1), 
some useful relations between the masses of leptons and quarks are written
\begin{equation}
\label{eq: R23.10}
\begin{array}{l}
m_{d}=\FFr{\eta^{q}}{\eta^{l}}m_{\tau},\quad
m_{s}=\FFr{m_{d}}{m_{\tau}}\left(m_{e}+\bar{m}_{s}\right),\quad
m_{b}=\FFr{m_{d}}{m_{\tau}}\left(m_{\mu}+\bar{m}_{b}\right),\\ 
m_{u}=\FFr{m_{d}}{m_{\tau}}\left\{
\left(  m_{e}+m_{\mu} \right)\cos^{2} \theta_{2}
+\left(m_{\mu}+m_{\tau}+\bar{m}_{c}\right)\sin^{2} 
\theta_{2} -
\FFr{\bar{f}_{2}}{2}
\left( m_{e}+m_{\tau}+\right.\right.\\
\left.\left.
2m_{\mu}\right)\sin 2\theta_{2}
\right\}=
\FFr{m_{d}}{m_{\tau}}\left\{
\left( m_{e}+m_{\mu}  \right)\cos^{2} \theta_{1}
+\left( m_{\tau}+m_{e}+\bar{m}_{t}\right)\sin^{2} 
\theta_{1} +
\FFr{\bar{f}_{1}}{2}
\left( m_{\tau}+m_{\mu}+\right.\right.\\
\left.\left.
2m_{e}\right)\sin 2\theta_{1}
\right\},\\
m_{c}=\FFr{m_{d}}{m_{\tau}}\left\{
\left( m_{\mu}+m_{\tau}+\bar{m}_{c} \right)\cos^{2} 
\theta_{2}
+\left(   m_{e}+m_{\mu} \right)\sin^{2} 
\theta_{2} +
\FFr{\bar{f}_{2}}{2}
\left(  m_{e}+m_{\tau}+\right.\right.\\
\left.\left.
2m_{\mu}\right)\sin 2\theta_{2}
\right\}=
\FFr{m_{d}}{m_{\tau}}\left\{
\left( m_{\mu}+m_{\tau}+\bar{m}_{c} \right)\cos^{2} 
\theta_{3}
+\left(  m_{\tau}+m_{e}+\bar{m}_{t}\right)\sin^{2} 
\theta_{3} -
\FFr{\bar{f}_{3}}{2}
\left(  m_{\mu}+\right.\right.\\
\left.\left.
m_{e}+2m_{\tau}\right)\sin 2\theta_{3}
\right\},\\
m_{t}=\FFr{m_{d}}{m_{\tau}}\left\{
\left( m_{\tau}+m_{e}+\bar{m}_{t}\right)\cos^{2} 
\theta_{3}
+\left(m_{\mu}+m_{\tau}+\bar{m}_{c}\right)\sin^{2} 
\theta_{3} +
\FFr{\bar{f}_{3}}{2}
\left(  m_{\mu}+m_{e}+\right.\right.\\
\left.\left.
2m_{\tau}\right)\sin 2\theta_{3}
\right\}=
\FFr{m_{d}}{m_{\tau}}\left\{
\left(  m_{\tau}+m_{e}+\bar{m}_{t}\right)\cos^{2} 
\theta_{1}
+\left( m_{e}+m_{\mu}  \right)\sin^{2} 
\theta_{1} -
\FFr{\bar{f}_{1}}{2}
\left(  m_{\tau}+m_{\mu}+\right.\right.\\
\left.\left.
2m_{e}\right)\sin 2\theta_{1}
\right\},
\end{array}
\end{equation}
where 
$\bar{f}_{i}\equiv \bar{f}(\tau_{i}, y, f_{i})\quad (i=1,2,3)$.
From the known values of the 
$\eta^{l}$,
angles $\theta_{i}$ and
masses of quarks one may obtain the values of $12$ quantities
$\tau_{i}, y, f_{i}, \eta^{q},\bar{m}_{s}, \bar{m}_{c}, \bar{m}_{b}, 
\bar{m}_{t}$,
while the explicit form of the functions
$\bar{f}(\tau_{i}, y, f_{i})$ can be inferred from the equation of 
renormalization group. For instance, in approximation to the lowest order 
(sec.16) this
equation reduced to Gell-Mann-Low's equation, which in the scope of
perturbation theory
$
\bar{f}(\tau, f)=f\,d(\tau, f)=f+\FFr{f^{2}\tau}{3\pi}+
\FFr{f^{3}}{9\pi^{2}}\left( \tau^{2}+\FFr{9}{4}\tau\right)+\cdots
$
gives the leading terms as follows: 
$
d(\tau, f)^{-1}=1-\FFr{f\tau}{3\pi}+
\FFr{3f}{4\pi}\ln\,\left(1-\FFr{f\tau}{3\pi}\right)+\cdots.
$ [17].
\section{Concluding Remarks}
\label{Conc}
Continuing the program developed in a previous paper [1]
we attempt to suggest a microscopic approach to the properties of particles 
and interactions. Within this approach the fields have composite nontrivial 
internal structure. The condition of
realization of multiworld connections is arisen due to the symmetry of 
Q-world of electric charge and embodied in the Gell-Mann-Nishijima relation. 
During a realization of multiworld structure the symmetries of 
corresponding internal worlds are unified into more higher symmetry 
including also the operators of isospin and hypercharge.
Such approach enables to conclude that
possible three lepton generations consist of six lepton fields
with integer electric and leptonic charges and being free of confinement
condition. Also three quark 
generations exist composed of six possible quark fields. They
carry fractional electric and baryonic charges and obey confinement 
condition. The global group unifying all global symmetries of the internal
worlds of quarks is the flavour group $SU_{f}(6)$.
The whole complexity of leptons, quarks and other composite particles, 
and their interactions arises from the primary
field, which has nontrivial multiworld internal structure
and involves nonlinear fermion self-interaction of the components. 
This Lagrangian
contains only two free parameters, which are the coupling constants of
nonlinear fermion and gauge interactions.
Due to specific structure of W-world of weak interactions implying the 
condition of realization of multiworld connections, the spanning eq.(12.1) 
takes place, which underlies the P-violation in W-world. 
It is expressed in the reduction of initial symmetry of the right-handed 
subquarks. Such reduction is characterized by the Weinberg mixing angle 
with the value fixed at $30^{0}$. It gives rise to 
the expanded local symmetry $SU(2)\otimes U(1)$, under which the 
left-handed fermions transform as six independent $SU(2)$ doublets, while
the right-handed fermions transform as twelve independent singlets.
Due to vacuum rearrangement in Q-world the Yukawa couplings arise between 
the fermion fields and corresponding isospinor-scalar $\varphi$-meson in
conventional form. We suggest the microscopic approach to Higgs 
bosons with self-interaction and Yukawa couplings. It involves the Higgs 
bosons as the collective excitations of bound quasi-particle iso-pairs.
In the framework of local gauge invariance of the theory incorporated with 
the P-violation in weak interactions we propose a mechanism providing the
Bose-condensation of iso-pairs, which is due to effective attraction
between the relativistic fermions caused by the exchange of the mediating
induced gauge quanta in W-world. The relativism qualitatively affects 
the value of mass of Higgs boson, coupling constant and transition 
temperature. The latter decreases inversely by relativistic factor.
We consider the four-component Bose-condensate, where
due to self-interaction its spin part is vanished.
Based on it we show that the field of 
symmetry-breaking Higgs boson always must be counted 
off from the gap symmetry restoring value as
the point of origin. Then the Higgs boson describes the excitations in the
neighbourhood of stable vacuum of the W-world. 
The resulting Lagrangian of unified electroweak 
interactions of leptons and quarks ensues, which 
in lowest order approximation
leads to the Lagrangian of phenomenological standard model. In general,
the self-energy operator underlies the Yukawa 
coupling constant, which takes into account a mass-spectrum of all expected
collective excitations of bound quasi-particle pairs arising as the
poles.
The implication of quarks into this scheme is carried out in the same manner
except that of appearance of quark mixing with Cabibbo angle
and the existence of CP-violating complex phase in unitary matrix of
quark mixing. The
Q-components of the quarks $u',c'$ and $t'$ contain at least 
one identical subquark, due to which the partial formfactors 
gain nonzero values. This underlies the quark mixing with 
Cabibbo angles. In lepton's case these formfactors
are vanished  and lepton mixing is absent.
The CP-violation stems from the spanning
eq.(18.2) incorporated with the expanded group of global rotations.
With a simple viewpoint on Higgs sector the masses of leptons and quarks 
are given in sec.19, which lead to some useful relations
between the masses of leptons and quarks. 
From these relations
one may define the values of $12$ quantities
$\tau_{i}, y, f_{i},\eta^{q}, \bar{m}_{s}, \bar{m}_{c}, \bar{m}_{b}, 
\bar{m}_{t}$,
where the explicit form of the functions
$\bar{f}(\tau_{i}, y, f_{i})$ can be inferred from the equation of 
renormalization group.\\
Although some key problems of particle physics are elucidated within outlined
approach, nevertheless the numerous issues still remain to be solved. We
hope that this approach will be an attractive basis for the future theory.

\centerline {\bf\large Acknowledgements}
\vskip 0.1\baselineskip
\noindent
It is pleasure to express my gratitude to G.Jona-Lasinio for fruitful
comments and valuable suggestions on the various issues treated in this paper.
I'm also indebted to V.Gurzadyan, A.M.Vardanian and K.L.Yerknapetian
for support.


\begin {thebibliography}{99}
\bibitem {A1} G.T.Ter-Kazarian, {\em hep-th/9812181}.
\bibitem {A10} W.Heisenberg, {\em Annual Intl.Conf. on High Energy Physics
at CERN}, CERN, Scientific Information Service, Geneva, 1958.
\bibitem {A11} H.P.D\"{u}rr, W.Heisenberg , H.Mitter, S.Schlieder,
R.Yamayaki , {\em Z. Naturforsch.}, {\bf 14a} 441 (1959); {\bf 16a}
726 (1961).
\bibitem {A12} S.L.Glashow, {\em Nucl. Phys.}, {\bf 22}, 579 (1961). 
\bibitem {A13} S.Weinberg, {\em Phys.Rev.Lett.}, {\bf 19} 1264 (1967).
\bibitem {A14} A.Salam, `{\em Elemenmtary Particle Theory}, p.367,Ed.N.
Svartholm.-Almquist and Wiksell, 1968.
\bibitem {A15} J.L.Lopes , {\em Nucl. Phys.}, {\bf 8} 234 (1958). 
\bibitem {A16} J.Schwinger, {\em Ann.of Phys.}, {\bf 2} 407 (1957). 
\bibitem {A17} J.Bardeen, L.N.Cooper, J.R.Schriefer, {\em Phys.Rev.}, 
{\bf 106} 162 (1957); {\bf 108} 5 (1957).
\bibitem {A18} N.N.Bogoliubov, {\em Zh.Eksperim. i Teor.Fiz}, 
{\bf 34} 735 (1958).
\bibitem {A19} N.N.Bogoliubov, V.V.Tolmachev, D.V.Shirkov,
{\em A New Method in the Theory of Superconductivity}, Akd. of Science
of U.S.S.R., Moscow, 1958.
\bibitem{A23} D.V.Shirkov, ``Invariance-Autoduality-Renormgroup'',pp.457-463
in {\em Principle of Invariance and its Applications}, Yerevan 1989
\bibitem {A24} M.Gell-Mann, F.Low., {\em Phys.Rew.}, {\bf 95} 1300 (1954).
\bibitem {A25} N.N.Bogoliubov, D.V.Shirkov, {\em DAN SSSR}, {\bf 103} 
203 (1955).
\bibitem {A26} M.Mnatsakanian, {\em Comm. Byurakan Obs.} {\bf 50} 69 (1978);
{\em DAN SSSR}, {\bf 262} 856 (1982).
\bibitem {A27} V.Ambartsumian, ``The Theory of Stellar Spectra'', Nauka, Moscow,
1966.
\bibitem {A28} N.N.Bogoliubov, D.V.Shirkov D.V.,{\em Introduction to the 
theory of quantized fields}, Moscow, 1978.

\bibitem {A29} Y.Nambu Y, G.Jona-Lasinio, {\em Phys.Rev.} {\bf 122} 345 (1961); 
{\bf 124} 246 (1961).
\bibitem {A30} V.Vaks, A.J.Larkin, {\em Zh.Eksperim. i Teor.Fiz}, 
{\bf 40} 282 (1961).
\bibitem {A31} L.P.Gor'kov, {\em Zh.Eksperim. i Teor.Fiz}, 
{\bf 34} 735 (1958).
\bibitem {A32} L.P.Gor'kov, {\em Zh.Eksperim. i Teor.Fiz}, 
{\bf 36}, 1918 (1959).
\bibitem{A33} D.Pines D., J.R.Schrieffer, {\em Nuovo Cimento},
{\bf 10} 496 (1958).
\bibitem {A34} P.W.Anderson, {\em Phys.Rev.}, {\bf 110}, 827, 1900 (1958);
{\bf 114} 1002 (1959). 
\bibitem {A35} G.Rickayzen, {\em Phys.Rev.}{\bf 115} 795 (1959).
\bibitem {A36} Y.Nambu, {\em Phys.Rev.} {\bf 117} 648 (1960).
\bibitem {A37} V.M.Galitzki' , {\em Zh.Eksperim. i Teor.Fiz}, 
{\bf 34} 1011 (1958).
\bibitem {A38} H.Fritzsh, M.Gell-Mann, In {\em 16th Intl. Conf.
on High Energy Physic}s, Chikago-Batawia, {\bf 2}, 135 1972.
\bibitem {A39} S.Weinberg, {\em Phys.Rev.Lett.} {\bf 31} 494 (1973);
{\em Phys.Rev.}, {\bf D8} 4482 (1973); {\bf D5} 1962 (1972).
\bibitem {A40} D.J.Gross, F.Wilczek, {\em Phys.Rev.}, 
{\bf 8} 3633 (1973).
\bibitem {A41} W.Marciano, H.Pagels, {\em Phys. Rep.}, {\bf 36} 137 (1978).
\bibitem {A42} S.D.Drell, {\em Am.J.Phys.}, {\bf 46} 597 (1978).
\bibitem {A43} D.J.Gross, R.D.Pisarski, L.G.Yaffe, {\em Rev.Mod.Phys.}, 
{\bf 53} 43 (1981).
\bibitem {A44} A.J.Buras, {\em Rev.Mod.Phys.}, 
{\bf 52} 199 (1980).
\bibitem {A45} G.Altarelli, {\em Phys. Rep.}, {\bf 81} 1 (1982).
\bibitem {A46} A.H.Mueller, {\em Phys. Rep.},  {\bf 73} 237 (1981).
\bibitem {A47} E.Reya, {\em Phys. Rep.}, {\bf 69} 195 (1981).
\bibitem{A44} J.Goldstone, {\em Nuovo Cimento}, {\bf 19} 154 (1961).
\bibitem {A48} J.Goldstone, S.Weinberg, A.Salam,
{\em Phys.Rev.},  {\bf 127} 965 (1962).
\bibitem {A50} S.A.Bludman, {\em Phys.Rev.},{\bf 131} 2364 (1963).
\bibitem {A51} P.W.Higgs, {\em Phys.Letters}, {\bf 12} 132 (1964).
\bibitem {A52} P.W.Higgs, {\em Phys.Rev.}, {\bf 140} B911 (1965);
1966, {\bf 145}, 1156.
\bibitem {A53} F.Bloch, {\em Z.Physik}, {\bf 61} 206 (1930);
{\bf 74} 295 (1932).
\bibitem {A54} L.D.Landau, E.M.Lifschitz, {\em Statistical Physics}, Part II,
Nauka, Moscow 1978.
\bibitem {A55} L.N.Cooper, {\em Phys.Rev.}, 1956, {\bf 104} 1189 (1956).
\bibitem {A56} H.Fr\"{o}hlich, {\em Phys.Rev.}, {\bf 79} 845 (1956).
\bibitem {A57} V.L.Ginzburg, L.D.Landau, {\em Zh.Eksperim. i Teor.Fiz}, 
{\bf 20} 1064 (1950).
\bibitem {A58} N.R.Werthamer, {\em Phys.Rev.}, {\bf 132}, 663 (1963).
\bibitem {A59} T.Tsuzuki, {\em Progr.Theoret.Phys.} (Kyoto), 
{\bf 31} 388 (1964).
\bibitem {A60} L.Tewordt, {\em Phys.Rev.}, {\bf 132}, 595 (1963).
\bibitem {A61} V.V.Shmidt, {\em Introduction to the Physics of
Superconductors}, Nauka, Moscow 1982.
\bibitem {A62} T.Matsubara, {\em Progr.Theoret.Phys.} (Kyoto), 
{\bf 14} 352 (1954).
\bibitem {A63} A.A.Abrikosov, L.P.Gor'kov, J.E.Dzyaloshinski, 
{\em Zh.Eksperim. i Teor.Fiz}, {\bf 36} 900 (1959).
\bibitem{A64} J.G.Valatin, {\em Nuovo Cimento}, {\bf 7} 843 (1958).
\bibitem {A65} G.A.Baraff, S.Borowitz, {\em Phys.Rev.},  {\bf 121} 1704 (1961).
\bibitem {A66} D.F.DuBois, M.G.Kivelson, {\em Phys.Rev.}, {\bf 127} 1182
(1962).
\bibitem {A67} G.'t Hooft, {\em Nucl.Phys.}, {\bf B35}, 167 (1971).
\bibitem {A68} B.W.Lee, {\em Phys.Rev.}, {\bf D5} 823 (1972).
\bibitem {A69} L.B.Okun, {\em Leptons and Quarks}, Nauka, Moscow, 1989.
\bibitem {A70} K.Gotow, S.Okubo, {\em Phys.Rev.}, {\bf 128} 1921 (1962).
\bibitem {A71} J.Leitner, S.Okubo, {\em Phys.Rev.}, {\bf 136B} 1542 (1964).
\bibitem {A72} H.Georgi, {\em Weak Interactions and Modern Particle
Theory}-The Benjamin/Cummings Co.1984
\bibitem {A73} T-P Cheng, L-F Li, {\em Gauge Theory of Elementary Particle
Physics}, Clarendon Press, Oxford, 1984.
\bibitem {A74} R.G.Sachs, {\em Phys.Rev.}, {\bf 129} 2280 (1963).
\bibitem {A75} J.H.Christenson, J.M.Cronin, V.L.Fitch, R.Turlay,
{\em Phys.Rev.Letters}, {\bf 13} 138 (1964).
\bibitem {A76} M.Kobayashi, T.Maskawa, {\em Progr.Theoret.Phys.} (Kyoto), 
{\bf 49} 652 (1973).
\end {thebibliography}
\end{document}